\documentclass[aps,prd,floats,floatfix,nofootinbib,superscriptaddress]{revtex4-1}
\usepackage{graphicx,amsmath,amssymb,amsfonts,hyperref,mathdots,caption,subcaption, setspace, wrapfig,slashed,dsfont}
\usepackage[usenames,dvipsnames,svgnames,table]{xcolor}
\usepackage{graphicx,etoolbox}

\makeatletter
\patchcmd{\Gin@setfile}
{\ProvidesFile}
{\xdef\imgfilename{#3}\ProvidesFile}
{}{}
\makeatother

\newcommand{\Tr}[1]{\ensuremath{\,\mathrm{Tr}\left[{#1}\right]}}
\newcommand{\ct}{\ensuremath{^{\dagger}}}
\newcolumntype{L}[1]{>{\raggedright\let\newline\\\arraybackslash\hspace{0pt}}m{#1}}
\newcolumntype{C}[1]{>{\centering\let\newline\\\arraybackslash\hspace{0pt}}m{#1}}
\newcolumntype{R}[1]{>{\raggedleft\let\newline\\\arraybackslash\hspace{0pt}}m{#1}}
\begin{document}

\title{Custodial symmetry violation in the Georgi-Machacek model}

\author{Ben Keeshan}
\email{BenKeeshan@cmail.carleton.ca}
\affiliation{Ottawa-Carleton Institute for Physics, Carleton University, Ottawa, Ontario K1S 5B6, Canada}

\author{Heather E.~Logan}
\email{logan@physics.carleton.ca}
\affiliation{Ottawa-Carleton Institute for Physics, Carleton University, Ottawa, Ontario K1S 5B6, Canada}

\author{Terry Pilkington}
\email{Terence.Pilkington@fuw.edu.pl}
\affiliation{Ottawa-Carleton Institute for Physics, Carleton University, Ottawa, Ontario K1S 5B6, Canada}
\affiliation{Institute of Theoretical Physics, Faculty of Physics, University of Warsaw, ul.~Pasteura 5, PL--02--093 Warsaw, Poland}

\date{January 13, 2020}

\begin{abstract}
We study the effects of custodial symmetry violation in the Georgi-Machacek (GM) model. The GM model adds isospin-triplet scalars to the Standard Model in a way that preserves custodial symmetry at tree level; however, this custodial symmetry has long been known to be violated at the one-loop level by hypercharge interactions.  We consider the custodial-symmetric GM model to arise at some high scale as a result of an unspecified ultraviolet completion, and quantify the custodial symmetry violation induced as the model is run down to the weak scale.  The measured value of the eletroweak $\rho$ parameter (along with perturbative unitarity) lets us constrain the scale of the ultraviolet completion to lie below tens to hundreds of TeV over almost all of the parameter space. Subject to this constraint, we quantify the size of other custodial-symmetry-violating effects at the weak scale, including custodial symmetry violation in the couplings of the 125~GeV Higgs boson to $W$ and $Z$ boson pairs and mixings and mass splittings among the additional Higgs bosons in the theory.  We find that these effects are small enough that they are unlikely to be probed by the Large Hadron Collider (LHC), but may be detectable at a future $e^+e^-$ collider. We note that the upper bound on the scale of the ultraviolet completion is large enough that virtual effects from the ultraviolet completion will also be undetectable at the LHC. This means that the GM model is a valid effective theory for LHC physics.
\end{abstract}

\maketitle 

\section{Introduction}

With the discovery of a Standard Model (SM)-like Higgs boson at the CERN Large Hadron Collider (LHC) in 2012~\cite{Aad:2012tfa}, we have the first direct access to the dynamics of electroweak symmetry breaking.  The simplest implementation of this dynamics is through a single complex scalar field transforming as a doublet under the weak SU(2)$_L$ gauge symmetry; this is consistent with experimental data to date~\cite{Khachatryan:2016vau}.

While at least one SU(2)$_L$ doublet is required to generate the masses of the SM fermions in a gauge-invariant way, the masses of the $W$ and $Z$ bosons can in principle also receive contributions from scalars in larger representations of SU(2)$_L$.  Such an extension to the Higgs sector is severely constrained by measurements of the $\rho$ parameter~\cite{Ross:1975fq}, defined as the ratio of the strengths of the neutral and charged weak currents in the low-energy limit and measured to very high precision via the global electroweak fit~\cite{Olive:2016xmw}.  Indeed, unless the vacuum expectation values (vevs) of the larger representations are negligibly small, the only viable models are those that preserve $\rho = 1$ at tree level:
\begin{enumerate}
	\item[\it i.] models with extra SU(2)$_L$ doublet(s) and/or singlet(s);
	\item[\it ii.] a model with an extra SU(2)$_L$ septet with appropriately-chosen hypercharge~\cite{Hisano:2013sn,Kanemura:2013mc}; and
	\item[\it iii.] the Georgi-Machacek (GM) model~\cite{Georgi:1985nv,Chanowitz:1985ug} and its generalizations to larger SU(2)$_L$ representations~\cite{Galison:1983qg,Robinett:1985ec,Logan:1999if,Chang:2012gn,Logan:2015xpa}.
\end{enumerate}

In this paper we consider the GM model.  In addition to the usual SU(2)$_L$ doublet, this model contains two SU(2)$_L$-triplet scalar fields, arranged in such a way that the scalar potential is invariant under a global SU(2)$_L \times$SU(2)$_R$ symmetry; upon electroweak symmetry breaking, this global symmetry breaks down to its diagonal subgroup [known as the custodial SU(2)] and $\rho = 1$ is thereby preserved.  The GM model gives rise to a rich and exotic phenomenology, including singly- and doubly-charged scalars that couple to vector boson pairs at tree level and the possibility that the SM-like Higgs boson's couplings to $WW$ and $ZZ$ could be larger than in the SM.  It has been used as a benchmark by the LHC experiments for interpreting searches for singly-charged Higgs bosons decaying into vector boson pairs~\cite{Aad:2015nfa,Sirunyan:2017sbn}.

However, it has been known since the early '90s that the custodial symmetry in the GM model holds only at tree level~\cite{Gunion:1990dt}: the global SU(2)$_R$ symmetry is explicitly violated by the gauging of hypercharge, which leads to an uncontrolled violation of the custodial symmetry at one loop.  The most obvious manifestation of this is that the standard calculation of the Peskin-Takeuchi $T$ parameter~\cite{Peskin:1990zt} yields an infinite result; this infinity is to be cancelled by a counterterm that is absent in the SU(2)$_L \times$SU(2)$_R$-invariant potential of the GM model but appears in the full gauge-invariant but custodial-symmetry-violating theory~\cite{Gunion:1990dt}.  

A further manifestation, most relevant for our purposes, is that it is not possible to compute a consistent set of renormalization group equations (RGEs) for the Lagrangian parameters of the custodial-symmetric GM model unless one sets the hypercharge gauge coupling to zero~\cite{Blasi:2017xmc}.  This implies that it is possible to choose the Lagrangian parameters to preserve the custodial symmetry, but \emph{only at one energy scale}.  In order to run away from that special scale, one must use the RGEs computed in the full gauge-invariant but custodial-symmetry-violating potential; the hypercharge contribution then causes custodial symmetry violation to build up as one runs.  Reference~\cite{Blasi:2017xmc} studied this effect by assuming that the theory is custodial-symmetric at the weak scale and quantifying the amount of custodial symmetry violation that develops as one runs to higher scales.

In this paper we take a different approach.  We imagine that the custodial-symmetric GM model arises at some high scale, for example as a theory of composite scalars with an accidental global SU(2)$_L \times$SU(2)$_R$ symmetry in the scalar sector.  (Such models have been constructed in the context of little Higgs theories in Refs.~\cite{Chang:2003un,Chang:2003zn}.)  Below the compositeness scale, custodial symmetry violation accumulates through the running of the Lagrangian parameters down to the weak scale.  Weak-scale measurements of the $\rho$ parameter can then be used to constrain how high the custodial-symmetric scale can be.  Subject to this constraint, we can also quantify the physical effects of custodial symmetry violation in Higgs-sector observables, such as the ratio of the SM-like Higgs boson couplings to $WW$ and $ZZ$ and custodial-violating mixings and mass splittings among the additional scalars in the GM model.  We will show that the custodial-symmetric scale can be as high as tens to hundreds of TeV, and that the effects of custodial symmetry violation at the weak scale are typically too small to be detected at the LHC.  The custodial-violation-induced mass splittings may however be detectable at a future $e^+e^-$ collider.  The fermiophobic scalars of the GM model acquire small fermion couplings due to custodial-violation-induced mixing, but the resulting branching ratios remain subdominant even for scalar masses below about 160~GeV, where fermionic decays could compete against the loop-induced diphoton decays that otherwise put strong experimental constraints on such light scalars.

Because our main objective is to quantify the custodial symmetry violation allowed in the model given the stringent experimental constraints on the $\rho$ parameter, we find it sufficient to work in the leading log approximation---i.e., we use one-loop RGEs and tree-level matching.  This is justified by the tiny size of the custodial-violating effects that we find over most of the parameter space.  Larger custodial-violating effects arise when scalar masses in the custodial-symmetric theory are tuned to be nearly degenerate, so that custodial symmetry violation induces resonant mixing among mass eigenstates.  We handle these situations by exactly diagonalizing the resulting mass matrices; nevertheless, in the small regions of parameter space around these resonances our perturbative calculation remains unstable.

This paper is organized as follows.  In Sec.~\ref{sec:model} we review the GM model with exact custodial symmetry in order to set our notation.  In Sec.~\ref{sec:potential} we write down the most general gauge invariant scalar potential for the custodial-violating theory with the same field content.  In Sec.~\ref{sec:physmasses} we compute the masses and mixing angles of the physical scalars in the custodial-violating theory and derive formulas for the most interesting custodial-violating couplings.  In Sec.~\ref{sec:numerics} we describe our calculational procedure and give our numerical results, using full scans of the parameter space as well as a convenient benchmark plane for ease of interpretation.  In Sec.~\ref{sec:conclusions} we conclude.  In Appendix~\ref{sec:rge} we collect the one-loop RGEs for the custodial-violating theory and give a translation between our notation and that of Ref.~\cite{Blasi:2017xmc}.  In Appendix~\ref{ap:SCOUP} we collect the expressions for triple scalar couplings in the custodial-violating theory.  Finally in Appendix~\ref{app:CEL} we give some details of our calculation method for the RGEs.

\section{Georgi-Machacek model with exact custodial symmetry}
\label{sec:model}
 
The scalar sector of the GM model~\cite{Georgi:1985nv,Chanowitz:1985ug} consists of the usual complex doublet $(\phi^+,\phi^0)$ with hypercharge\footnote{We use $Q = T^3 + Y/2$.} $Y = 1$, a real triplet $(\xi^+,\xi^0,\xi^-)$ with $Y = 0$, and  a complex triplet $(\chi^{++},\chi^+,\chi^0)$ with $Y=2$.  The doublet is responsible for the fermion masses as in the SM.
In order to make the global SU(2)$_L \times$SU(2)$_R$ symmetry explicit, we write the doublet in the form of a bidoublet $\Phi$ and combine the triplets to form a bitriplet $X$:
\begin{equation}
	\Phi = \left( \begin{array}{cc}
	\phi^{0*} &\phi^+  \\
	-\phi^{+*} & \phi^0  \end{array} \right), \qquad
	X =
	\left(
	\begin{array}{ccc}
	\chi^{0*} & \xi^+ & \chi^{++} \\
	 -\chi^{+*} & \xi^{0} & \chi^+ \\
	 \chi^{++*} & -\xi^{+*} & \chi^0  
	\end{array}
	\right).
	\label{eq:PX}
\end{equation}
The vevs are defined by $\langle \Phi  \rangle = \frac{ v_{\phi}}{\sqrt{2}} I_{2\times2}$  and $\langle X \rangle = v_{\chi} I_{3 \times 3}$, where $I$ is the appropriate identity matrix and the $W$ and $Z$ boson masses constrain
\begin{equation}
	v_{\phi}^2 + 8 v_{\chi}^2 \equiv v^2 = \frac{1}{\sqrt{2} G_F} \approx (246~{\rm GeV})^2,
	\label{eq:vevrelation}
\end{equation} 
where $G_F$ is the Fermi constant. 

Upon electroweak symmetry breaking, the global SU(2)$_L \times $SU(2)$_R$ symmetry breaks down to the diagonal subgroup, which is the custodial SU(2) symmetry.

The most general gauge-invariant scalar potential involving these fields that conserves custodial SU(2) is given, in the conventions of Ref.~\cite{Hartling:2014zca}, by\footnote{A translation table to other parameterizations in the literature has been given in the appendix of Ref.~\cite{Hartling:2014zca}.}
\begin{eqnarray}
	V(\Phi,X) &= & \frac{\mu_2^2}{2}  \text{Tr}(\Phi^\dagger \Phi) 
	+  \frac{\mu_3^2}{2}  \text{Tr}(X^\dagger X)  
	+ \lambda_1 [\text{Tr}(\Phi^\dagger \Phi)]^2  
	+ \lambda_2 \text{Tr}(\Phi^\dagger \Phi) \text{Tr}(X^\dagger X)   \nonumber \\
          & & + \lambda_3 \text{Tr}(X^\dagger X X^\dagger X)  
          + \lambda_4 [\text{Tr}(X^\dagger X)]^2 
           - \lambda_5 \text{Tr}( \Phi^\dagger \tau^a \Phi \tau^b) \text{Tr}( X^\dagger t^a X t^b) 
           \nonumber \\
           & & - M_1 \text{Tr}(\Phi^\dagger \tau^a \Phi \tau^b)(U X U^\dagger)_{ab}  
           -  M_2 \text{Tr}(X^\dagger t^a X t^b)(U X U^\dagger)_{ab}.
           \label{eq:potential}
\end{eqnarray} 
Here the SU(2) generators for the doublet representation are $\tau^a = \sigma^a/2$ with $\sigma^a$ being the Pauli matrices, the generators for the triplet representation are
\begin{equation}
	t^1= \frac{1}{\sqrt{2}} \left( \begin{array}{ccc}
	 0 & 1  & 0  \\
	  1 & 0  & 1  \\
	  0 & 1  & 0 \end{array} \right), \qquad  
	  t^2= \frac{1}{\sqrt{2}} \left( \begin{array}{ccc}
	 0 & -i  & 0  \\
	  i & 0  & -i  \\
	  0 & i  & 0 \end{array} \right), \qquad 
	t^3= \left( \begin{array}{ccc}
	 1 & 0  & 0  \\
	  0 & 0  & 0  \\
	  0 & 0 & -1 \end{array} \right),
\end{equation}
and the matrix $U$, which rotates $X$ into the Cartesian basis, is given by~\cite{Aoki:2007ah}
\begin{equation}
	 U = \left( \begin{array}{ccc}
	- \frac{1}{\sqrt{2}} & 0 &  \frac{1}{\sqrt{2}} \\
	 - \frac{i}{\sqrt{2}} & 0  &   - \frac{i}{\sqrt{2}} \\
	   0 & 1 & 0 \end{array} \right).
	 \label{eq:U}
\end{equation}

The minimization conditions for the scalar potential read
\begin{eqnarray}
	0 = \frac{\partial V}{\partial v_{\phi}} &=& 
	v_{\phi} \left[ \mu_2^2 + 4 \lambda_1 v_{\phi}^2 
	+ 3 \left( 2 \lambda_2 - \lambda_5 \right) v_{\chi}^2 - \frac{3}{2} M_1 v_{\chi} \right], 
		\nonumber \\
	0 = \frac{\partial V}{\partial v_{\chi}} &=& 
	3 \mu_3^2 v_{\chi} + 3 \left( 2 \lambda_2 - \lambda_5 \right) v_{\phi}^2 v_{\chi}
	+ 12 \left( \lambda_3 + 3 \lambda_4 \right) v_{\chi}^3
	- \frac{3}{4} M_1 v_{\phi}^2 - 18 M_2 v_{\chi}^2.
	\label{eq:mincond}
\end{eqnarray}

The physical fields can be organized by their transformation properties under the custodial SU(2) symmetry into a fiveplet, a triplet, and two singlets.  The fiveplet and triplet states are given by
\begin{eqnarray}
	&&H_5^{++} = \chi^{++}, \qquad
	H_5^+ = \frac{\left(\chi^+ - \xi^+\right)}{\sqrt{2}}, \qquad
	H_5^0 = \sqrt{\frac{2}{3}} \xi^{0,r} - \sqrt{\frac{1}{3}} \chi^{0,r}, \nonumber \\
	&&H_3^+ = - s_H \phi^+ + c_H \frac{\left(\chi^++\xi^+\right)}{\sqrt{2}}, \qquad
	H_3^0 = - s_H \phi^{0,i} + c_H \chi^{0,i},
\end{eqnarray}
where the vevs are parameterized by
\begin{equation}
	c_H \equiv \cos\theta_H = \frac{v_{\phi}}{v}, \qquad
	s_H \equiv \sin\theta_H = \frac{2\sqrt{2}\,v_\chi}{v},
\end{equation}
and we have decomposed the neutral fields into real and imaginary parts according to
\begin{equation}
	\phi^0 \to \frac{v_{\phi}}{\sqrt{2}} + \frac{\phi^{0,r} + i \phi^{0,i}}{\sqrt{2}},
	\qquad
	\chi^0 \to v_{\chi} + \frac{\chi^{0,r} + i \chi^{0,i}}{\sqrt{2}}, 
	\qquad
	\xi^0 \to v_{\chi} + \xi^{0,r}.
	\label{eq:decomposition}
\end{equation}
The masses within each custodial multiplet are degenerate at tree level and can be written (after eliminating $\mu_2^2$ and $\mu_3^2$ in favor of the vevs) as\footnote{Note that the ratio $M_1/v_{\chi}$ is finite in the limit $v_{\chi} \to 0$, 
\begin{equation}
	\frac{M_1}{v_{\chi}} = \frac{4}{v_{\phi}^2} 
	\left[ \mu_3^2 + (2 \lambda_2 - \lambda_5) v_{\phi}^2 
	+ 4(\lambda_3 + 3 \lambda_4) v_{\chi}^2 - 6 M_2 v_{\chi} \right],
\end{equation}
which follows from the minimization condition $\partial V/\partial v_{\chi} = 0$.}
\begin{eqnarray}
	m_5^2 &=& \frac{M_1}{4 v_{\chi}} v_\phi^2 + 12 M_2 v_{\chi} 
	+ \frac{3}{2} \lambda_5 v_{\phi}^2 + 8 \lambda_3 v_{\chi}^2, \nonumber \\
	m_3^2 &=&  \frac{M_1}{4 v_{\chi}} (v_\phi^2 + 8 v_{\chi}^2) 
	+ \frac{\lambda_5}{2} (v_{\phi}^2 + 8 v_{\chi}^2) 
	= \left(  \frac{M_1}{4 v_{\chi}} + \frac{\lambda_5}{2} \right) v^2.
\end{eqnarray}

The two custodial SU(2)--singlet mass eigenstates are given by
\begin{equation}
	h = \cos \alpha \, \phi^{0,r} - \sin \alpha \, H_1^{0\prime},  \qquad
	H = \sin \alpha \, \phi^{0,r} + \cos \alpha \, H_1^{0\prime},
	\label{mh-mH}
\end{equation}
where 
\begin{equation}
	H_1^{0 \prime} = \sqrt{\frac{1}{3}} \xi^{0,r} + \sqrt{\frac{2}{3}} \chi^{0,r}.
\end{equation}
Their mixing angle and masses are given by
\begin{eqnarray}
	&&\sin 2 \alpha =  \frac{2 \mathcal{M}^2_{12}}{m_H^2 - m_h^2},    \qquad
	\cos 2 \alpha =  \frac{ \mathcal{M}^2_{22} - \mathcal{M}^2_{11}  }{m_H^2 - m_h^2}, 
	\nonumber \\
	&&m^2_{h,H} = \frac{1}{2} \left[ \mathcal{M}_{11}^2 + \mathcal{M}_{22}^2
	\mp \sqrt{\left( \mathcal{M}_{11}^2 - \mathcal{M}_{22}^2 \right)^2 
	+ 4 \left( \mathcal{M}_{12}^2 \right)^2} \right],
	\label{eq:hmass}
\end{eqnarray}
where we choose $m_h < m_H$, and 
\begin{eqnarray}
	\mathcal{M}_{11}^2 &=& 8 \lambda_1 v_{\phi}^2, \nonumber \\
	\mathcal{M}_{12}^2 &=& \frac{\sqrt{3}}{2} v_{\phi} 
	\left[ - M_1 + 4 \left(2 \lambda_2 - \lambda_5 \right) v_{\chi} \right], \nonumber \\
	\mathcal{M}_{22}^2 &=& \frac{M_1 v_{\phi}^2}{4 v_{\chi}} - 6 M_2 v_{\chi} 
	+ 8 \left( \lambda_3 + 3 \lambda_4 \right) v_{\chi}^2.
\end{eqnarray}

We will later apply constraints on the parameters of the custodial-symmetric scalar potential from perturbative unitarity of two-to-two scalar scattering amplitudes and bounded-from-belowness of the scalar potential.
Perturbative unitarity requires that the $\lambda_i$ obey the following relations~\cite{Aoki:2007ah,Hartling:2014zca}:
\begin{eqnarray}
	\sqrt{(6 \lambda_1 -7 \lambda_3 -11 \lambda_4)^2 +36 \lambda_2^2} + \left| 6 \lambda_1 + 7 \lambda_3 +11 \lambda_4 \right| &<& 4 \pi, \nonumber \\
	\sqrt{(2 \lambda_1 + \lambda_3 -2 \lambda_4)^2 + \lambda_5^2} + \left| 2 \lambda_1 - \lambda_3 +2 \lambda_4 \right| &<& 4 \pi, \nonumber \\
	\left| 2 \lambda_3 + \lambda_4 \right| &<& \pi, \nonumber \\
	\left| \lambda_2 - \lambda_5 \right| &<& 2 \pi.
	\label{eqn:pertUnitarity}
\end{eqnarray}
Requiring that the scalar potential is bounded from below imposes the following constraints~\cite{Hartling:2014zca}:
\begin{eqnarray}
	\lambda_1 &>& 0, \nonumber \\
	\lambda_4 &>& \left\{ \begin{array}{ll}
		-\frac{1}{3} \lambda_3 & \quad {\rm for} \ \lambda_3 \geq 0, \\
		-\lambda_3 & \quad {\rm for} \ \lambda_3 < 0, \end{array} \right. \nonumber \\
	\lambda_2 &>& \left\{ \begin{array}{ll}
		\frac{1}{2} \lambda_5 - 2 \sqrt{\lambda_1 (\frac{1}{3}\lambda_3 + \lambda_4)} &
			\quad {\rm for} \ \lambda_5 \geq 0 \ {\rm and} \ \lambda_3 \geq 0, \\
		\omega_+(\zeta) \lambda_5 - 2 \sqrt{\lambda_1 (\zeta \lambda_3 + \lambda_4)} &
			\quad {\rm for} \ \lambda_5 \geq 0 \ {\rm and} \ \lambda_3 < 0, \\
		\omega_-(\zeta) \lambda_5 - 2 \sqrt{\lambda_1 (\zeta \lambda_3 + \lambda_4)} &
			\quad {\rm for} \ \lambda_5 < 0,
		\end{array} \right. 
	\label{eq:BFB}
\end{eqnarray}
where
\begin{eqnarray}
	\omega_{\pm}(\zeta) &=& \frac{1}{6} (1 - B) \pm \frac{\sqrt{2}}{3} 
		\left[ (1-B) \left( \frac{1}{2} + B \right) \right]^{1/2}, \nonumber \\
	B &\equiv& \sqrt{\frac{3}{2} \left( \zeta - \frac{1}{3} \right)} \in [0,1],
\end{eqnarray}
and Eq.~(\ref{eq:BFB}) must be satisfied for all values of $\zeta \in [\frac{1}{3}, 1]$.

\section{Custodial violation and the most general gauge-invariant scalar potential}
\label{sec:potential}

In order to allow for custodial symmetry violation, we rewrite the scalar potential in Eq.~(\ref{eq:potential}) in the most general SU(2)$_L \times$U(1)$_Y$ gauge invariant form, following Ref.~\cite{Gunion:1990dt}.
We define the scalar fields in SU(2)$_L$ vector notation as
\begin{equation}
	\phi = \left( \begin{array}{c} \phi^+ \\ \phi^0 \end{array} \right), \qquad
	\chi = \left( \begin{array}{c} \chi^{++} \\ \chi^+ \\ \chi^0 \end{array} \right), \qquad
	\xi = \left( \begin{array}{c} \xi^+ \\ \xi^0 \\ -\xi^{+*} \end{array} \right),
\end{equation}
with vevs given by [compare Eq.~(\ref{eq:decomposition})],
\begin{equation}
	\phi^0 \to \frac{\tilde v_{\phi}}{\sqrt{2}} + \frac{\phi^{0,r} + i \phi^{0,i}}{\sqrt{2}},
	\qquad
	\chi^0 \to \tilde v_{\chi} + \frac{\chi^{0,r} + i \chi^{0,i}}{\sqrt{2}}, 
	\qquad
	\xi^0 \to \tilde v_{\xi} + \xi^{0,r}.
\end{equation}
We use tildes to denote the vevs, parameters, and mass eigenstates of the custodial-violating theory.
The vevs of these three fields will be determined by $G_F$ according to [compare Eq.~(\ref{eq:vevrelation})]
\begin{equation}
	\tilde v_{\phi}^2 + 4 \tilde v_{\chi}^2 + 4 \tilde v_{\xi}^2 \equiv \tilde v^2 = \frac{1}{\sqrt{2}G_F} = v^2,
\end{equation}
and will be constrained by the $\rho$ parameter, 
\begin{equation}
	\rho = \frac{\tilde v_{\phi}^2 + 4 \tilde v_{\chi}^2 + 4 \tilde v_{\xi}^2}
		{\tilde v_{\phi}^2 + 8 \tilde v_{\chi}^2}
		= \frac{v^2}{v^2 + 4 (\tilde v_{\chi}^2 - \tilde v_{\xi}^2)}.
	\label{eq:rhoCV}
\end{equation}

For convenience, we define the conjugate multiplets,
\begin{eqnarray}
	\tilde \phi &\equiv& C_2 \phi^* 
		= \left( \begin{array}{cc} 0 & 1 \\ -1 & 0 \end{array} \right) \phi^*
		= \left( \begin{array}{c} \phi^{0*} \\ -\phi^{+*} \end{array} \right) \nonumber \\
	\tilde \chi &\equiv& C_3 \chi^* 
		= \left( \begin{array}{ccc} 0 & 0 & 1 \\ 0 & -1 & 0 \\ 1 & 0 & 0 \end{array} \right) \chi^*
		= \left( \begin{array}{c} \chi^{0*} \\ -\chi^{+*} \\ \chi^{++*} \end{array} \right).
\end{eqnarray}
We also define the following matrix forms of the triplet fields,
\begin{eqnarray}
	\Delta_2 &\equiv& \sqrt{2} \tau^a U_{ai} \chi_i 
		= \left( \begin{array}{cc} \chi^+/\sqrt{2} & -\chi^{++} \\ 
			\chi^0 & -\chi^+/\sqrt{2} \end{array} \right), \nonumber \\
	\Delta_0 &\equiv& \sqrt{2} \tau^a U_{ai} \xi_i
		= \left( \begin{array}{cc} \xi^0/\sqrt{2} & -\xi^+ \\
			-\xi^{+*} & -\xi^0/\sqrt{2} \end{array} \right), \nonumber \\
	\overline \Delta_0 &\equiv& -t^a U_{ai} \xi_i
		= \left( \begin{array}{ccc} -\xi^0 & \xi^+ & 0 \\
			\xi^{+*} & 0 & \xi^+ \\ 0 & \xi^{+*} & \xi^0 \end{array} \right).
\end{eqnarray}

The most general gauge invariant scalar potential can then be written as
\begin{eqnarray}
	V(\phi, \chi, \xi) &=& \tilde \mu_2^2 \phi^{\dagger} \phi + \tilde \mu_3^{\prime 2} \chi^{\dagger} \chi 
		+ \frac{\tilde \mu_3^2}{2} \xi^{\dagger} \xi \nonumber \\
	&& + \tilde \lambda_1 (\phi^{\dagger} \phi)^2 
		+ \tilde \lambda_2 | \tilde \chi^{\dagger} \chi |^2 
		+ \tilde \lambda_3 (\phi^{\dagger} \tau^a \phi) (\chi^{\dagger} t^a \chi)
		+ \left[ \tilde \lambda_4 (\tilde \phi^{\dagger} \tau^a \phi)(\chi^{\dagger} t^a \xi) 
			+ {\rm h.c.} \right] \nonumber \\
	&&	+ \tilde \lambda_5 (\phi^{\dagger} \phi)(\chi^{\dagger} \chi)
		+ \tilde \lambda_6 (\phi^{\dagger} \phi)(\xi^{\dagger} \xi) 
		+ \tilde \lambda_7 (\chi^{\dagger} \chi)^2
		+ \tilde \lambda_8 (\xi^{\dagger} \xi)^2 
		+ \tilde \lambda_9 | \chi^{\dagger} \xi |^2
		+ \tilde \lambda_{10} (\chi^{\dagger} \chi) (\xi^{\dagger} \xi) \nonumber \\
	&& - \frac{1}{2} \left[ \tilde M_1^{\prime} \phi^{\dagger} \Delta_2 \tilde \phi + {\rm h.c.} \right]
		+ \frac{\tilde M_1}{\sqrt{2}} \phi^{\dagger} \Delta_0 \phi
		- 6 \tilde M_2 \chi^{\dagger} \overline \Delta_0 \chi.
	\label{eq:potential3}
\end{eqnarray}
Note that $\tilde \lambda_4$ and $\tilde M_1^{\prime}$ are complex in general, while the rest of the parameters are real.  
We have adopted the same notation as in Eq.~(3.2) of Ref.~\cite{Gunion:1990dt} for the coefficients of the quartic terms, and we have added the trilinear terms that were eliminated in Ref.~\cite{Gunion:1990dt} by the imposition of a $Z_2$ symmetry.  This scalar potential has also been written down (for real $\tilde \lambda_4$ and $\tilde M_1^{\prime}$) in Ref.~\cite{Blasi:2017xmc}; we give a translation table to their notation in Appendix~\ref{sec:rge}.

We note that the last term in Eq.~(\ref{eq:potential3}) can also be written as
\begin{equation}
	- 6 \tilde M_2 \chi^{\dagger} \overline \Delta_0 \chi = -6 \tilde M_2 \epsilon_{ijk} \tilde \chi_i \xi_j \chi_k,
\end{equation}
where $\epsilon_{ijk}$ is the totally antisymmetric tensor with $\epsilon_{123} = +1$. 

In the custodially-symmetric limit, the Lagrangian parameters in the gauge-invariant scalar potential in Eq.~(\ref{eq:potential3}) reduce to those in the custodially-symmetric potential in Eq.~(\ref{eq:potential}) according to
\begin{eqnarray}
	\tilde \mu_2^2 &=& \mu_2^2 \nonumber \\
	\tilde \mu_3^{\prime 2} &=& \mu_3^2 \nonumber \\
	\tilde \mu_3^2 &=& \mu_3^2 \nonumber \\
	\tilde \lambda_1 &=& 4 \lambda_1 \nonumber \\
	\tilde \lambda_2 &=& 2 \lambda_3 \nonumber \\
	\tilde \lambda_3 &=& -2 \lambda_5 \nonumber \\
	\tilde \lambda_4 &=& - \sqrt{2} \lambda_5 \nonumber \\
	\tilde \lambda_5 &=& 4 \lambda_2 \nonumber \\
	\tilde \lambda_6 &=& 2 \lambda_2 \nonumber \\
	\tilde \lambda_7 &=& 2 \lambda_3 + 4 \lambda_4 \nonumber \\
	\tilde \lambda_8 &=& \lambda_3 + \lambda_4 \nonumber \\
	\tilde \lambda_9 &=& 4 \lambda_3 \nonumber \\
	\tilde \lambda_{10} &=& 4 \lambda_4 \nonumber \\
	\tilde M_1^{\prime} &=& M_1 \nonumber \\
	\tilde M_1 &=& M_1 \nonumber \\
	\tilde M_2 &=& M_2,
	\label{eq:nogprime}
\end{eqnarray}
where the relations among the quadratic and quartic couplings are in agreement with Ref.~\cite{Gunion:1990dt}.

Replacing the fields with their vevs and assuming CP conservation, the most general scalar potential becomes
\begin{eqnarray}
	V(v_{\phi}, v_{\chi}, v_{\xi}) &=& \frac{\tilde \mu_2^2}{2} \tilde v_{\phi}^2  + \tilde \mu_3^{\prime 2} \tilde v_{\chi}^2 
		+ \frac{\tilde \mu_3^2}{2} \tilde v_{\xi}^2  \nonumber \\
	&& + \frac{\tilde \lambda_1}{4} \tilde v_\phi^4 
		+ \frac{\tilde \lambda_3}{4} \tilde v_{\phi}^2 \tilde v_{\chi}^2
		+ \frac{\tilde \lambda_4}{\sqrt{2}} \tilde v_{\phi}^2 \tilde v_{\chi} \tilde v_{\xi} \nonumber \\
	&&	+ \frac{\tilde \lambda_5}{2} \tilde v_{\phi}^2 \tilde v_{\chi}^2
		+ \frac{\tilde \lambda_6}{2} \tilde v_{\phi}^2 \tilde v_{\xi}^2 
		+ \tilde \lambda_7 \tilde v_{\chi}^4
		+ \tilde \lambda_8 \tilde v_{\xi}^4 
		+ \tilde \lambda_{10} \tilde v_{\chi}^2 \tilde v_{\xi}^2 \nonumber \\
	&& - \frac{\tilde M_1^{\prime}}{2} \tilde v_{\phi}^2 \tilde v_{\chi}
		- \frac{\tilde M_1}{4} \tilde v_{\phi}^2 \tilde v_{\xi}
		- 6 \tilde M_2 \tilde v_{\chi}^2 \tilde v_{\xi}.
	\label{eq:potential4}
\end{eqnarray}
Minimizing this potential yields three equations:

\begin{eqnarray}
	0 = \frac{\partial V}{\partial \tilde v_{\phi}} &=& \tilde v_{\phi} \left[ \tilde \mu_2^2 
		+ \tilde \lambda_1 \tilde v_{\phi}^2 
		+ \frac{\tilde \lambda_3}{2} \tilde v_{\chi}^2
		+ \sqrt{2} \tilde \lambda_4 \tilde v_{\chi} \tilde v_{\xi} 
		+ \tilde \lambda_5 \tilde v_{\chi}^2
		+ \tilde \lambda_6 \tilde v_{\xi}^2 
		- \tilde M_1^{\prime} \tilde v_{\chi} 
		- \frac{\tilde M_1}{2} \tilde v_{\xi} \right], 
	\label{eq:potential5} \\
	0 = \frac{\partial V}{\partial \tilde v_{\chi}} &=& 2 \tilde \mu_3^{\prime 2} \tilde v_{\chi} 
	+ \frac{\tilde \lambda_3}{2} \tilde v_{\phi}^2 \tilde v_{\chi} 
	+ \frac{\tilde \lambda_4}{\sqrt{2}} \tilde v_{\phi}^2 \tilde v_{\xi}
	+ \tilde \lambda_5 \tilde v_{\phi}^2 \tilde v_{\chi}
	+ 4 \tilde \lambda_7 \tilde v_{\chi}^3 
	+ 2 \tilde \lambda_{10} \tilde v_{\chi} \tilde v_{\xi}^2 	 
	- \frac{\tilde M_1^{\prime}}{2} \tilde v_{\phi}^2
	- 12 \tilde M_2 \tilde v_{\chi} \tilde v_{\xi}, 
	\label{eq:potential6} \\
	0 = \frac{\partial V}{\partial \tilde v_{\xi}} &=& \tilde \mu_3^2 \tilde v_{\xi} 
	+ \frac{\tilde \lambda_4}{\sqrt{2}} \tilde v_{\phi}^2 \tilde v_{\chi} 
	+ \tilde \lambda_6 \tilde v_{\phi}^2 \tilde v_{\xi} 
	+ 4 \tilde \lambda_8 \tilde v_{\xi}^3 
	+ 2 \tilde \lambda_{10} \tilde v_{\chi}^2 \tilde v_{\xi} 
	- \frac{\tilde M_1}{4} \tilde v_{\phi}^2  
	- 6 \tilde M_2 \tilde v_{\chi}^2.
	\label{eq:potential7}
\end{eqnarray}
When the SU(2)$_L \times$SU(2)$_R$ symmetry is imposed, these conditions reduce to those in Eq.~(\ref{eq:mincond}).

The one-loop RGEs for the parameters of the most general gauge invariant potential are given in Appendix~\ref{sec:rge} for completeness.

\section{Physical masses and mixing in the custodial symmetry violating theory}
\label{sec:physmasses}

Isolating all terms quadratic in scalar fields from the potential and using Eqs.~(\ref{eq:potential5}--\ref{eq:potential7}) to eliminate $\tilde \mu_2^2$, $\tilde \mu_3^{\prime 2}$ and $\tilde \mu_3^2$ in favour of the vevs yields the following mass matrices for the physical scalars.

There is only one doubly-charged scalar, $\tilde H_5^{++} = \chi^{++} = H_5^{++}$, and its mass is given by
\begin{eqnarray}
	m_{\tilde H_5^{++}}^2 &=& 4 \tilde \lambda_2 \tilde v_{\chi}^2 
		- \frac{\tilde \lambda_3 \tilde v_{\phi}^2}{2} 
		- \frac{\tilde \lambda_4 \tilde v_{\phi}^2 \tilde v_{\xi}}{2 \sqrt{2} \tilde v_{\chi}} 
		+ \frac{\tilde M_1^{\prime}}{4 \tilde v_{\chi}} \tilde v_\phi^2 
		+ 12 \tilde M_2 \tilde v_{\xi}.
	\label{eq:potential8}
\end{eqnarray}

There are two CP-odd neutral scalars (one of which becomes the neutral Goldstone boson), whose mass-squared matrix in the basis $(\chi^{0,i}, \phi^{0,i})$ is given by
\begin{eqnarray}
	\mathcal{M}_{i}^2 = \left(
	\begin{array}{cc}
	\mathcal{M}^2_{i,11} & \mathcal{M}^2_{i,12} \\
	\mathcal{M}^2_{i,12} & \mathcal{M}^2_{i,22}
	\end{array}
	\right),
\end{eqnarray}
where
\begin{eqnarray}
	\mathcal{M}_{i,11}^2 &=& -\frac{\tilde \lambda_4 \tilde v_{\phi}^2 \tilde v_{\xi}}{2 \sqrt{2} \tilde v_{\chi}} + \frac{\tilde M_1^{\prime}}{4 \tilde v_{\chi}} \tilde v_\phi^2, \nonumber \\
        \mathcal{M}_{i,22}^2 &=& - 2 \sqrt{2} \tilde \lambda_4 \tilde v_{\chi} \tilde v_{\xi} + 2 \tilde M_1^{\prime} \tilde v_{\chi}, \nonumber \\ 
	\mathcal{M}_{i,12}^2 &=& \tilde \lambda_4 \tilde v_{\phi} \tilde v_{\xi} - \frac{\tilde M_1^{\prime}}{\sqrt{2}} \tilde v_{\phi}.
\end{eqnarray}

Note that the mass-squared matrix for the neutral imaginary states can be written as
\begin{equation}
	\mathcal{M}^2_{i} =  \left[\frac{\tilde M_1^{\prime}}{4 \tilde v_{\chi}} 
		- \frac{\tilde \lambda_4 \tilde v_{\xi}}{2 \sqrt{2} \tilde v_{\chi}} \right]
		\left( \begin{array}{cc} 
		\tilde v_{\phi}^2 & -\sqrt{8} \tilde v_{\phi} \tilde v_{\chi} \\
		-\sqrt{8} \tilde v_{\phi} \tilde v_{\chi} & 8 \tilde v_{\chi}^2
		\end{array} \right).
	\label{eq:MniAform}
\end{equation}
This matrix is easily diagonalized, yielding exact mass eigenstates
\begin{equation}
	\tilde G^0 = \frac{\tilde v_{\phi} \phi^{0,i} + \sqrt{8} \tilde v_{\chi} \chi^{0,i}}
		{\sqrt{\tilde v_{\phi}^2 + 8 \tilde v_{\chi}^2}}, \qquad \qquad
	\tilde H_3^0 = \frac{-\sqrt{8} \tilde v_{\chi} \phi^{0,i} + \tilde v_{\phi} \chi^{0,i}}
		{\sqrt{\tilde v_{\phi}^2 + 8 \tilde v_{\chi}^2}},
\end{equation}
where $\tilde G^0$ is the (massless) neutral Goldstone boson and the mass of $\tilde H_3^0$ is given by
\begin{equation}
	m^2_{\tilde H_3^0} = \left[\frac{\tilde M_1^{\prime}}{4 \tilde v_{\chi}} 
		- \frac{\tilde \lambda_4 \tilde v_{\xi}}{2 \sqrt{2} \tilde v_{\chi}} \right] 
		(\tilde v_{\phi}^2 + 8 \tilde v_{\chi}^2).
\end{equation}

There are three singly-charged scalars (one of which becomes the charged Goldstone boson), whose mass-squared matrix in the basis $(\chi^+,\xi^+,\phi^+)$ is given by
\begin{eqnarray}
	\mathcal{M}_{+}^{2} = \left(
	\begin{array}{ccc}
	\mathcal{M}^2_{+,11} & \mathcal{M}^2_{+,12} & \mathcal{M}^2_{+,13} \\
	\mathcal{M}^2_{+,12} & \mathcal{M}^2_{+,22} & \mathcal{M}^2_{+,23} \\
	\mathcal{M}^2_{+,13} & \mathcal{M}^2_{+,23} & \mathcal{M}^2_{+,33} 
	\end{array}
	\right),
\end{eqnarray}
where
\begin{eqnarray}
	\mathcal{M}_{+,11}^2 &=& 
		- \frac{\tilde \lambda_3 \tilde v_{\phi}^2}{4} 
		- \frac{\tilde \lambda_4 \tilde v_{\phi}^2 \tilde v_{\xi}}{2 \sqrt{2} \tilde v_{\chi}} 
		+ \tilde \lambda_9 \tilde v_{\xi}^2 
		+ \frac{\tilde M_1^{\prime}}{4 \tilde v_{\chi}} \tilde v_\phi^2 
		+ 6 \tilde M_2 \tilde v_{\xi}, \nonumber \\
        \mathcal{M}_{+,22}^2 &=& 
		- \frac{\tilde \lambda_4 \tilde v_{\phi}^2 \tilde v_{\chi}}{\sqrt{2} \tilde v_{\xi}} 
        		+ \tilde \lambda_9 \tilde v_{\chi}^2 
		+ \frac{\tilde M_1}{4 \tilde v_{\xi}} \tilde v_\phi^2 
		+ 6 \tilde M_2 \frac{\tilde v_{\chi}^2}{\tilde v_{\xi}}, \nonumber \\
	\mathcal{M}_{+,33}^2 &=& 
		- \tilde \lambda_3 \tilde v_{\chi}^2 
		- \sqrt{2} \tilde \lambda_4 \tilde v_{\chi} \tilde v_{\xi} 
		+ \tilde M_1 \tilde v_{\xi} 
		+ \tilde M_1^{\prime}\tilde v_{\chi}, \nonumber \\ 
	\mathcal{M}_{+,12}^2 &=& 
		\frac{\tilde \lambda_4 \tilde v_{\phi}^2}{2 \sqrt{2}} 
		- \tilde \lambda_9 \tilde v_{\chi} \tilde v_{\xi} 
		- 6 \tilde M_2 \tilde v_{\chi}, \nonumber \\
	\mathcal{M}_{+,13}^2 &=& 
		\frac{\tilde \lambda_3 \tilde v_{\phi} \tilde v_{\chi}}{2}  
		- \frac{\tilde M_1^{\prime}}{2} \tilde v_{\phi}, \nonumber \\
	\mathcal{M}_{+,23}^2 &=& 
		\frac{\tilde \lambda_4 \tilde v_{\phi} \tilde v_{\chi} }{\sqrt{2}} 
		- \frac{\tilde M_1}{2} \tilde v_{\phi} .
\end{eqnarray}
We first transform this mass-squared matrix into the basis of custodial-symmetric states $(H_5^+, H_3^+, G^+)$ using
\begin{equation}
	\mathcal{M}^{\prime 2}_+ = R_+ \mathcal{M}^2_+ R_+^T,
\end{equation}
where the orthogonal matrix $R_+$ is defined according to
\begin{equation}
	\left( \begin{array}{c} H_{5}^+ \\ H_{3}^{+} \\ G^+ \end{array} \right) = 
	R_+ \left( \begin{array}{c} \chi^{+} \\ \xi^{+} \\ \phi^{+} \end{array} \right),
\end{equation}
with
\begin{equation}
	R_+ = \left( \begin{array}{ccc} \frac{1}{\sqrt{2}} & -\frac{1}{\sqrt{2}} & 0 \\ 
		\frac{c_H}{\sqrt{2}} & \frac{c_H}{\sqrt{2}} & -s_H \\ 
		\frac{s_H}{\sqrt{2}} & \frac{s_H}{\sqrt{2}} & c_H \end{array} \right).
\end{equation}
Because the custodial-symmetry-violating effects will be small, we can diagonalize the mass-squared matrix $\mathcal{M}^{\prime 2}_+$ using first-order perturbation theory over most of the parameter space, as detailed below.  This gives some analytic insight into the structure of the custodial-symmetry-violating effects.  Of course, the perturbative diagonalization only works well when the diagonal elements of the mass-squared matrix are not too degenerate; this condition is satisfied over the parameter space of the H5plane benchmark, but is violated in some regions of parameter space in our general scans.  For this reason, in the next section we will use the first-order perturbative formulas below in our scans of the H5plane benchmark, but exact numerical diagonalization for our general scans.  We have checked numerically that the perturbative diagonalization is a very good approximation where we use it.

To first order in the custodial violation, the masses of the singly-charged physical mass eigenstates $\tilde H_5^+$ and $\tilde H_3^+$ are just given by the diagonal elements of the mass-squared matrix,
\begin{equation}
	m^2_{\tilde H_5^+} = \mathcal{M}^{\prime 2}_{+,11}, 
	\qquad \qquad
	m^2_{\tilde H_3^+} = \mathcal{M}^{\prime 2}_{+, 22}.
\end{equation}
The compositions of the mass eigenstates are given to first order using
\begin{equation}
	\tilde H_n = H_n + \sum\limits_{m \neq n} \frac{\mathcal{M}_{nm}^2}{\mathcal{M}_{nn}^2 - \mathcal{M}_{mm}^2} H_m,
	\label{eq:firstodercorr}
\end{equation}
where $\mathcal{M}^2$ is the mass-squared matrix in the appropriate basis.  
Applying this to the singly-charged states and using the fact that $\mathcal{M}_{+,33}^2 = 0$, we get,
\begin{eqnarray}
	\tilde H_5^+ &=& H_5^+ 
		+ \frac{\mathcal{M}_{+,12}^{\prime 2}}
			{\mathcal{M}_{+,11}^{\prime 2} - \mathcal{M}_{+,22}^{\prime 2}} H_3^+ 
		+ \frac{\mathcal{M}_{+,13}^{\prime 2}}{\mathcal{M}_{+,11}^{\prime 2}} G^+ \nonumber \\
	&=& \frac{\chi^+ - \xi^+}{\sqrt{2}}
		+ \left[ c_H \frac{\mathcal{M}^{\prime 2}_{+,13}}{\mathcal{M}^{\prime 2}_{+,11}}
			- s_H \frac{\mathcal{M}^{\prime 2}_{+,12}}
				{\mathcal{M}^{\prime 2}_{+,11} - \mathcal{M}^{\prime 2}_{+,22}} \right] \phi^+
		+ \left[ s_H \frac{\mathcal{M}^{\prime 2}_{+,13}}{\mathcal{M}^{\prime 2}_{+,11}}
			+ c_H \frac{\mathcal{M}^{\prime 2}_{+,12}}
				{\mathcal{M}^{\prime 2}_{+,11} - \mathcal{M}^{\prime 2}_{+,22}} \right]
			\frac{\chi^+ + \xi^+}{\sqrt{2}}, \\
	\tilde H_3^+ &=& H_3^+ 
		+ \frac{\mathcal{M}_{+,12}^{\prime 2}}
			{\mathcal{M}_{+,22}^{\prime 2} - \mathcal{M}_{+,11}^{\prime 2}} H_5^+ 
		+ \frac{\mathcal{M}_{+,23}^{\prime 2}}{\mathcal{M}_{+,11}^{\prime 2}} G^+,  \\
	\tilde G^+ &=& G^+ 
		+ \frac{\mathcal{M}_{+,13}^{\prime 2}}{-\mathcal{M}_{+,11}^{\prime 2}} H_5^+ 
		+\frac{\mathcal{M}_{+,23}^{\prime 2}}{-\mathcal{M}_{+,22}^{\prime 2}} H_3^+.
\end{eqnarray}
We highlight the composition of $\tilde H_5^+$ in particular because the custodial symmetry violation results in an admixture of $\phi^+$ into this state.  This allows $\tilde H_5^+$ to couple to fermions, which does not occur in the custodial-symmetric GM model.  Indeed, we can write the Feynman rule for the $\tilde H_5^+ \bar u d$ vertex as
\begin{equation}
	\tilde H_5^+ \bar u d: \quad i \frac{\sqrt{2}}{v} V_{ud} 
		\kappa^{\tilde H_5^+}_f (m_u P_L - m_d P_R),
	\label{eq:kappaH5pdef}
\end{equation}
where the coupling to fermions induced by the custodial symmetry violation is, to first order,
\begin{equation}
	\kappa^{\tilde H_5^+}_f = \frac{\mathcal{M}^{\prime 2}_{+,13}}{\mathcal{M}^{\prime 2}_{+,11}}
	- \tan \theta_H \frac{\mathcal{M}^{\prime 2}_{+,12}}
		{\mathcal{M}^{\prime 2}_{+,11} - \mathcal{M}^{\prime 2}_{+,22}}.
	\label{eq:kappaH5+}
\end{equation}
For comparison, in the custodial-symmetric GM model we can write the analogous coupling of $H_3^+$ to fermion pairs as $\kappa^{H_3^+}_f = -\tan\theta_H$.

Finally, there are three CP-even neutral scalars, whose mass-squared matrix in the basis $(\chi^{0,r}, \xi^{0,r}, \phi^{0,r})$ is given by
\begin{eqnarray}
	\mathcal{M}_{r}^2 = \left(
	\begin{array}{ccc}
	\mathcal{M}^2_{r,11} & \mathcal{M}^2_{r,12} & \mathcal{M}^2_{r,13} \\
	\mathcal{M}^2_{r,12} & \mathcal{M}^2_{r,22} & \mathcal{M}^2_{r,23} \\
	\mathcal{M}^2_{r,13} & \mathcal{M}^2_{r,23} & \mathcal{M}^2_{r,33} 
	\end{array} \right),
\end{eqnarray}
where
\begin{eqnarray}
	\mathcal{M}_{r,11}^2 &=& 
		- \frac{\tilde \lambda_4 \tilde v_{\phi}^2 \tilde v_{\xi}}{2 \sqrt{2} \tilde v_{\chi}} 
		+ 4 \tilde \lambda_7 \tilde v_{\chi}^2 
		+ \frac{\tilde M_1^{\prime}}{4 \tilde v_{\chi}} \tilde v_\phi^2, \nonumber \\
        \mathcal{M}_{r,22}^2 &=& 
		- \frac{\tilde \lambda_4 \tilde v_{\phi}^2 \tilde v_{\chi}}{\sqrt{2} \tilde v_{\xi}} 
        		+ 8 \tilde \lambda_8 \tilde v_{\xi}^2 
		+ \frac{\tilde M_1}{4 \tilde v_{\xi}} \tilde v_\phi^2 
		+ 6 \tilde M_2 \frac{\tilde v_{\chi}^2}{\tilde v_{\xi}}, \nonumber \\
	\mathcal{M}_{r,33}^2 &=& 2 \tilde \lambda_1 \tilde v_{\phi}^2, \nonumber \\ 
	\mathcal{M}_{r,12}^2 &=& 
		\frac{\tilde \lambda_4 \tilde v_{\phi}^2}{2}  
		+ 2 \sqrt{2} \tilde \lambda_{10} \tilde v_{\chi} \tilde v_{\xi} 
		- 6 \sqrt{2} \tilde M_2 \tilde v_{\chi}, \nonumber \\
	\mathcal{M}_{r,13}^2 &=& 
		\frac{\tilde \lambda_3 \tilde v_{\phi} \tilde v_{\chi}}{\sqrt{2}} 
		+ \tilde \lambda_4 \tilde v_{\phi} \tilde v_{\xi} 
		+ \sqrt{2} \tilde \lambda_5 \tilde v_{\phi} \tilde v_{\chi} 
		- \frac{\tilde M_1^{\prime} \tilde v_{\phi}}{\sqrt{2} }, \nonumber \\
	\mathcal{M}_{r,23}^2 &=& 
		\sqrt{2} \tilde \lambda_4 \tilde v_{\phi} \tilde v_{\chi}  
		+ 2 \tilde \lambda_6 \tilde v_{\phi} \tilde v_{\xi} 
		- \frac{\tilde M_1 \tilde v_{\phi}}{2} .
\end{eqnarray}
We first transform this mass-squared matrix into the basis of custodial-symmetric states $(H_5^0, H_1^{0\prime}, \phi^{0,r})$ using
\begin{equation}
	\mathcal{M}^{\prime 2}_r = R_r \mathcal{M}^2_r R_r^T,
\end{equation}
where the orthogonal matrix $R_r$ is defined according to
\begin{equation}
	\left( \begin{array}{c} H_{5}^0 \\ H_{1}^{0 \prime} \\ \phi^{0,r} \end{array} \right) 
	= R_r \left( \begin{array}{c} \chi^{0,r} \\ \xi^{0,r} \\ \phi^{0,r} \end{array} \right),
\end{equation}
with
\begin{equation}
	R_r = \left( \begin{array}{ccc} 
	-\sqrt{\frac{1}{3}} & \sqrt{\frac{2}{3}} & 0 \\ 
	\sqrt{\frac{2}{3}} & \sqrt{\frac{1}{3}} & 0 \\ 
	0 & 0 & 1 \end{array} \right).
\end{equation}
To first order in the custodial symmetry violation, the mass of $\tilde H_5^0$ is given by
\begin{equation}
	m_{\tilde H_5^0}^2 = \mathcal{M}_{r,11}^{\prime 2}.
\end{equation}
It is most straightforward to find the masses of $\tilde h$ and $\tilde H$ by diagonalizing the remaining $2\times 2$ block of $\mathcal{M}^{\prime 2}_r$ as follows:
\begin{eqnarray}
	m^2_{\tilde h, \tilde H} = \frac{1}{2} 
		\left[ \mathcal{M}^{\prime 2}_{r,33} + \mathcal{M}^{\prime 2}_{r,22} 
			\mp \sqrt{\left( \mathcal{M}^{\prime 2}_{r,33} - \mathcal{M}^{\prime 2}_{r,22} \right)^2
			+ 4 \left( \mathcal{M}^{\prime 2}_{r,23} \right)^2} \right].
\end{eqnarray}
The mixing angle that achieves this diagonalization is given by
\begin{equation}
	\sin 2 \tilde \alpha = \frac{2 \mathcal{M}^{\prime 2}_{r,23}}{m^2_{\tilde H} - m^2_{\tilde h}},
	\qquad \qquad
	\cos 2 \tilde \alpha = \frac{\mathcal{M}^{\prime 2}_{r,22} - \mathcal{M}^{\prime 2}_{r, 33}}
			{m^2_{\tilde H} - m^2_{\tilde h}},
\end{equation}
where the states are given in terms of $\tilde \alpha$ by
\begin{equation}
	h_{\tilde \alpha} = c_{\tilde \alpha} \phi^{0,r} - s_{\tilde \alpha} H_1^{0 \prime},
	\qquad \qquad 
	H_{\tilde \alpha} = s_{\tilde \alpha} \phi^{0,r} + c_{\tilde \alpha} H_1^{0 \prime},
\end{equation}
and we have defined $c_{\tilde \alpha} = \cos \tilde \alpha$, $s_{\tilde \alpha} = \sin \tilde \alpha$.
(Note that these are not yet the mass eigenstates: there is still a small mixing with $H_5^0$ to be dealt with below.)
We introduce a second orthogonal rotation matrix $R_{\tilde \alpha}$, defined according to
\begin{equation}
	\left( \begin{array}{c} H_5^0 \\ H_{\tilde \alpha} \\ h_{\tilde \alpha} \end{array} \right)
	= R_{\tilde \alpha} \left( \begin{array}{c} H_5^0 \\ H_1^{0 \prime} \\ \phi^{0,r} \end{array} \right),
\end{equation}
with
\begin{equation}
	R_{\tilde \alpha} = \left( \begin{array}{ccc} 
		1 & 0 & 0 \\
		0 & c_{\tilde \alpha} & s_{\tilde \alpha} \\
		0 & -s_{\tilde \alpha} & c_{\tilde \alpha} \end{array} \right).
\end{equation}
The mass-squared matrix in the basis $(H_5^0, H_{\tilde \alpha}, h_{\tilde \alpha})$ is then given by
\begin{equation}
	\mathcal{M}^{\prime \prime 2}_r = R_{\tilde \alpha} \mathcal{M}^{\prime 2}_r R_{\tilde \alpha}^T
	= \left( \begin{array}{ccc}
		\mathcal{M}^{\prime \prime 2}_{r, 11} & \mathcal{M}^{\prime \prime 2}_{r, 12} & \mathcal{M}^{\prime \prime 2}_{r, 13} \\
		\mathcal{M}^{\prime \prime 2}_{r, 12} & \mathcal{M}^{\prime \prime 2}_{r, 22} & 0 \\
		\mathcal{M}^{\prime \prime 2}_{r, 13} & 0 & \mathcal{M}^{\prime \prime 2}_{r, 33} 
		\end{array} \right).
\end{equation}
Note that $\mathcal{M}_{r,11}^{\prime \prime 2} = \mathcal{M}_{r,11}^{\prime 2}$.
The masses of $\tilde h$ and $\tilde H$ can then be written (to first order in the custodial symmetry violation) in terms of the diagonal elements of this matrix as
\begin{equation}
	m_{\tilde h}^2 = \mathcal{M}^{\prime \prime 2}_{r, 33}, 
	\qquad \qquad
	m_{\tilde H}^2 = \mathcal{M}^{\prime \prime 2}_{r, 22}.
\end{equation}

We now use Eq.~(\ref{eq:firstodercorr}) to write the compositions of the CP-even neutral mass eigenstates to first order in the custodial violation as
\begin{eqnarray}
	\tilde H_5^0 &=& H_5^0 
		+ \frac{\mathcal{M}^{\prime \prime 2}_{r, 12}}
			{\mathcal{M}^{\prime \prime 2}_{r, 11} - \mathcal{M}^{\prime \prime 2}_{r, 22}} 
			H_{\tilde \alpha}
		+ \frac{\mathcal{M}^{\prime \prime 2}_{r, 13}}
			{\mathcal{M}^{\prime \prime 2}_{r, 11} - \mathcal{M}^{\prime \prime 2}_{r, 33}} 
			h_{\tilde \alpha}, \nonumber \\
		&=& \left( \sqrt{\frac{2}{3}} \xi^{0,r} - \sqrt{\frac{1}{3}} \chi^{0,r} \right)
		+ \left[ s_{\tilde \alpha} \frac{\mathcal{M}^{\prime \prime 2}_{r,12}}
			{\mathcal{M}^{\prime \prime 2}_{r,11} - \mathcal{M}^{\prime \prime 2}_{r,22}}
			+ c_{\tilde \alpha} \frac{\mathcal{M}^{\prime \prime 2}_{r,13}}
			{\mathcal{M}^{\prime \prime 2}_{r,11} - \mathcal{M}^{\prime \prime 2}_{r,33}} \right] 
			\phi^{0,r} \nonumber \\
	&&	+ \left[ c_{\tilde \alpha} \frac{\mathcal{M}^{\prime \prime 2}_{r,12}}
			{\mathcal{M}^{\prime \prime 2}_{r,11} - \mathcal{M}^{\prime \prime 2}_{r,22}}
			- s_{\tilde \alpha} \frac{\mathcal{M}^{\prime \prime 2}_{r,13}}
			{\mathcal{M}^{\prime \prime 2}_{r,11} - \mathcal{M}^{\prime \prime 2}_{r,33}} \right]
			\left( \sqrt{\frac{1}{3}} \xi^{0,r} + \sqrt{\frac{2}{3}} \chi^{0,r} \right), \\
	\tilde H &=& H_{\tilde \alpha} 
		+ \frac{\mathcal{M}^{\prime \prime 2}_{r, 12}}
			{\mathcal{M}^{\prime \prime 2}_{r, 22} - \mathcal{M}^{\prime \prime 2}_{r, 11}}
			H_5^0, \\
	\tilde h &=& h_{\tilde \alpha} 
		+ \frac{\mathcal{M}^{\prime \prime 2}_{r, 13}}
			{\mathcal{M}^{\prime \prime 2}_{r, 33} - \mathcal{M}^{\prime \prime 2}_{r, 11}}
			H_5^0.
			\label{eq:H50inth}
\end{eqnarray}
We highlight the composition of $\tilde H_5^0$ in particular because the custodial symmetry violation results in an admixture of $\phi^{0,r}$ into this state.  This allows $\tilde H_5^0$ to couple to fermions, which does not occur in the custodial-symmetric GM model.  The coupling of $\tilde H_5^0$ to $\bar f f$, normalized to the corresponding coupling of the SM Higgs boson, is then given to first order in the custodial symmetry violation by
\begin{equation}
	\kappa^{\tilde H_5^0}_f = \frac{1}{c_H} \left[ 
		s_{\tilde \alpha} \frac{\mathcal{M}^{\prime \prime 2}_{r,12}}
			{\mathcal{M}^{\prime \prime 2}_{r,11} - \mathcal{M}^{\prime \prime 2}_{r,22}}
		+ c_{\tilde \alpha} \frac{\mathcal{M}^{\prime \prime 2}_{r,13}}
			{\mathcal{M}^{\prime \prime 2}_{r,11} - \mathcal{M}^{\prime \prime 2}_{r,33}} \right].
	\label{eq:kappaH50}
\end{equation}


Finally, the mixing of a small amount of custodial-fiveplet $H_5^0$ into the physical Higgs boson $\tilde h$, together with $\tilde v_{\chi} \neq \tilde v_{\xi}$, leads to a violation of custodial symmetry in the couplings of $\tilde h$ to $WW$ and $ZZ$.  This is parameterized in terms of the physical observable
\begin{equation}
	\lambda_{WZ}^{\tilde h} \equiv \frac{\kappa_W^{\tilde h}}{\kappa_Z^{\tilde h}},
	\label{eq:lambdaWZ}
\end{equation}
where $\kappa_W^{\tilde h}$ and $\kappa_Z^{\tilde h}$ are the couplings of $\tilde h$ to $WW$ and $ZZ$, respectively, normalized to the corresponding couplings of the SM Higgs boson.  We can write this in terms of the vevs and the mixing with $H_5^0$ as follows:
\begin{equation}
	\lambda_{WZ}^{\tilde h} = 
		\frac{\tilde \kappa_W^{h_{\tilde \alpha}} + \epsilon \tilde \kappa_W^{H_5^0}}
		{\tilde \kappa_Z^{h_{\tilde \alpha}} + \epsilon \tilde \kappa_Z^{H_5^0}},
\end{equation}
where the couplings of $h_{\tilde \alpha}$ to $W$ and $Z$ boson pairs, including the effects of $\tilde v_{\chi} \neq \tilde v_{\xi}$, are given by
\begin{equation}
	\tilde \kappa_W^{h_{\tilde \alpha}} = c_{\tilde \alpha} \frac{\tilde v_{\phi}}{v}
		- s_{\tilde \alpha} \frac{4}{\sqrt{3}} \frac{\tilde v_{\chi} + \tilde v_{\xi}}{v},
	\qquad \qquad
	\tilde \kappa_Z^{h_{\tilde \alpha}} = c_{\tilde \alpha} \frac{\tilde v_{\phi}}{v}
		- s_{\tilde \alpha} \frac{8}{\sqrt{3}} \frac{\tilde v_{\chi}}{v},
\end{equation}
the couplings of $H_5^0$ to $W$ and $Z$ boson pairs are given by
\begin{equation}
	\tilde \kappa_W^{H_5^0} = \sqrt{\frac{2}{3}} \frac{4 \tilde v_{\xi} - 2 \tilde v_{\chi}}{v} 
		\simeq \frac{1}{\sqrt{3}} s_H,
	\qquad \qquad
	\tilde \kappa_Z^{H_5^0} = - \sqrt{\frac{2}{3}} \frac{4 \tilde v_{\chi}}{v}
		\simeq -\frac{2}{\sqrt{3}} s_H,
\end{equation}
and the mixing of $H_5^0$ into $\tilde h$ from Eq.~(\ref{eq:H50inth}) is
\begin{equation}
	\epsilon = \frac{\mathcal{M}^{\prime \prime 2}_{r, 13}}
			{\mathcal{M}^{\prime \prime 2}_{r, 33} - \mathcal{M}^{\prime \prime 2}_{r, 11}}.
\end{equation}

\section{Numerical results}
\label{sec:numerics}

\subsection{Calculational procedure}

In this paper we imagine that the custodially-symmetric GM model emerges at some scale $\Lambda$ as an effective theory of some unspecified ultraviolet (UV) completion.  For example, the scalars in the GM model could be composites and the custodial symmetry an accidental global symmetry resulting from the particle content of the UV theory.  The running of the scalar potential parameters down to the weak scale induces custodial symmetry violation.  We can then use the experimental constraint on the $\rho$ parameter at the weak scale to set an upper bound on the scale $\Lambda$.  Subject to this constraint, we can also predict the size of other custodial symmetry violating effects such as mass splittings among the members of the custodial fiveplet and triplet scalars, mixing between scalars in different custodial-symmetry representations (which, for example, can induce fermionic decays of the otherwise fermiophobic $H_5$ states), and the value of the ratio $\lambda_{WZ} \equiv \kappa_W/\kappa_Z$ of the 125~GeV Higgs boson (predicted as $\lambda_{WZ} = 1$ in custodial-symmetric theories).

For concreteness, we start our analysis within the context of the so-called H5plane benchmark, which is a two-dimensional slice through the custodial-symmetric GM model parameter space as defined in Table~\ref{tab:H5plane} at the weak scale.  This benchmark was introduced in Ref.~\cite{deFlorian:2016spz} for interpretation of LHC searches for $H_5^{\pm}$ and $H_5^{\pm\pm}$, and its phenomenology was studied in some detail in Ref.~\cite{Logan:2017jpr}.  The H5plane benchmark takes $m_5$ and $s_H$ as its two free parameters: this will allow us to plot our results as contours in the $m_5$--$s_H$ plane. The benchmark is defined for $m_5$ values of 200~GeV and higher. To test the generality of our results in the H5plane benchmark we then perform a general parameter scan and compare the results to the benchmark region.  The parameter points are generated and checked for theoretical consistency using the public code GMCALC~\cite{Hartling:2014xma}.  The input parameters used in the general scan are given in Table~\ref{tab:generalscan}.  We finally perform a second dedicated parameter scan focusing on $m_5$ masses below 200~GeV, to cover the region in which the H5plane benchmark is not defined.  A dedicated scan is needed in this region because only a small fraction of scanned points satisfy the theoretical constraints.  The input parameters used in this dedicated low-$m_5$ scan are given in Table~\ref{tab:lowm5scan}.  Results of this dedicated scan are collected in Sec.~\ref{sec:lowm5}.

\begin{table}
\begin{center}
    \begin{tabular}{lll}
    \hline \hline
    Fixed Parameters & Variable Parameters & Dependent Parameters \\ \hline
    $G_F= 1.1663787 \times 10^{-5}~\rm{GeV}^{-2}$ & $m_5 \in [200,3000]~\rm{GeV} $ & $\lambda_2 =0.4 m_5 / (1000~{\rm GeV})$ \\
    $m_h = 125~{\rm GeV}$ & $s_H \in (0,1)$ & $M_1 = \sqrt{2} s_H(m^2_5 +v^2)/v$\\
    $\lambda_3 = -0.1$ &  & $M_2 = M_1/6$\\
    $\lambda_4 = 0.2$ &  & \\ \hline\hline
    \end{tabular}
\end{center}
\caption{Input parameters for the H5plane benchmark scenario~\cite{deFlorian:2016spz} in the custodial-symmetric GM model.}
\label{tab:H5plane}
\end{table}

\begin{table}
\begin{center}
\begin{tabular}{ll}
\hline\hline
Fixed Parameters & Variable Parameters \\ \hline
$G_F= 1.1663787 \times 10^{-5}~\rm{GeV}^{-2}$ & $\mu_3^2 \in [-(200~{\rm GeV})^2, (4200~{\rm GeV})^2]$ \\
$m_h = 125~{\rm GeV}$ & $\lambda_2$, $\lambda_3$, $\lambda_4$, $\lambda_5$ \\
  & $M_1$, $M_2$ \\
\hline \hline
\end{tabular}
\end{center}
\caption{Input parameters for the general scan in the custodial-symmetric GM model.}
\label{tab:generalscan}
\end{table}

\begin{table}
\begin{center}
\begin{tabular}{ll}
\hline\hline
Fixed Parameters & Variable Parameters \\ \hline
$G_F= 1.1663787 \times 10^{-5}~\rm{GeV}^{-2}$ & $m_5 < 200~{\rm GeV}$ \\
$m_h = 125~{\rm GeV}$ & $s_H \in (0,1)$ \\
  & $\lambda_2$, $\lambda_3$, $\lambda_4$, $\lambda_5$ \\
  & $M_2 \in [-1200~{\rm GeV}, 1200~{\rm GeV}]$ \\
\hline \hline
\end{tabular}
\end{center}
\caption{Input parameters for the dedicated low-$m_5$ scan in the custodial-symmetric GM model.}
\label{tab:lowm5scan}
\end{table}

We perform the calculations as follows. We start by specifying an input point in the custodial-symmetric GM model at the weak scale.  Because it is not possible to separate the scale of the GM model states from the SM weak scale so long as the triplets contribute to electroweak symmetry breaking, for the purposes of renormalization group running we will define the ``weak scale'' to be $m_5$ as defined in the custodial-symmetric low-scale input parameter set.  We define the electroweak gauge couplings at the weak scale in terms of the inputs $G_F$, $M_W$, and $M_Z$, and we take $\alpha_s(M_Z) = 0.118$ to define the strong coupling at the weak scale (we ignore the running of the strong coupling between $M_Z$ and $m_5$; this is a small effect because the strong coupling only enters in the running of the top Yukawa coupling).  We extract the value of the top Yukawa coupling using the relation $y_t = \sqrt{2} m_t/v_{\phi}$ evaluated in terms of the custodial-symmetric input parameters at the weak scale.  For simplicity, we set $y_b = y_{\tau} = 0$; their effects would be very small.  

We then run the parameters of the custodial-symmetric scalar potential up to a scale $\Lambda$ using the RGEs in Eqs.~(\ref{eq:RGEtmu2sq}--\ref{eq:RGEtM2}) but with $g_1$ set to zero.  We also run the gauge couplings (including the actual value of $g_1$) and the top Yukawa coupling from $m_5$ to $\Lambda$ using Eqs.~(\ref{eq:RGEg1}--\ref{eq:RGEyt}).  For the running we use fourth-order Runge-Kutta with a small step size.
The result of this is a custodial-symmetric scalar potential at the scale $\Lambda$.  At this stage we can check whether any of the quartic scalar couplings has grown large enough to violate perturbative unitarity (indicating that we have almost run into a Landau pole).  This allows us to determine the maximum scale allowed by perturbativity.  We also check whether the potential has become unbounded from below; this turns out not to happen for any of our scan points in the H5plane benchmark or in the general parameter scans.  Because the potential is still custodial-symmetric, we can use the requirements for perturbative unitarity and boundedness-from-below as derived for the custodial-symmetric theory~\cite{Hartling:2014zca} as given at the end of Sec.~\ref{sec:model}.

From the custodial-symmetric scalar potential at scale $\Lambda$, we then run back down to the scale $m_5$ using the full RGEs in Eqs.~(\ref{eq:RGEtmu2sq}--\ref{eq:RGEyt}) with $g_1 \neq 0$.  The nonzero hypercharge coupling induces custodial symmetry violation in the scalar potential, causing violation of the custodial-symmetry relations of Eq.~(\ref{eq:nogprime}) among the parameters of the most general gauge invariant scalar potential. Having determined the custodial violating parameters we can now solve the minimization conditions in Eqs.~(\ref{eq:potential5}), (\ref{eq:potential6}), and (\ref{eq:potential7}) for the custodial-violating vevs $\tilde v_{\phi}$, $\tilde v_{\chi}$, and $\tilde v_{\xi}$.  First we solve Eq.~(\ref{eq:potential5}) for $\tilde v_{\phi}^2$ in terms of the other vevs and plug this in to Eqs.~(\ref{eq:potential6}) and (\ref{eq:potential7}), which we then solve numerically using a two-dimensional Newton's method.  For the initial guess we take $\tilde v_{\chi} = \tilde v_{\xi} = v_{\chi}$, where $v_{\chi}$ is the custodial-symmetric triplet vev in our original weak-scale input point.

However, this procedure suffers from a complication. The definition of the original weak scale input point uses the measured $m_h$ and $G_F$ as input parameters.  These are used to fix $\lambda_1$ and $\mu_2^2$ in the weak-scale custodial-symmetric theory.  After running the parameters up to the scale $\Lambda$ using the custodial-symmetric RGEs (with $g^{\prime}$ set to zero) and then running them back down to the weak scale with the full custodial violating RGEs, the new weak-scale calculations of $m_{\tilde h}$ and $G_F^{-1} = \sqrt{2}(\tilde v_{\phi}^2 + 4 \tilde v_{\chi}^2 + 4 \tilde v_{\xi}^2)$ yield numbers that do not match the original input values. To address this, we need to adjust the custodially-symmetric weak-scale input values for $\lambda_1$ and $\mu_2^2$ (while keeping all the other weak-scale inputs fixed) until we obtain the correct experimental values of $m_{\tilde h}$ and $G_F$ \emph{after} implementing the custodial symmetry violation.  We do this by defining two functions, $f_1=m_{\tilde h}^{calc}(\lambda_1,\mu_2^2)-m_h^{expt}$ and $f_2=G_F^{calc}(\lambda_1,\mu_2^2)-G_F^{expt}$, where $\lambda_1$ and $\mu_2^2$ are the inputs at the weak scale, $m_{\tilde h}^{calc}$ and $G_F^{calc}$ are calculated using the procedure described above, and $m_h^{expt}$ and $G_F^{expt}$ are the desired (experimental) values.  The solution is the point at which $f_1 = f_2 = 0$, which we find iteratively using a two-dimensional Newton's method.  This involves running the full RGE machinery up and down multiple times and is the slowest part of our numerical work.  The same is generically true for $m_t$ (which we use to fix the top quark Yukawa coupling $y_t$ at the weak scale).  In the H5plane benchmark, the change to $m_t$ after running up and back down again is within the current experimental error so we ignore this effect. In the general parameter scans, however, the change in $m_t$ can be larger, so in these scans we extend the iterative procedure to include $y_t$.

Having solved for the appropriate input values of $\lambda_1$ and $\mu_2^2$, we now have a self-consistent set of scalar potential input parameters at the weak scale ($\mu = m_5$), corresponding to a custodial-symmetric theory at the high scale ($\mu = \Lambda$), which we then run back down to obtain the custodial-violating theory at the weak scale (again $m_5$) with the correct predictions for $m_h$ and $G_F$.  We then calculate our desired observables including the $\rho$ parameter, the mass splittings among the states of the would-be custodial multiplets, and the effects of the mixing among the would-be custodial eigenstates.
 
In the rest of this section we present our results as contour plots in the H5plane benchmark in the $m_5$--$s_H$ plane, and as scatter plots for the general scans.  We emphasize that $m_5$ and $s_H$ here are defined as part of the weak-scale custodial-symmetric input parameter point, and do not directly correspond to the physical masses, couplings, or vevs of the corresponding parameter point in the weak-scale custodial-violating theory.  However, as we will show in what follows, the deviations of these physical observables from the custodial-symmetric input parameters are small enough that the differences are unlikely to be observable at the LHC.

\subsection{Constraints on the cutoff scale from perturbativity and the $\rho$ parameter}

\begin{figure}
\resizebox{0.5\textwidth}{!}{\includegraphics{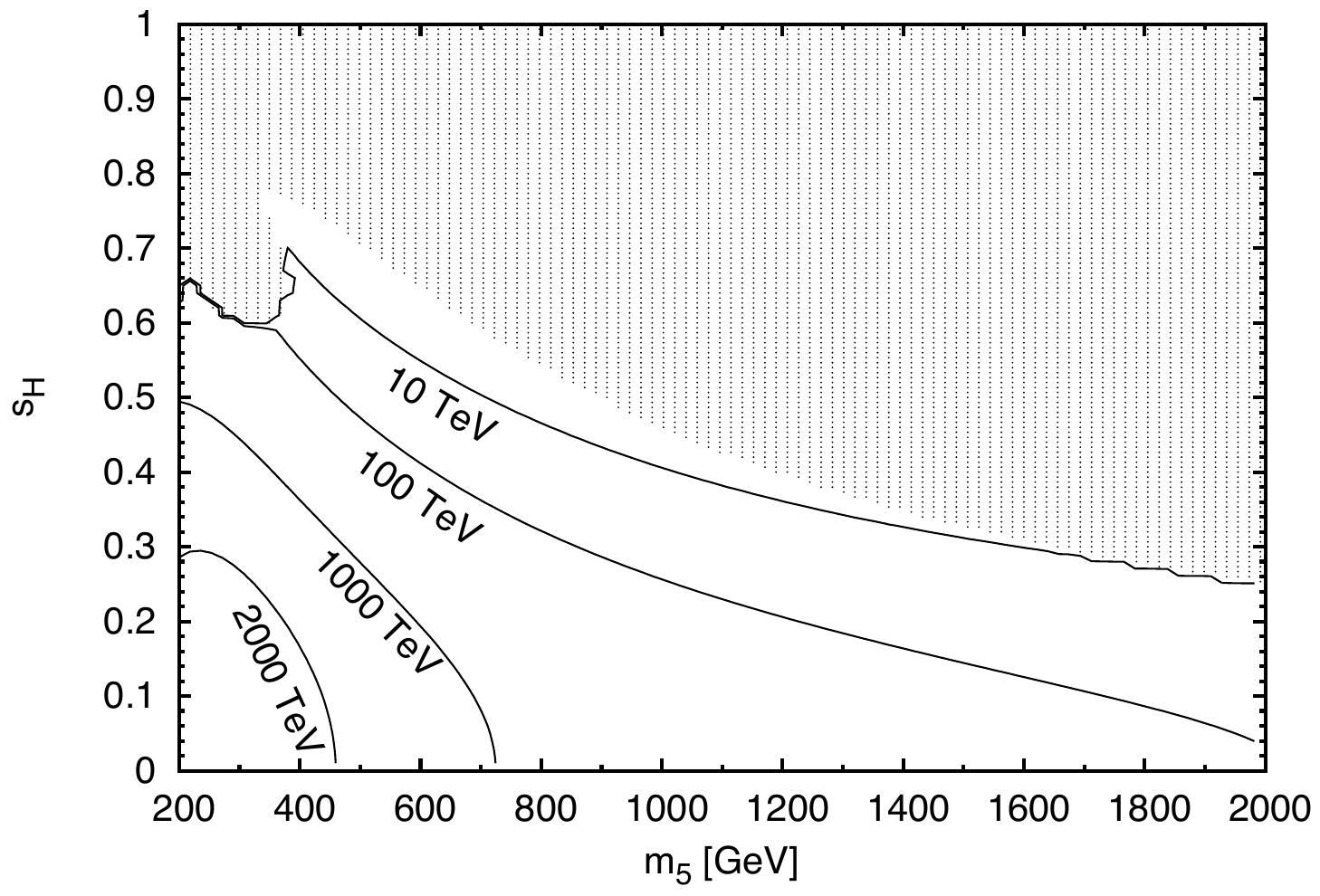}}%
\resizebox{0.5\textwidth}{!}{\includegraphics{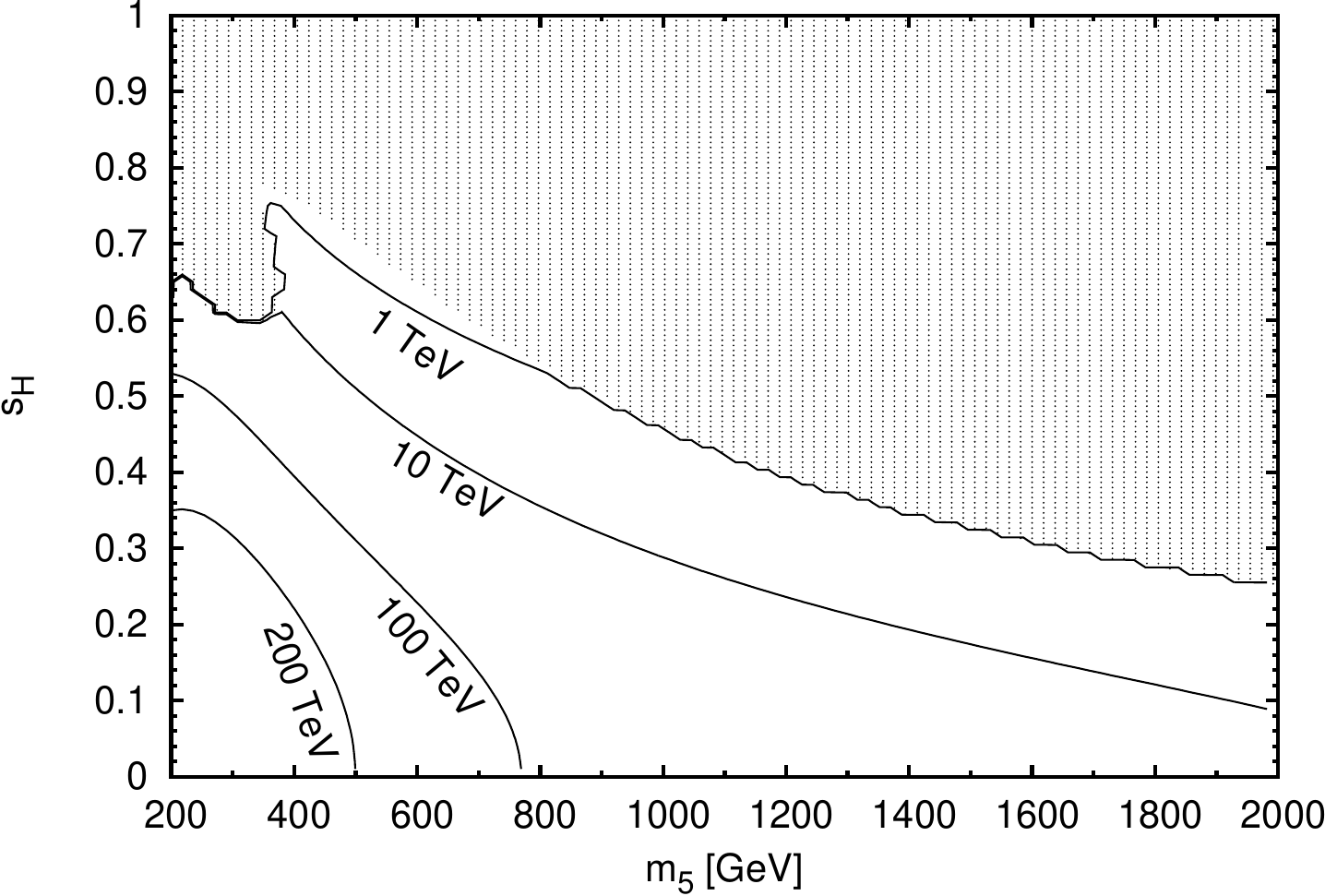}}
\caption{Constraints on the custodial-symmetric cutoff scale due to perturbativity of the model in the H5plane benchmark.  Left: the scale of the Landau pole, defined as the scale at which any of the $\lambda_i$ in the custodial-symmetric theory becomes larger than $10^3$. This scale varies between 2.5~TeV and 2594.2~TeV over the benchmark considered.  Right: the highest scale at which the perturbative unitarity constraints of Eq.~(\ref{eqn:pertUnitarity}) in the custodial-symmetric theory remain satisfied. This scale varies between 346.8~GeV and 291.1~TeV over the benchmark considered.}
\label{fig:CSLimit}
\end{figure}

We begin by determining the maximum scale allowed for the custodial-symmetric ultraviolet completion by running the custodial-symmetric model up until we hit a Landau pole.  This is shown in the left panel of Fig.~\ref{fig:CSLimit} in the H5plane benchmark.  The shaded region at large $s_H$ in these plots is excluded by theoretical constraints on the custodial-symmetric model.  We define the Landau pole as the scale at which any of the custodial-symmetric quartic couplings $\lambda_i$ becomes larger than $10^3$; the true divergence happens extremely close to this scale.  In the right panel of Fig.~\ref{fig:CSLimit} we also show the scale at which the quartic couplings in the custodial-symmetric theory violate any of the conditions for perturbative unitarity of two-to-two scattering amplitudes given in Eq.~(\ref{eqn:pertUnitarity}).  We can see that the scale at which perturbative unitarity is violated is roughly an order of magnitude below the scale of the Landau pole.  Within the H5plane benchmark, if the theory is to remain perturbative the ultraviolet completion has to appear at 290~TeV or below, and the maximum scale of the Landau pole in this benchmark is around 2600~TeV.  For $m_5 \gtrsim 400$~GeV, the upper bound on $s_H$ from theory constraints in the H5plane benchmark is due to the perturbative unitarity constraint; therefore along this boundary the scale of perturbative unitarity violation is essentially the same as $m_5$, and the Landau pole occurs around 10~TeV.

We also note that in the H5plane benchmark, the value of $\lambda_2$ at the weak scale grows linearly with $m_5$ (see Table~\ref{tab:H5plane}).  This is responsible for the decrease in the scale of perturbative unitarity violation and the subsequent Landau pole with increasing $m_5$ at small $s_H$ values, and is a quirk of the H5plane benchmark.

In all the scans that follow, we take the scale of perturbative unitarity violation to be an upper bound on the scale of the custodial-symmetric theory, and we do not run above this scale.

\begin{figure}
\resizebox{0.5\textwidth}{!}{\includegraphics{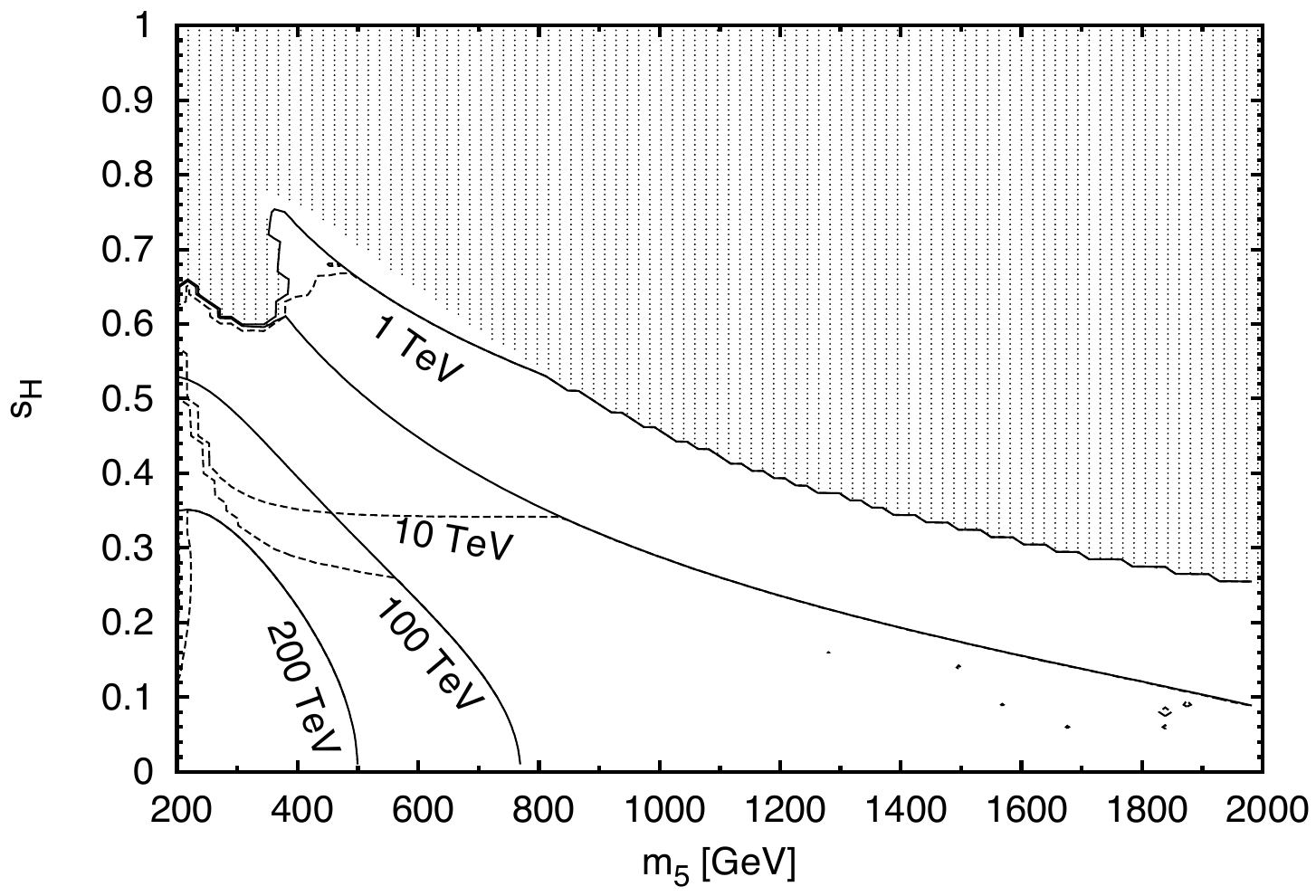}}%
\resizebox{0.5\textwidth}{!}{\includegraphics{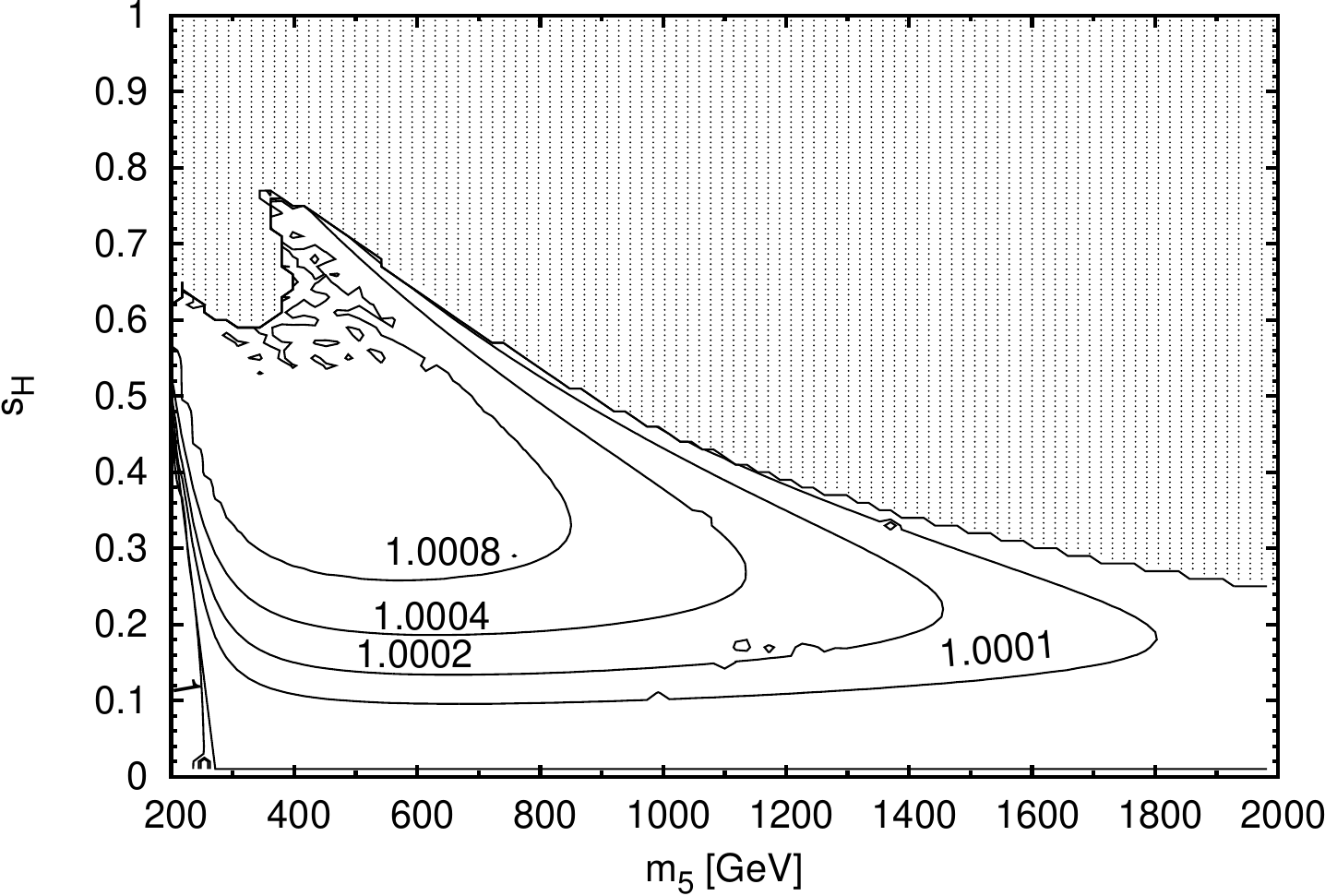}}
	\caption{Values of and constraints due to the $\rho$ parameter in the H5plane benchmark.
Left: the highest scale at which the perturbative unitarity constraints of Eq.~(\ref{eqn:pertUnitarity}) in the custodial-symmetric theory remain satisfied as in the right panel of Fig.~\ref{fig:CSLimit} (solid lines), showing also the highest allowed custodial-symmetric scale after requiring that the $\rho$ parameter remain within $\pm 2 \sigma$ of its experimental value [Eq.~(\ref{eq:rho2016PDG})] in the custodial-violating weak-scale theory (dashed lines).  The range of scales allowed after imposing the $\rho$ parameter constraint remains the same as in Fig.~\ref{fig:CSLimit}.  Right: the value of $\rho$ in the weak-scale custodial-violating theory when the custodial-symmetric scale is taken as large as possible subject to perturbative unitarity at the high scale and the experimental limits on $\rho$.  The values of $\rho$ range between the $\pm 2 \sigma$ limits of 0.99991 and 1.00083.}
	\label{fig:rhoLimits}
\end{figure}

The maximum allowed scale of the custodial-symmetric ultraviolet completion can also be constrained by the stringent experimental limits on the $\rho$ parameter, as defined in Eq.~(\ref{eq:rhoCV}).  For this calculation (and those that follow), we bring to bear the full computational machinery described in the previous section, including adjusting the input values of $\lambda_1$ and $\mu_2^2$ to obtain the correct measured values of $G_F$ and $m_h$ in the custodial-violating theory at the weak scale.  We take the current value of $\rho$ from the 2016 Particle Data Group electroweak fit~\cite{Olive:2016xmw},
\begin{equation}
	\rho = 1.00037 \pm 0.00023,
	\label{eq:rho2016PDG}
\end{equation}
and require that the value of $\rho$ in the weak-scale custodial-violating theory be within $2\sigma$ of this value; i.e., between $\rho_{lower} = 0.99991$ and $\rho_{upper} = 1.00083$.  Because the deviation in the $\rho$ parameter in the custodial-violating weak-scale theory grows as the scale of the custodial-symmetric ultraviolet completion increases, this constraint puts a stronger upper bound on the scale of the ultraviolet completion in part of the H5plane benchmark parameter space, as shown in the left panel of Fig.~\ref{fig:rhoLimits}, where we also plot the upper bound from requiring perturbative unitarity.  The $\rho$ parameter constraint is stronger than that from perturbative unitarity for moderate $s_H$ values and $m_5$ below about 850~GeV.  

In the right panel of Fig.~\ref{fig:rhoLimits} we plot contours of $\rho$ at the weak scale in the custodial-violating theory after running down from the maximum scale allowed by the stronger of the perturbative unitarity and $\rho$ parameter constraints.  $\rho > 1$ in almost all of the H5plane benchmark, except for a tiny sliver of parameter space at low $m_5 < 250$~GeV and $s_H$ below 0.4. 

\begin{figure}
	\resizebox{0.5\textwidth}{!}{\includegraphics{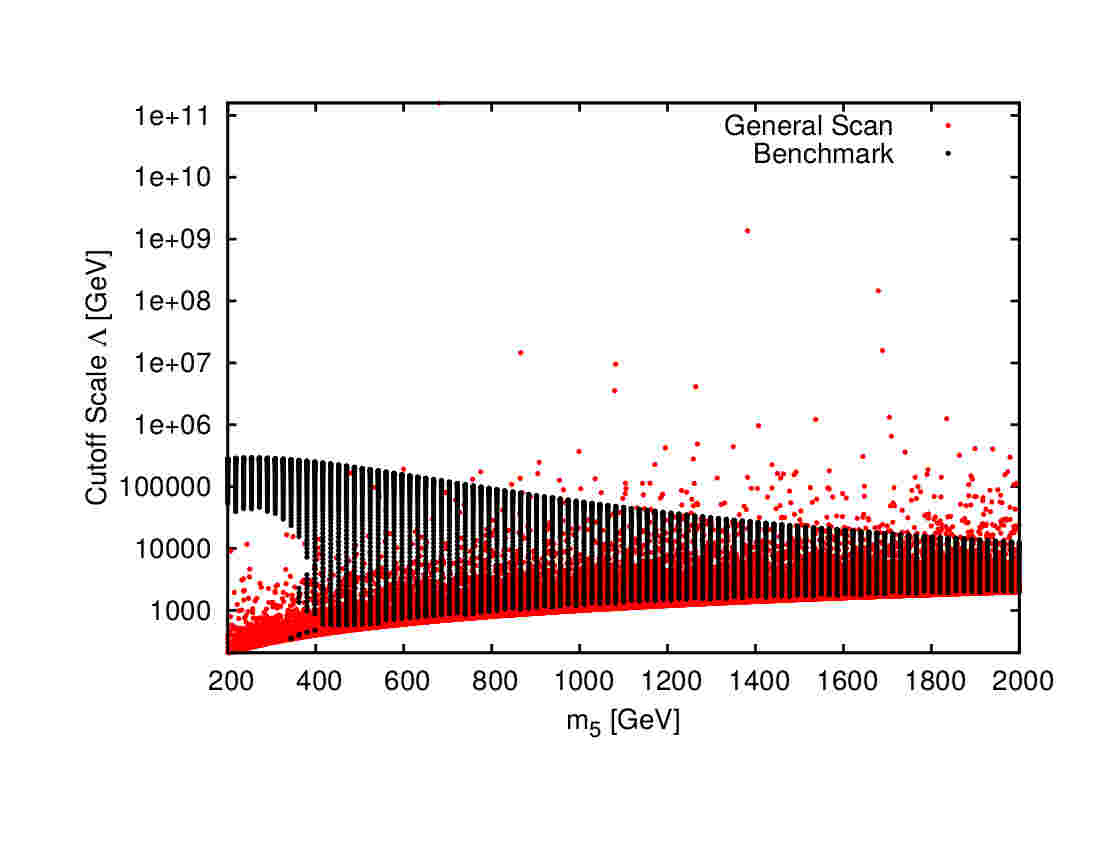}}%
	\resizebox{0.5\textwidth}{!}{\includegraphics{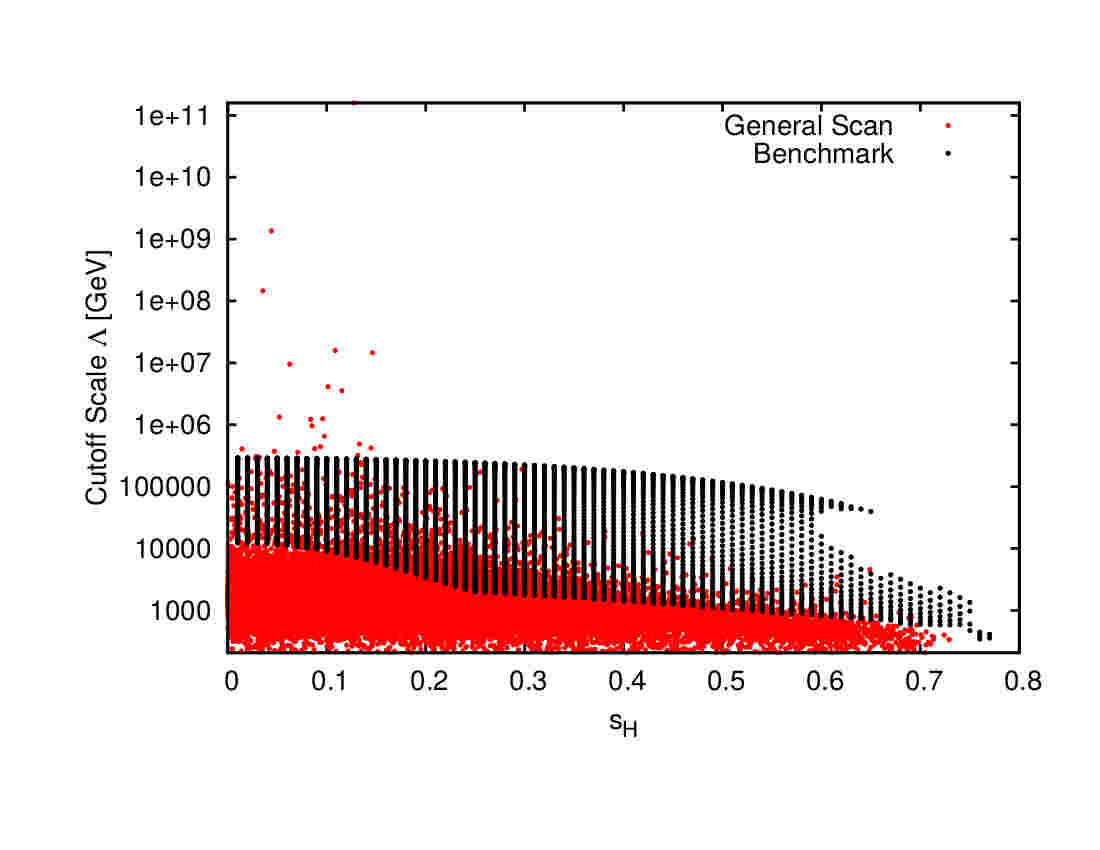}}
	\caption{The highest allowed custodial-symmetric cutoff scale due to perturbative unitarity of the quartic couplings in a general scan (red) and in the H5plane benchmark (black), as a function of $m_5$ (left) and $s_H$ (right). The highest allowed cutoff scale in the general scan ranges between 207~GeV and $1.6 \times 10^{11}$~GeV, though almost all points lie below $\sim 2 \times 10^9$~GeV.  (The point with the highest cutoff scale is at the upper edge of the plots at $m_5 = 681$~GeV and $s_H \simeq 0.12$).}
	\label{fig:GSpertLimits}
\end{figure}

In Fig.~\ref{fig:GSpertLimits} we show scatter plots comparing the perturbative unitarity constraints in the H5plane benchmark to the results of the general scan. In the left panel we plot the maximum cutoff scale $\Lambda$ versus $m_5$ while in the right panel we plot it versus $s_H$.  For high values of $m_5$ we find that the H5plane benchmark gives cutoff scales lower than are typical in the general scan. This is expected because the quartic coupling $\lambda_2$ grows with $m_5$ in the H5plane benchmark, putting those points closer to the limit from perturbative unitarity.  For lower $m_5$ values, the H5plane benchmark tends to give cutoff scales which are larger than typical in the general scan.  This is a statistical effect caused by the fact that much of the parameter space in the general scan tends to have one or more quartic coupling already moderately large, while in the H5plane benchmark specific (smaller) parameter values have been chosen by hand.  Similarly, points in the H5plane benchmark yield higher maximum cutoff scales than typical points in the general scan for all values of $s_H$.  The points in the general scan with very high maximum cutoff scales cluster at low $s_H$ below 0.2. 

\begin{figure}
	\resizebox{0.5\textwidth}{!}{\includegraphics{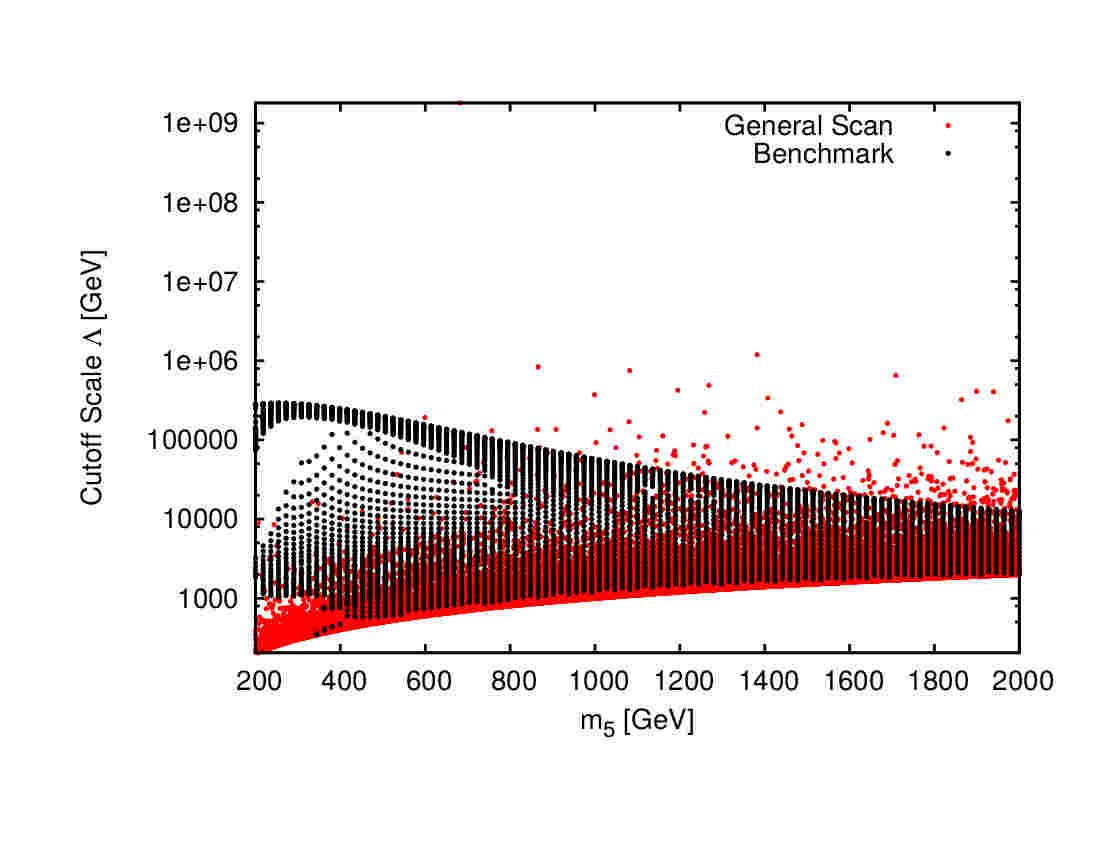}}%
	\resizebox{0.5\textwidth}{!}{\includegraphics{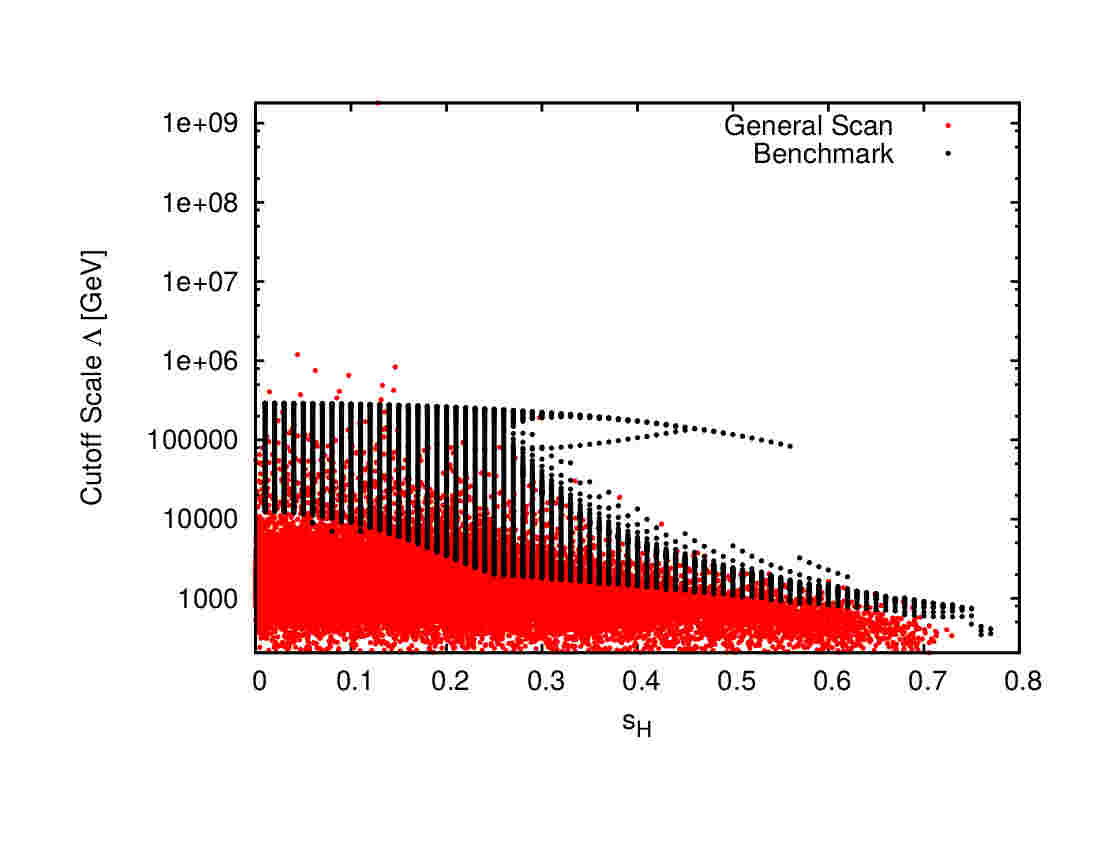}}
	\caption{The highest allowed custodial-symmetric cutoff scale imposing perturbative unitarity and the requirement that the weak-scale $\rho$ parameter lie within $\pm 2 \sigma$ of its measured value.  Red points are for a general scan and black are for the H5plane benchmark.  The highest allowed cutoff scale in the general scan ranges between 205~GeV and $1.8 \times 10^9$~GeV, though almost all the points lie below $\sim 2 \times 10^6$~GeV.  (The point with the highest cutoff scale is at the upper edge of the plots at $m_5 = 681$~GeV and $s_H \simeq 0.12$).}
	\label{fig:GSrhoLimits}
\end{figure}

In Fig.~\ref{fig:GSrhoLimits} we again show scatter plots comparing the maximum cutoff scale in the H5plane benchmark to the results of the general scan, now imposing the requirement that the $\rho$ parameter at the weak scale is within $2\sigma$ of its experimental value in addition to the perturbative unitarity requirement.  The $\rho$ parameter constraint lowers the maximum allowed cutoff scale in both the H5plane benchmark and the general scan, and brings the two distributions closer to each other.  At lower $m_5$ values the H5plane benchmark still permits somewhat atypically large cutoffs but it gives mostly typical cutoff values for higher values of $s_H$. The general scan still admits higher cutoff scales than the H5plane benchmark (particularly at large $m_5$), but the highest cutoff scale in the general scan is now less than an order of magnitude higher than that in the H5plane benchmark.

\begin{figure}
	\resizebox{0.5\textwidth}{!}{\includegraphics{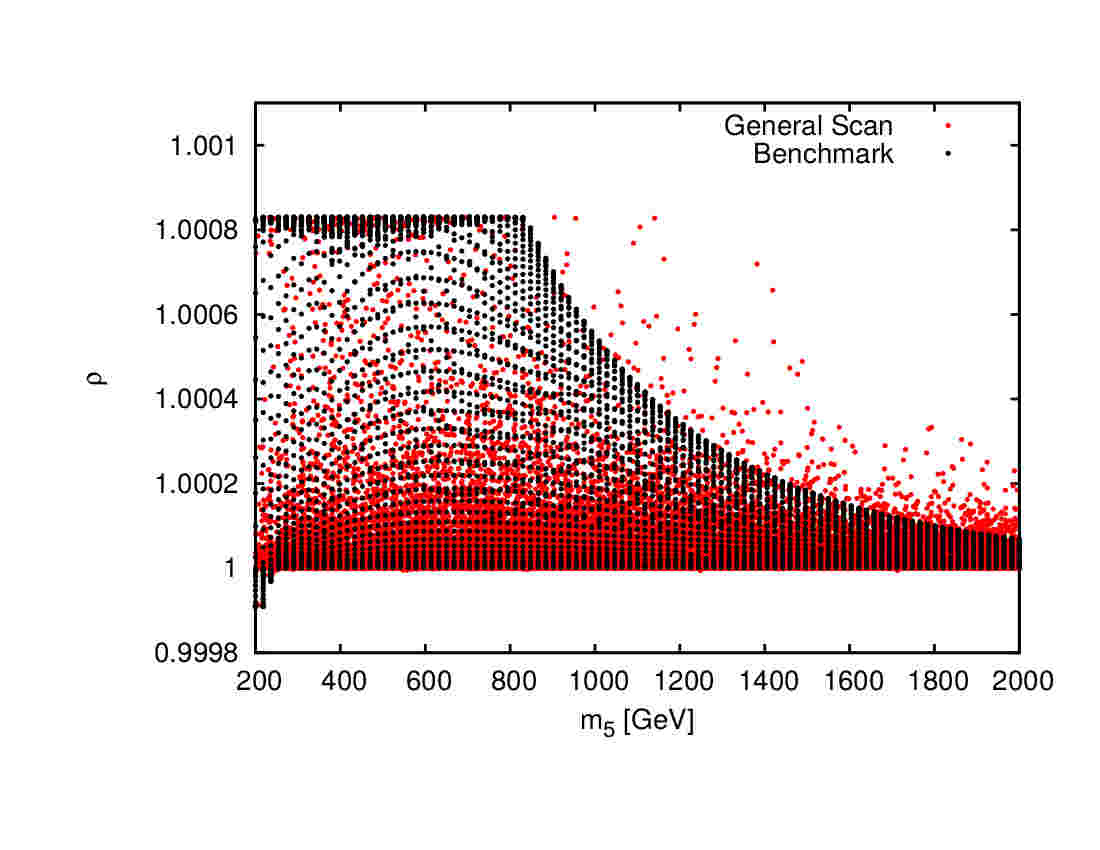}}%
	\resizebox{0.5\textwidth}{!}{\includegraphics{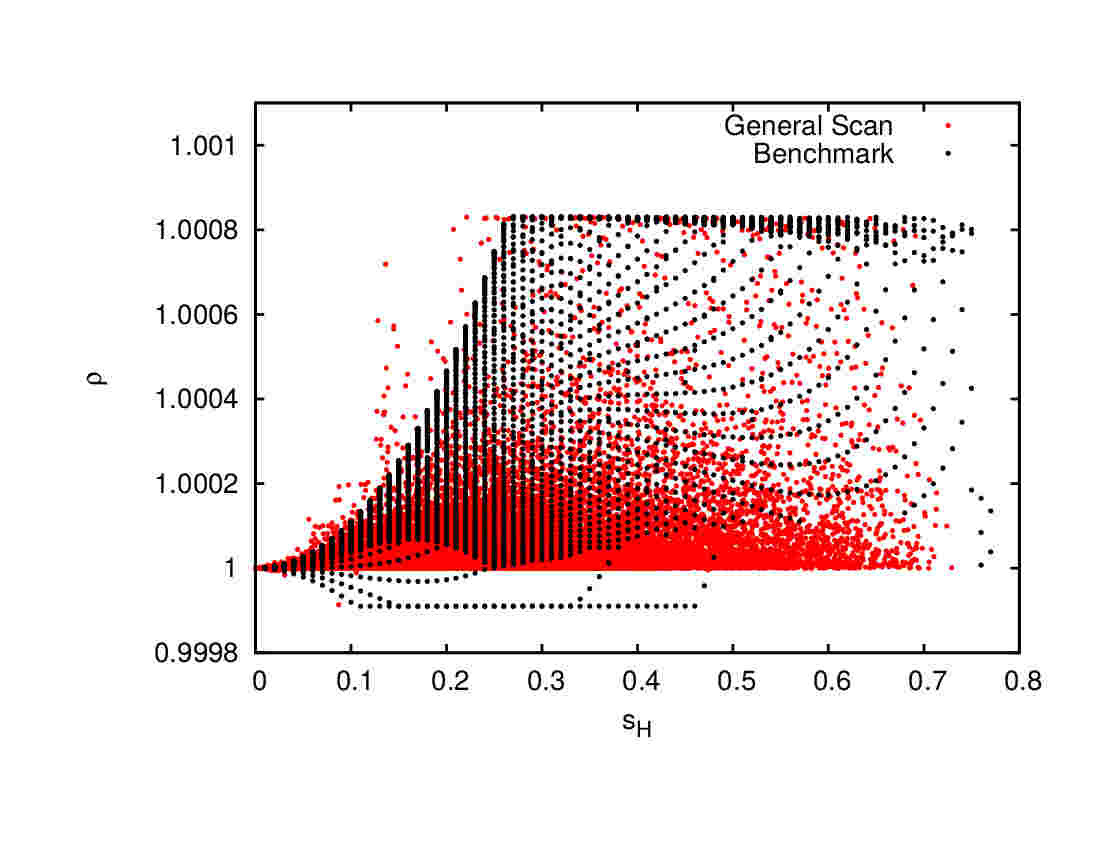}}
	\caption{The weak-scale $\rho$ parameter evaluated with the cutoff scale at its maximum allowed value in a general scan (red) and in the H5plane benchmark (black). Both the general scan and the H5plane benchmark populate the entire $\pm 2 \sigma$ allowed region of $\rho \in (0.99991, 1.00083)$.}
	\label{fig:GSrhoValue}
\end{figure}

In Fig.~\ref{fig:GSrhoValue} we plot the value of the weak-scale $\rho$ parameter when the cutoff scale is at its maximum value allowed by perturbative unitarity and the experimental constraints on $\rho$ in the H5plane benchmark and the general scan.  We see that, as in the H5plane benchmark, the general scan yields $\rho \geq 1$ in the overwhelming majority of parameter space (as mildly favoured by experiment).  Indeed, the region at very low $m_5$ in the H5plane benchmark in which $\rho < 1$ is quite atypical in the general scan.  The general scan also tends to give slightly larger values of $\rho$ at higher $m_5$, as one would expect given the higher maximum cutoff scales (and hence more custodial-symmetry-violation-inducing running) in this mass range in the general scan. 

\subsection{Custodial violation in couplings}

Custodial symmetry violation can modify the phenomenology of the GM model by changing the decay patterns of the physical Higgs bosons.  The most experimentally-interesting manifestations of this are in the ratio of the couplings of the SM-like Higgs boson mass eigenstate $\tilde h$ to $W$ boson and $Z$ boson pairs, $\lambda^{\tilde h}_{WZ} \equiv \kappa^{\tilde h}_W / \kappa^{\tilde h}_Z$ [Eq.~(\ref{eq:lambdaWZ})], and in the couplings of the otherwise-fermiophobic mass eigenstates $\tilde H_5^{\pm}$ and $\tilde H_5^0$ to fermion pairs induced by custodial-violating mixing among the custodial-symmetry eigenstates [Eqs.~(\ref{eq:kappaH5+}) and (\ref{eq:kappaH50})].  In what follows we maximize the custodial-violating effects by taking the scale of the custodial-symmetric theory as high as possible, subject to the constraints from perturbative unitarity and the $\rho$ parameter.  In what follows, we focus on the H5plane benchmark and its comparison to a general parameter scan.  We discuss the dedicated low-$m_5$ parameter scan in Sec.~\ref{sec:lowm5}.

\begin{figure}
	\includegraphics[scale=0.5]{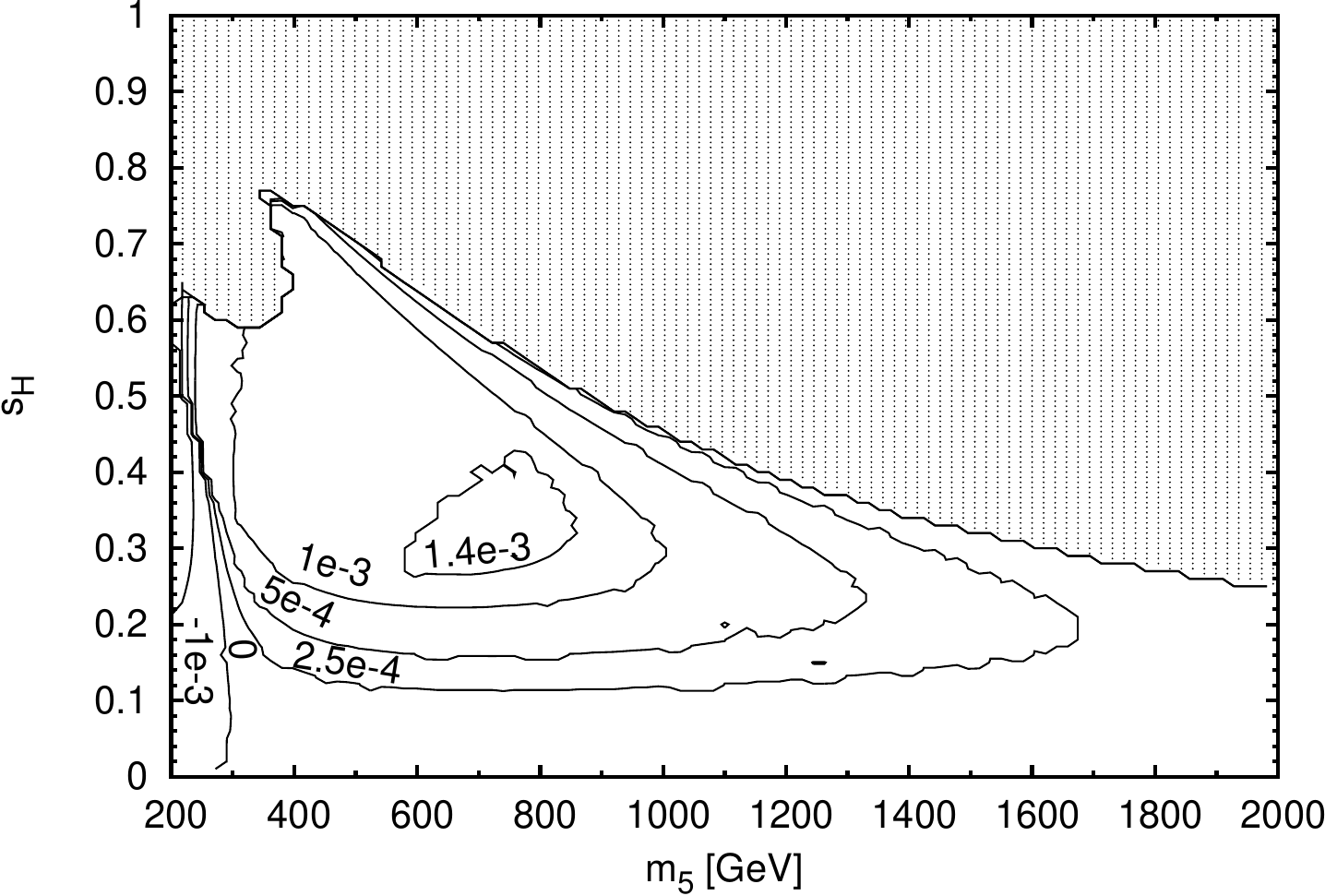}
	\caption{Contours of $\delta \lambda^{\tilde h}_{WZ} \equiv \lambda^{\tilde h}_{WZ} -1$ in the H5plane benchmark, taking the scale of the custodial-symmetric theory to be as large as possible subject to perturbative unitarity and the $\rho$ parameter constraint.  $\delta \lambda^{\tilde h}_{WZ}$ varies between $-5.1 \times 10^{-3}$ and $1.4 \times 10^{-3}$.  
	}
	\label{fig:lambdahwz}
\end{figure}

In Fig.~\ref{fig:lambdahwz} we plot the deviation of $\lambda^{\tilde h}_{WZ}$ from its SM value of 1 in the H5plane benchmark. The effect is tiny, reaching at most half a percent in a small region of the H5plane benchmark with $m_5 \lesssim 250$~GeV and moderate values of $s_H$; for larger $m_5$, the deviation is below two per mille.  This deviation is well below the sensitivity of the current experimental measurement at the LHC, $\lambda^{\tilde h}_{WZ} = 0.88^{+0.10}_{-0.09}$~\cite{Khachatryan:2016vau}.  It is also below the expected sensitivity obtained by combining the projections for the measurement precision of the SM Higgs couplings $\kappa_W$ and $\kappa_Z$ at the High-Luminosity LHC (a few percent) and the proposed International Linear $e^+e^-$ Collider (ILC) (roughly half a percent) as summarized in Ref.~\cite{Dawson:2013bba}.  The proposed Future Circular Collider (FCC-ee) could begin to reach the required precision, with projected sensitivity for $\kappa_W$ and $\kappa_Z$ of 1.5 to 2 per mille~\cite{dEnterria:2016fpc}.\footnote{Because these coupling extraction methods are based on measurements of Higgs production cross sections and decay branching ratios, they probe only the magnitude of $\lambda^{\tilde h}_{WZ}$, not the sign; a method involving the dependence of the $\tilde h \to 4 \ell$ decay distributions on the $\tilde h WW$ coupling at one loop provides sensitivity to the sign of $\lambda^{\tilde h}_{WZ}$, but can achieve a precision only of order 20--50\% at the High-Luminosity LHC~\cite{Chen:2016ofc}.}

\begin{figure}
	\resizebox{0.5\textwidth}{!}{\includegraphics{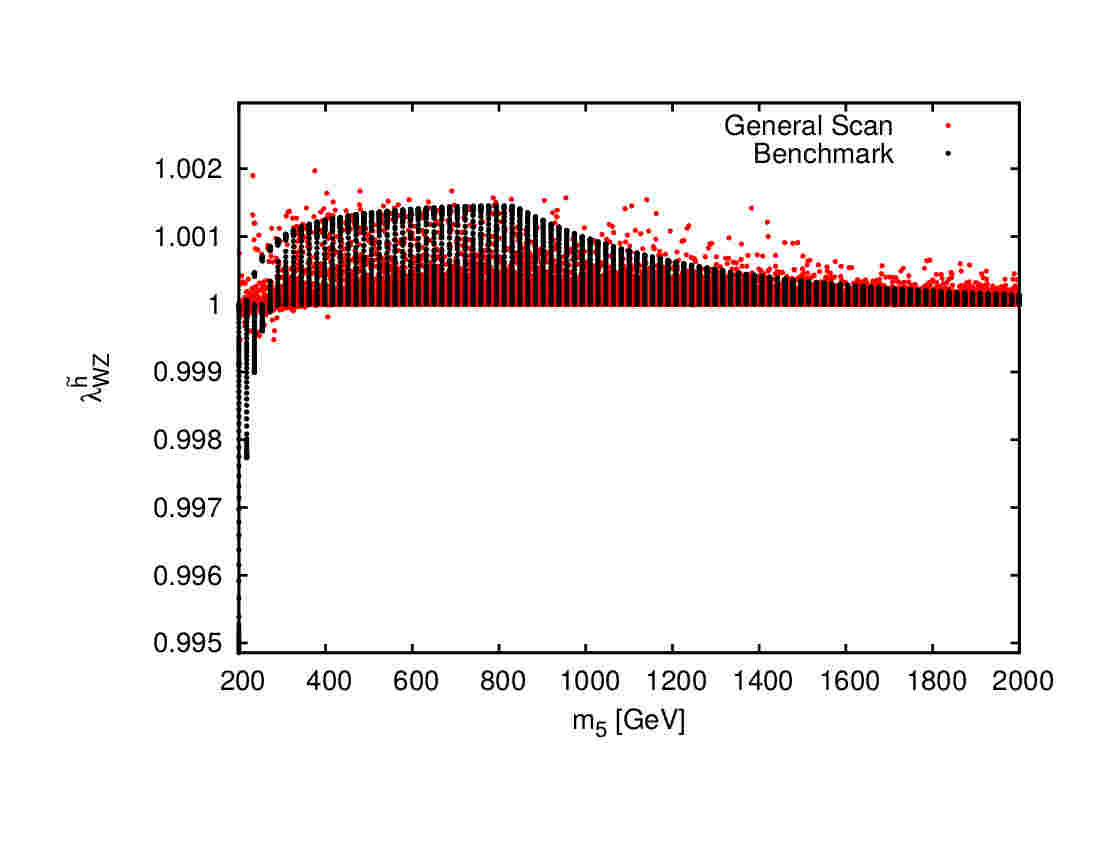}}%
	\resizebox{0.5\textwidth}{!}{\includegraphics{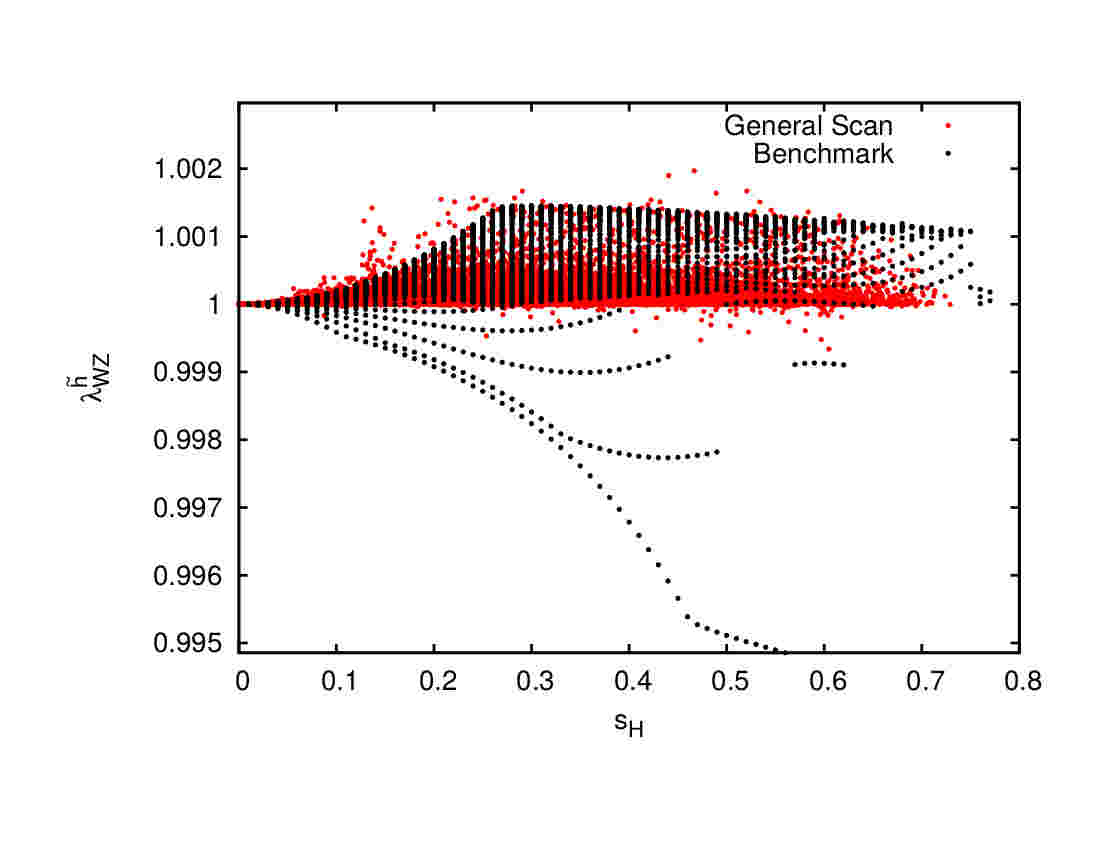}}
	\caption{$\lambda^{\tilde h}_{WZ} \equiv \kappa^{\tilde h}_W / \kappa^{\tilde h}_Z$ evaluated with the maximum allowed cutoff scale for a general parameter scan (red) and in the H5plane benchmark (black), as a function of $m_5$ (left) and $s_H$ (right). The minimum value in the general scan is 0.99934 and the maximum value is 1.00197.}
	\label{fig:GSlambdahwz}
\end{figure}

In Fig.~\ref{fig:GSlambdahwz} we compare the value of $\lambda^{\tilde h}_{WZ}$ evaluated with the maximum allowed cutoff scale in the H5plane benchmark (black points) to the results of a general parameter scan (red points).  The range of deviations in the H5plane benchmark is representative of that in the general scan, except for $m_5 \simeq 200$~GeV and moderate $s_H$ in which the H5plane benchmark probes a rather atypical region of parameter space in which negative deviations of up to half a percent are possible.  In the general scan the deviation is typically below 0.2\%.

\begin{figure} 
\resizebox{0.5\textwidth}{!}{\includegraphics{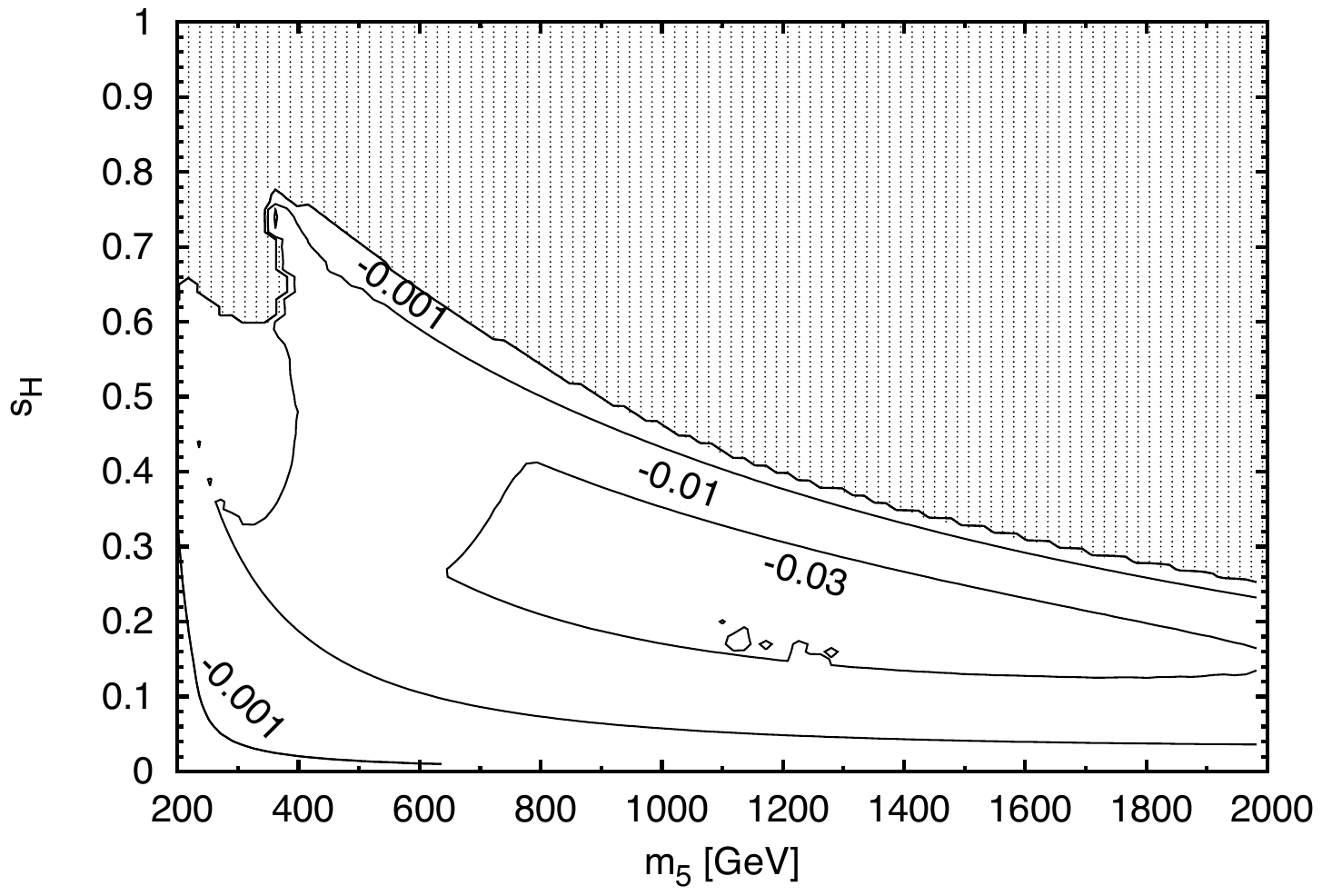}}%
\resizebox{0.5\textwidth}{!}{\includegraphics{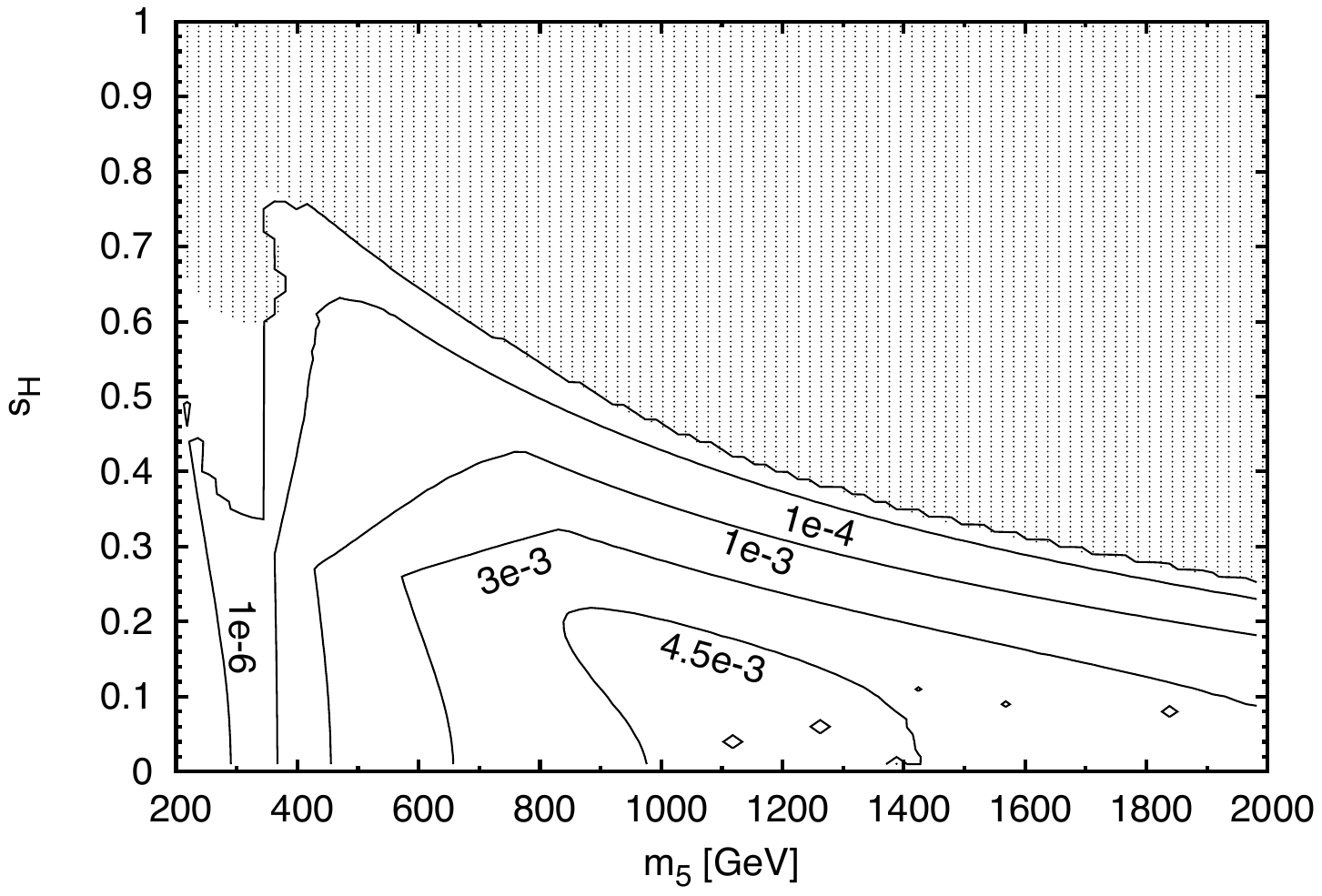}}
	\caption{The coupling of $\tilde H_5^0$ to fermions and the resulting fermionic branching ratio in the H5plane benchmark, taking the scale of the custodial-symmetric theory to be as large as possible subject to perturbative unitarity and the $\rho$ parameter constraint. Left: contours of $\kappa_f^{\tilde H_5^0}$ [defined above Eq.~(\ref{eq:kappaH50})].  The allowed values range between $-4.0 \times 10^{-2}$ and $2.0 \times 10^{-3}$.  Right: contours of the branching ratio of $\tilde H_5^0$ to fermions.  We compute only the partial width to the heaviest kinematically accessible pair of fermions; i.e., to $t \bar t$ for $m_{\tilde H_5^0} > 2 m_t$ and $b \bar b$ otherwise.  The branching ratio of $\tilde H_5^0$ to fermions ranges from $3.5 \times 10^{-11}$ to $4.8 \times 10^{-3}$.}
	\label{fig:kappaH5Nf}
\end{figure}

In Fig.~\ref{fig:kappaH5Nf} we plot the custodial-violation-induced coupling and branching ratio of $\tilde H_5^0$ to fermions in the H5plane benchmark.  The $\tilde H_5^0$ coupling to fermions $\kappa_f^{\tilde H_5^0}$ reaches a magnitude of at most 0.04 in the H5plane benchmark, leading to fermion-induced (e.g., via gluon fusion) production cross sections at most $(0.04)^2 = 1.6 \times 10^{-3}$ times that of a SM Higgs boson of the same mass.  Potentially more interesting is the effect of this coupling on the $\tilde H_5^0$ decays: as shown in the right panel of Fig.~\ref{fig:kappaH5Nf}, the branching ratio of $\tilde H_5^0$ to fermions can reach almost half a percent in the H5plane benchmark.  For $\tilde H_5^0$ masses above 350~GeV, these fermionic decays are overwhelmingly into $t \bar t$ pairs.

\begin{figure}
	\resizebox{0.5\textwidth}{!}{\includegraphics{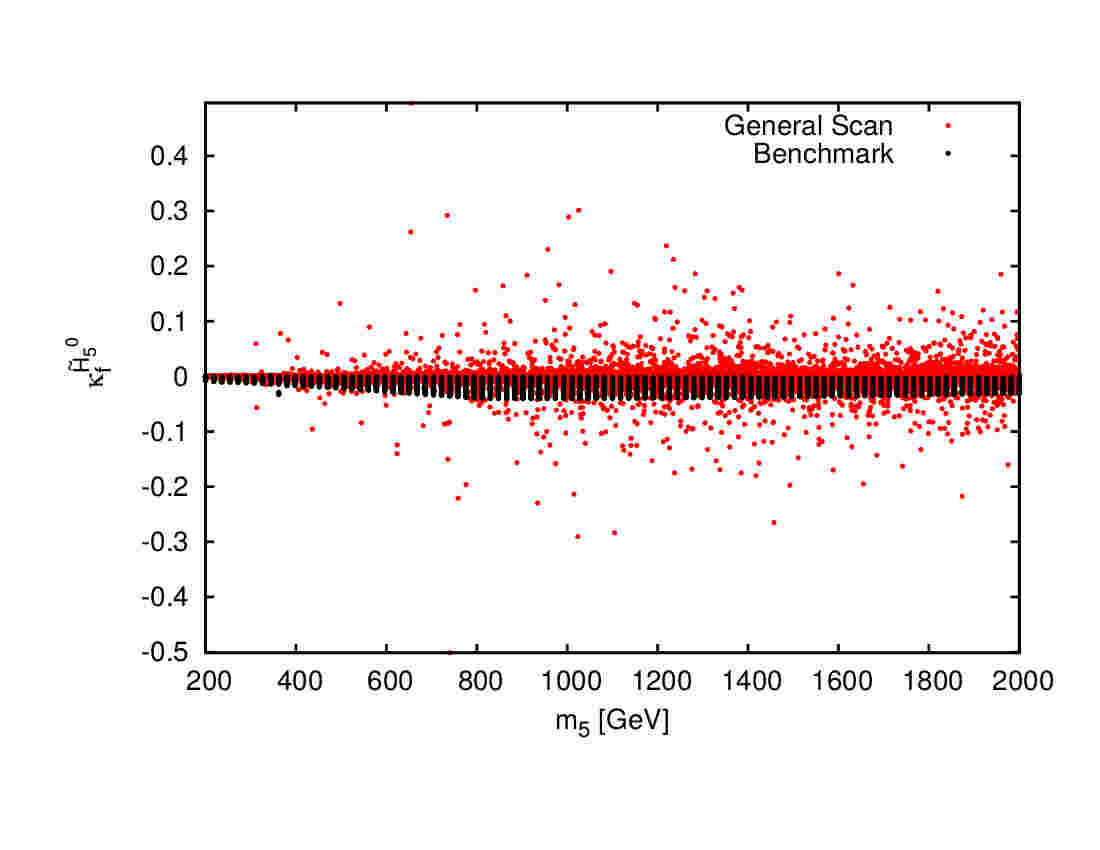}}%
	\resizebox{0.5\textwidth}{!}{\includegraphics{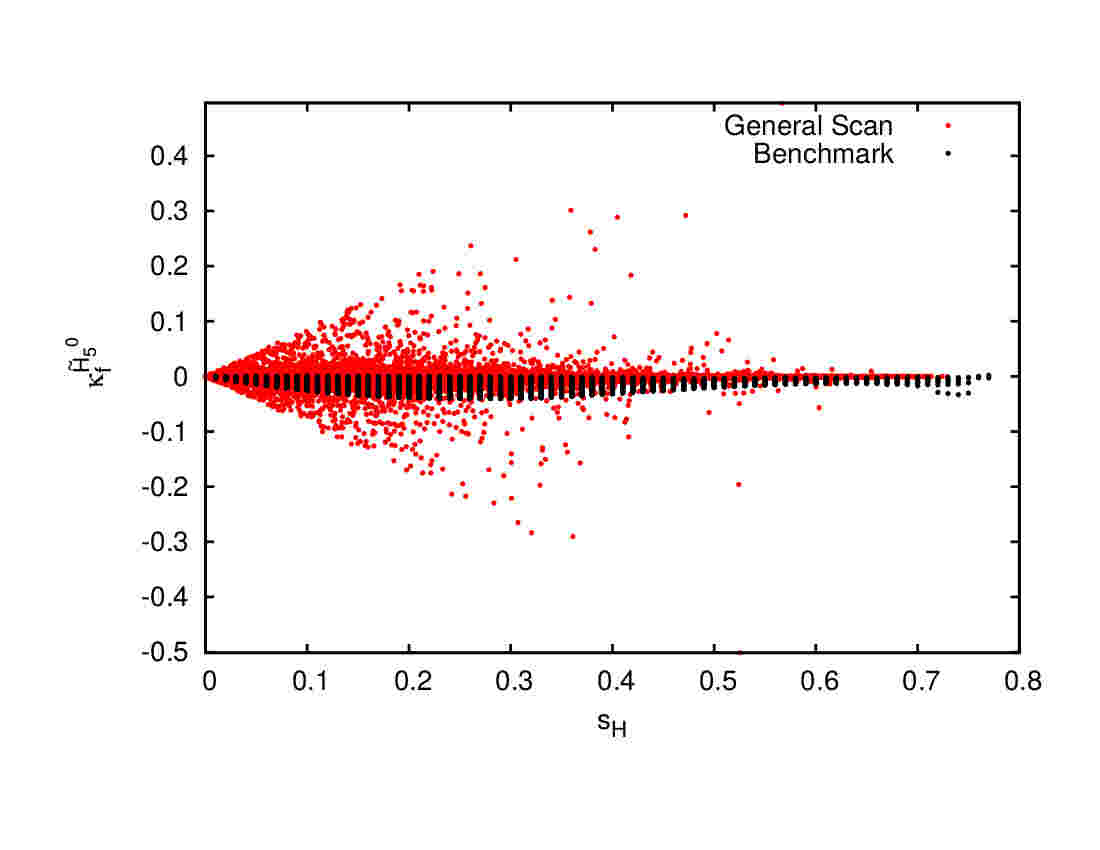}}
	\caption{The coupling $\kappa_f^{\tilde H_5^0}$ of $\tilde H_5^0$ to fermions evaluated with the maximum allowed cutoff scale in a general parameter scan (red) and in the H5plane benchmark (black), as a function of $m_5$ (left) and $s_H$ (right).  In the general scan the coupling ranges between $-0.50$ and $+0.50$ (for rare points at large $s_H$ between 0.5 and 0.6).  This coupling is zero in the custodial-symmetric model.}
	\label{fig:GSKappaH5Nf}
\end{figure}

\begin{figure}
	\resizebox{0.5\textwidth}{!}{\includegraphics{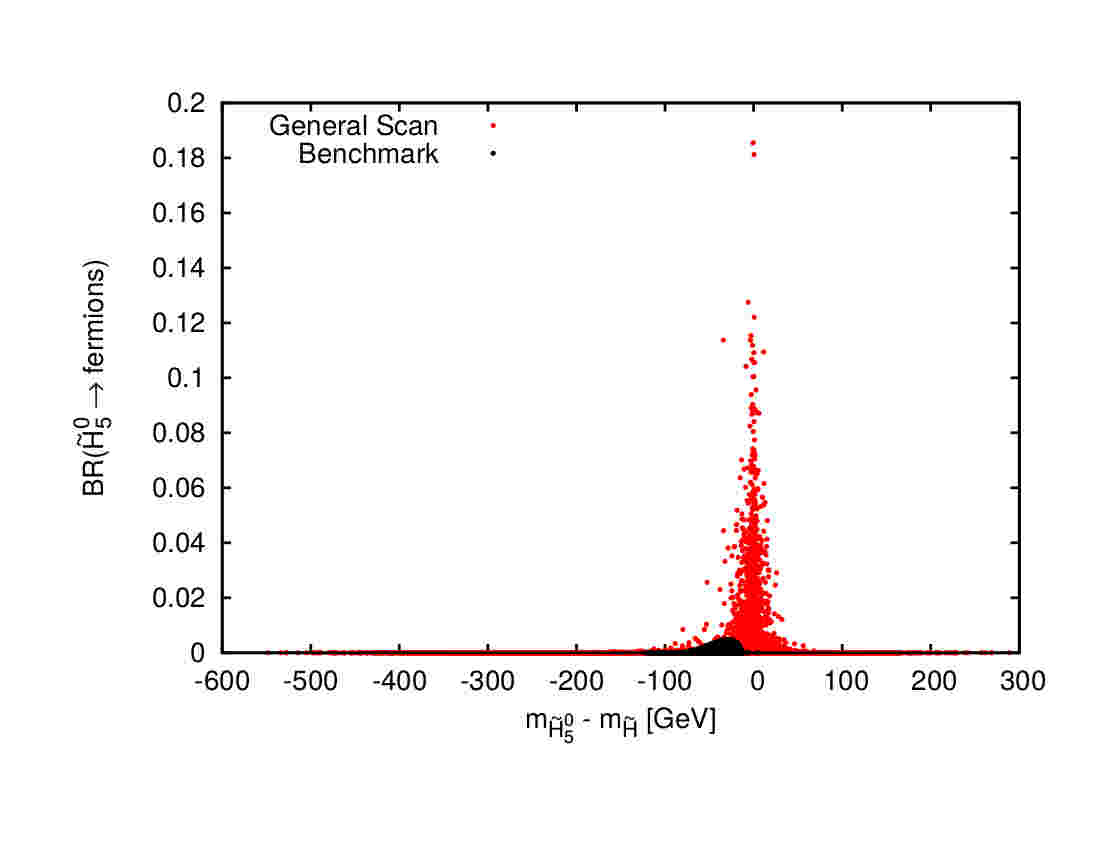}}%
	\resizebox{0.5\textwidth}{!}{\includegraphics{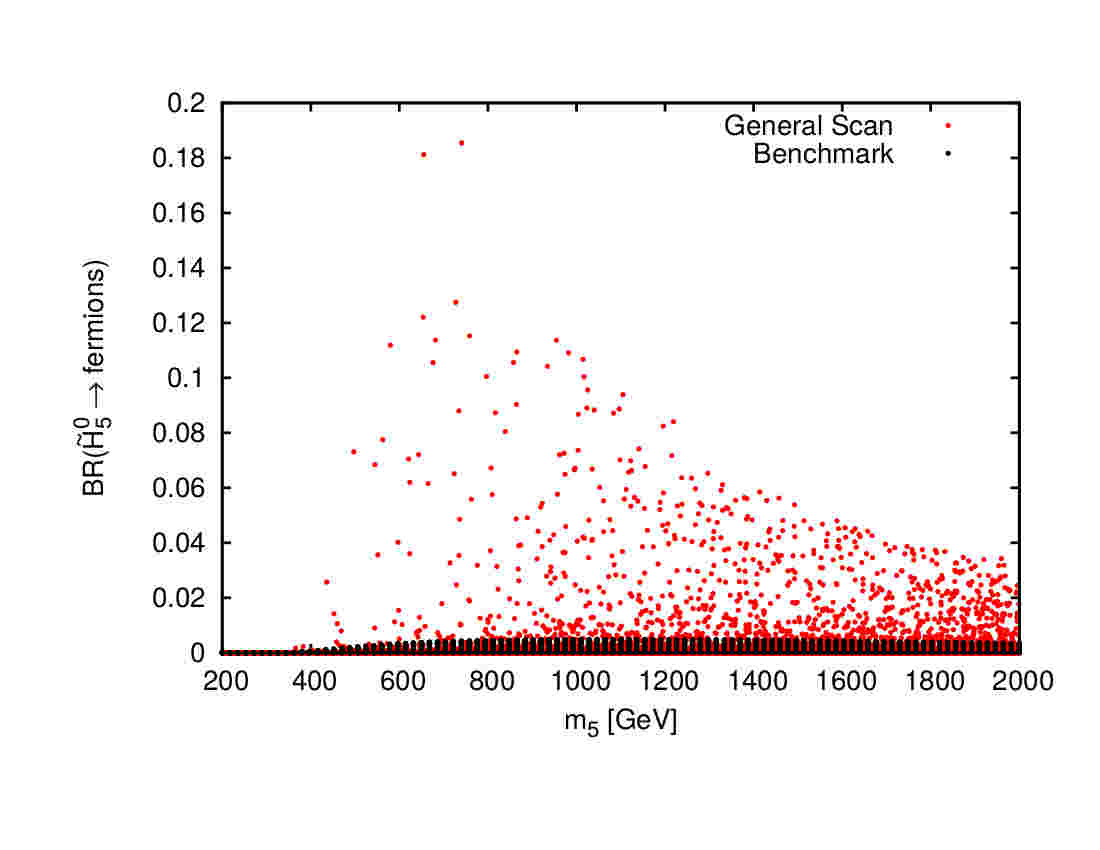}}
	\caption{Branching ratio of $\tilde H_5^0 \to f \bar f$ evaluated with the maximum allowed cutoff scale in a general parameter scan (red) and in the H5plane benchmark (black), as a function of the mass difference $m_{\tilde H_5^0} - m_{\tilde H}$ showing the resonant mixing effect (left) and $m_5$ (right). The maximum branching ratio to fermions in the general scan is 19\%.}
	\label{fig:GSH5NfBR}
\end{figure}

In Fig.~\ref{fig:GSKappaH5Nf} we compare these results to the range of $\kappa_f^{\tilde H_5^0}$ accessible in the general parameter scan.  The general scan can yield significantly larger values of this  custodial-violation-induced coupling, reaching as high as $\pm 0.5$ and populating both positive and negative values.  The maximum size of the coupling grows with $s_H$.  In contrast, the H5plane benchmark yields quite small couplings of magnitude at most $0.04$ and mainly negative values.
The large custodial-symmetry-violating coupling values in the general scan are due to resonant mixing between the $H_5^0$ and $H$ states when their masses are nearly degenerate.  We illustrate this in the left panel of Fig.~\ref{fig:GSH5NfBR}, where we plot BR($\tilde H_5^0 \to f \bar f$) as a function of the mass difference between $\tilde H_5^0$ and $\tilde H$.  In the mass-degenerate region the mixing is enhanced and fermionic branching ratios on the order of 10--20\% are possible.  We also show this branching ratio as a function of $m_5$ in the right panel of Fig.~\ref{fig:GSH5NfBR}; the branching ratio to fermions reaches its maximum for $m_5$ between 600 and 800~GeV and falls with increasing $m_5$.

\begin{figure} 
\resizebox{0.5\textwidth}{!}{\includegraphics{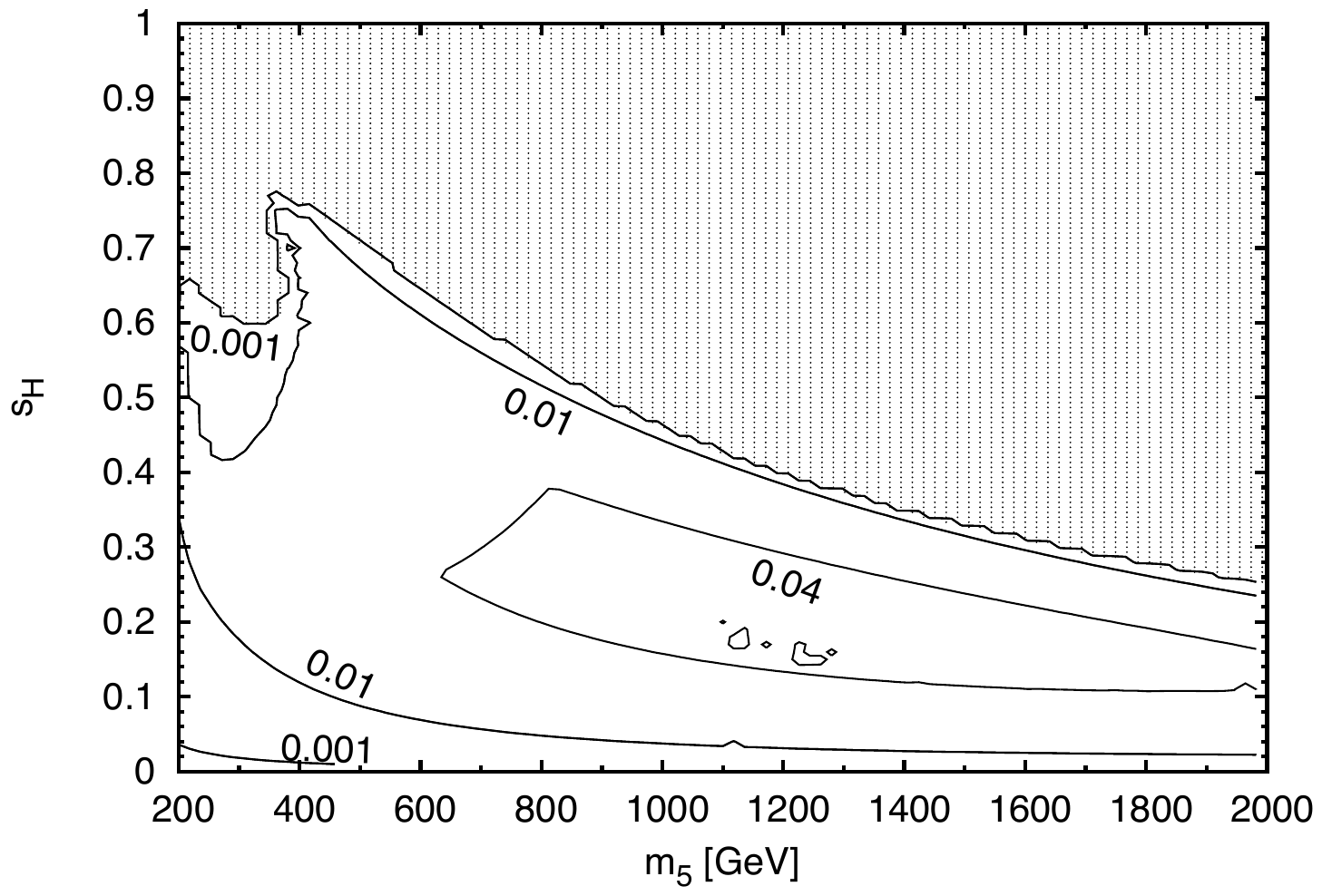}}%
\resizebox{0.5\textwidth}{!}{\includegraphics{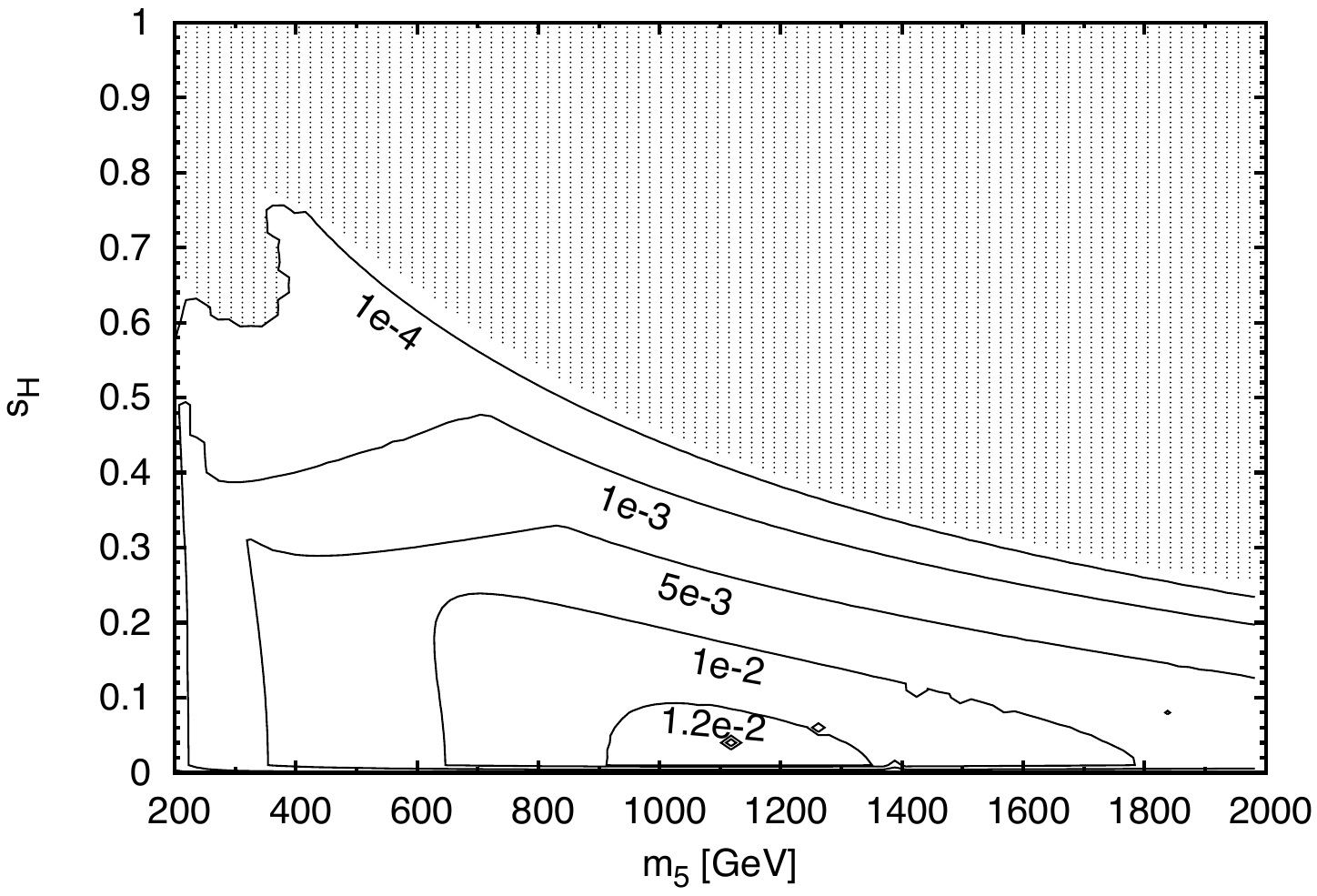}}
	\caption{The coupling of $\tilde H_5^+$ to fermions and the resulting fermionic branching ratio in the H5plane benchmark, taking the scale of the custodial-symmetric theory to be as large as possible subject to perturbative unitarity and the $\rho$ parameter constraint.  Left: contours of $\kappa_f^{\tilde H_5^+}$ [defined in Eq.~(\ref{eq:kappaH5pdef})].  The allowed values range between $1.0 \times 10^{-4}$ and $5.2 \times 10^{-2}$.  Right: contours of the branching ratio of $\tilde H_5^+$ to fermions, including only the decay to $t \bar b$. This branching ratio ranges from $2.0 \times 10^{-8}$ to $1.2 \times 10^{-2}$.}
	\label{fig:kappaH5Pf} 
\end{figure}

In Fig.~\ref{fig:kappaH5Pf} we plot the custodial-violation-induced coupling and branching ratio of $\tilde H_5^{\pm}$ to fermions in the H5plane benchmark.  The $\tilde H_5^{\pm}$ coupling to fermions $\kappa_f^{\tilde H_5^+}$ reaches a magnitude of at most 0.052 in the H5plane benchmark.  Again, production processes involving $\tilde H_5^+$ coupling to fermions, such as associated production with a top quark, will have cross sections that are far too small to be interesting at the LHC.  The branching ratio of $\tilde H_5^+ \to t \bar b$ can reach 1.2\%, as shown in the right panel of Fig.~\ref{fig:kappaH5Pf}.

\begin{figure}
	\resizebox{0.5\textwidth}{!}{\includegraphics{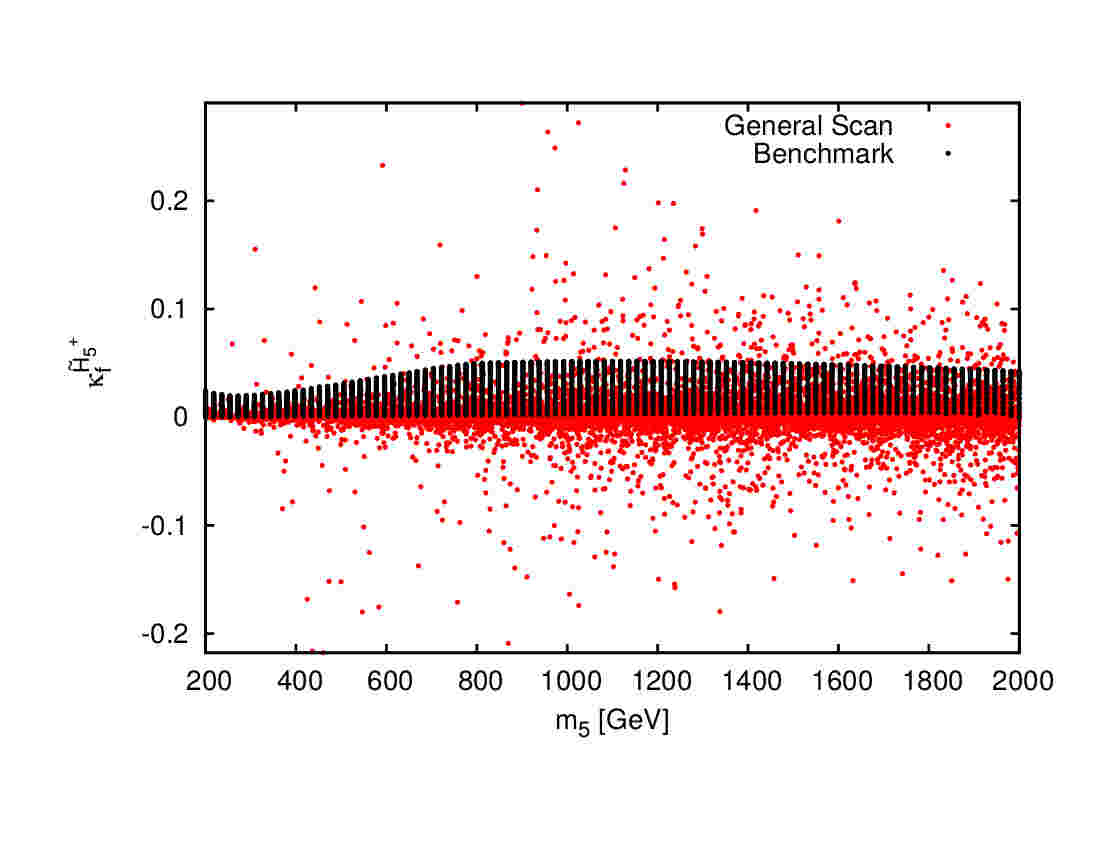}}%
	\resizebox{0.5\textwidth}{!}{\includegraphics{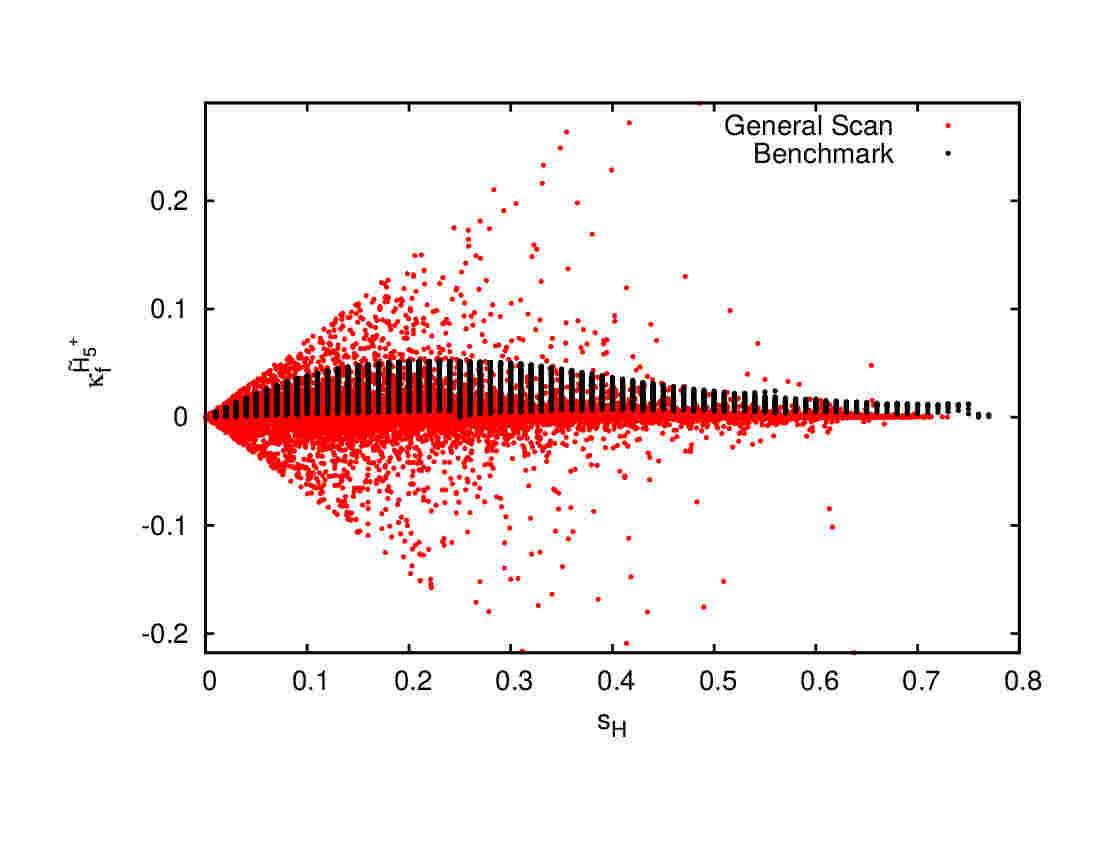}}
	\caption{The coupling $\kappa_f^{\tilde H_5^+}$ of $\tilde H_5^+$ to fermions evaluated with the maximum allowed cutoff scale in a general parameter scan (red) and in the H5plane benchmark (black), as a function of $m_5$ (left) and $s_H$ (right). In the general scan the coupling ranges between $-0.22$ and $+0.29$.}
	\label{fig:GSKappaH5Pf}
\end{figure}

\begin{figure}
	\resizebox{0.5\textwidth}{!}{\includegraphics{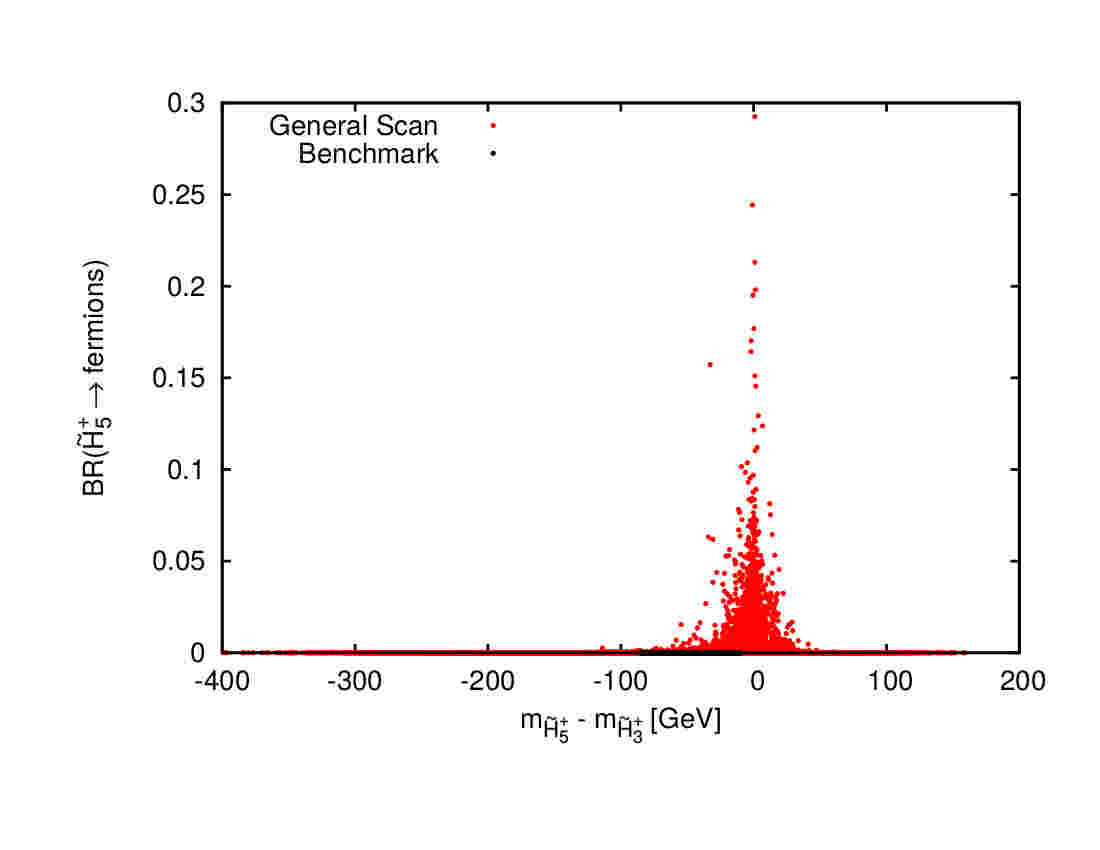}}%
	\resizebox{0.5\textwidth}{!}{\includegraphics{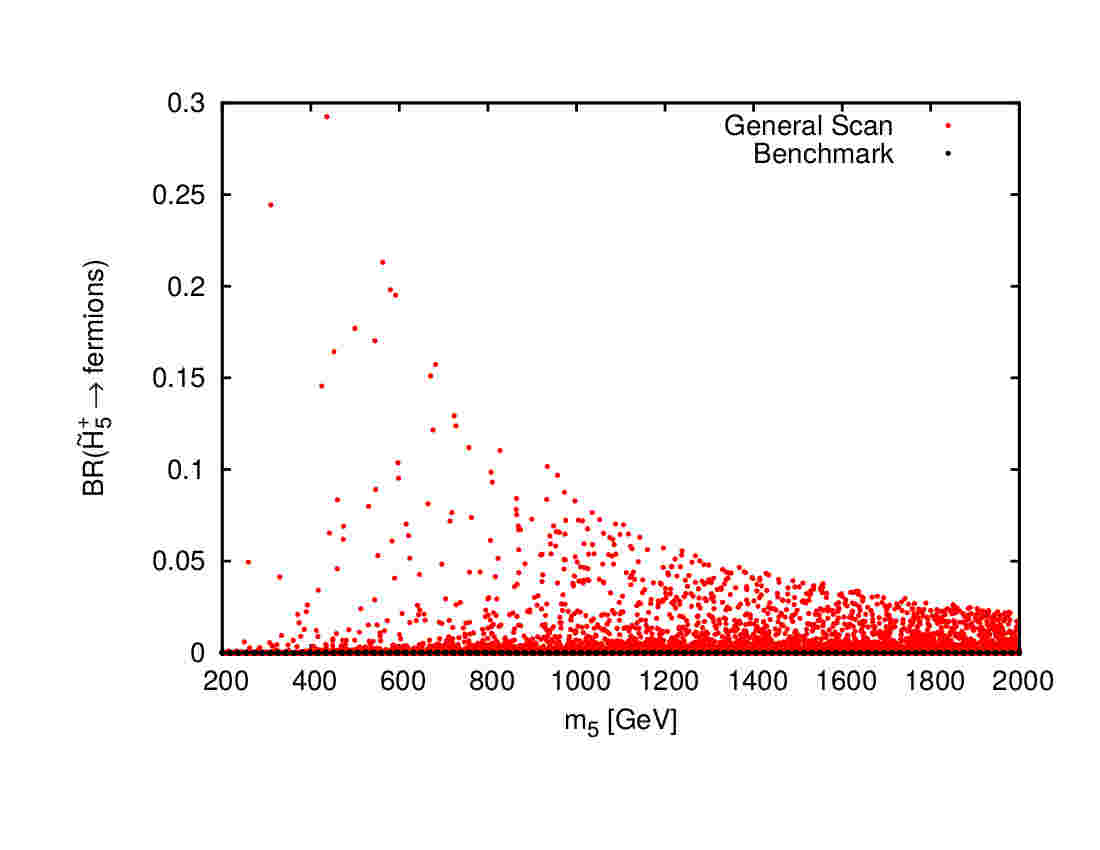}}
	\caption{Branching ratio of $\tilde H_5^+ \to t \bar b$ evaluated with the maximum allowed cutoff scale in a general parameter scan (red) and in the H5plane benchmark (black), as a function of the mass difference $m_{\tilde H_5^+} - m_{\tilde H_3^+}$ showing the resonant mixing effect (left) and $m_5$ (right). The maximum branching ratio to fermions in the general scan is 29\%.}
	\label{fig:GSH5PfBR}
\end{figure}

In Fig.~\ref{fig:GSKappaH5Pf} we compare these results to the range of $\kappa_f^{\tilde H_5^+}$ accessible in the general parameter scan.  The general scan can again yield larger values of this custodial-violation-induced coupling, reaching a magnitude of at most 0.3 and populating both positive and negative values.  The maximum size of the coupling again grows with $s_H$.  In contrast, the H5plane benchmark yields somewhat smaller values of this coupling of at most 0.052 and only populates positive values.  The large custodial-symmetry-violating coupling values that can be obtained in the general scan are again a consequence of resonant mixing, this time between the $H_5^+$ and the $H_3^+$ states when they are nearly degenerate.  We illustrate this in the left panel of Fig.~\ref{fig:GSH5PfBR}, where we plot BR($\tilde H_5^+ \to f \bar f^{\prime}$) as a function of the mass difference between $\tilde H_5^+$ and $\tilde H_3^+$.  In this mass-degenerate region the mixing is enhanced and fermionic branching ratios on the order of 20--30\% are possible.  We also show this branching ratio as a function of $m_5$ in the right panel of Fig.~\ref{fig:GSH5PfBR}; the branching ratio to fermions again falls with increasing $m_5$.

The custodial-violation-induced decays of $\tilde H_5^0$ and $\tilde H_5^{\pm}$ to fermion pairs do not dramatically alter the phenomenology within the H5plane benchmark, and do so in the general scans only when there are near mass degeneracies with the fermiophilic heavy Higgs bosons $H$ or $H_3^+$.  Potentially more interesting is the effect of fermionic decays of these particles for low masses below the $WW$ or $WZ$ thresholds, when the dominant diboson decays of these scalars go off shell.  In the custodial-symmetric GM model, $H_5^0$ decays to $\gamma\gamma$ and $H_5^+$ decays to $W^+ \gamma$ become interesting for these low masses~\cite{Delgado:2016arn,Degrande:2017naf,Logan:2018wtm}; competition from custodial-violation-induced fermionic decays could dramatically change the phenomenology in this mass region.  We perform a detailed study of this low $m_5$ region in Sec.~\ref{sec:lowm5}.

\subsection{Custodial-violating mass splittings}

Custodial symmetry violation also induces splittings between the masses of the otherwise-degenerate custodial fiveplet and triplet states.  These splittings follow a universal pattern everywhere within the H5plane benchmark and over the vast majority of the parameter space of our general scans.  We again maximize the custodial-violating effects in what follows by taking the scale of the custodial-symmetric theory as high as possible, subject to the constraints from perturbative unitarity and the $\rho$ parameter.

\begin{figure}
\resizebox{0.5\textwidth}{!}{\includegraphics{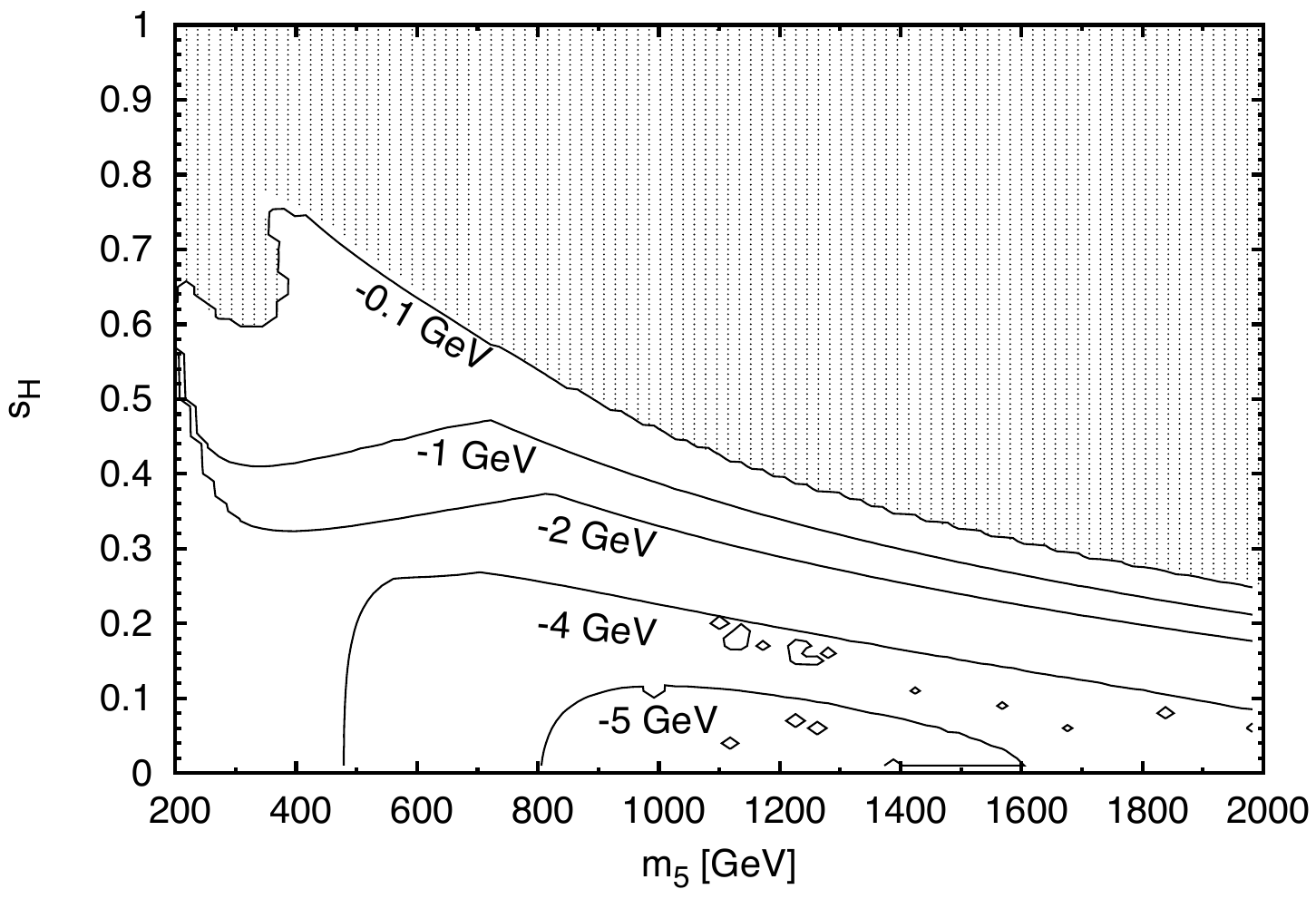}}%
\resizebox{0.5\textwidth}{!}{\includegraphics{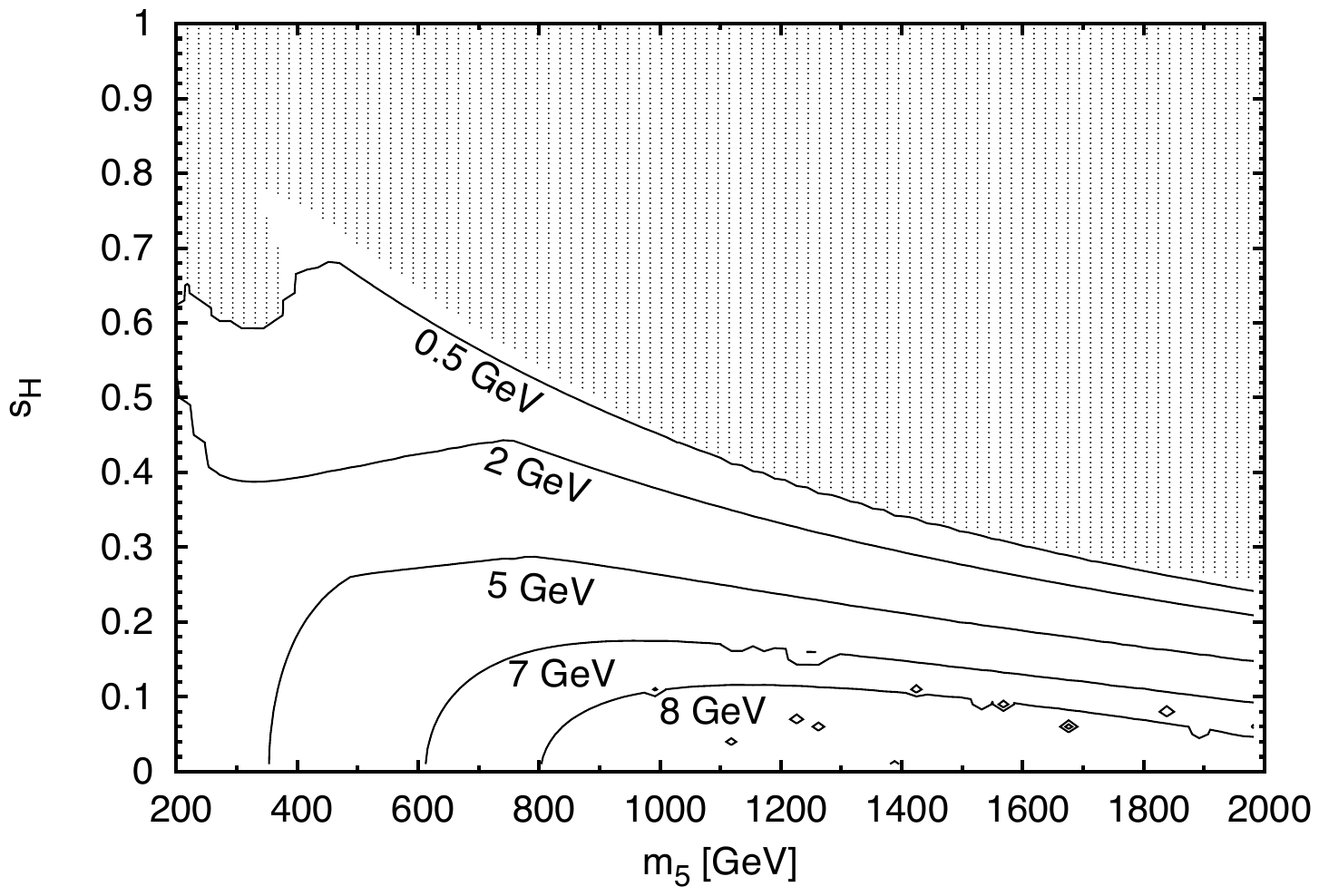}}
	\caption{The mass splittings within the custodial triplet in the H5plane benchmark, taking the scale of the custodial-symmetric theory to be as large as possible subject to perturbative unitarity and the $\rho$ parameter constraint.  Left: $m_{\tilde H_3^+} - m_{\tilde H_3^0}$.  This quantity is negative because $\tilde H_3^+$ is lighter than $\tilde H_3^0$.  The mass splitting ranges between zero and 5.3~GeV.  Right: $m_{\tilde H_3^0} - m_3$, where $m_3$ is the weak-scale custodial-symmetric input value of the custodial triplet mass.  $m_{\tilde H_3^0}$ and $m_{\tilde H_3^+}$ are both larger than $m_3$ over the entire benchmark.  In our numerical scan, the difference between $m_{\tilde H_3^0}$ and $m_3$ ranges between 4~MeV and 9.1~GeV.}
	\label{fig:H3MassSplittings}
\end{figure}

Among the custodial-triplet mass eigenstates, $\tilde H_3^0$ is almost always heavier than $\tilde H_3^+$, and both of these masses are shifted up relative to the weak-scale custodial-symmetric input value of $m_3$.  The splittings are small, as shown in Fig.~\ref{fig:H3MassSplittings} for the H5plane benchmark: the mass difference between $\tilde H_3^0$ and $\tilde H_3^+$ reaches at most 5.3~GeV (left panel of Fig.~\ref{fig:H3MassSplittings}).  The shift of the $\tilde H_3^0$ mass upward from the input value of $m_3$ is shown in the right panel of Fig.~\ref{fig:H3MassSplittings}, and is at most 9.1~GeV.  The shift of the $\tilde H_3^+$ mass from the input $m_3$ value is smaller, reaching at most 3.9~GeV in the benchmark.

\begin{figure}
	\resizebox{0.5\textwidth}{!}{\includegraphics{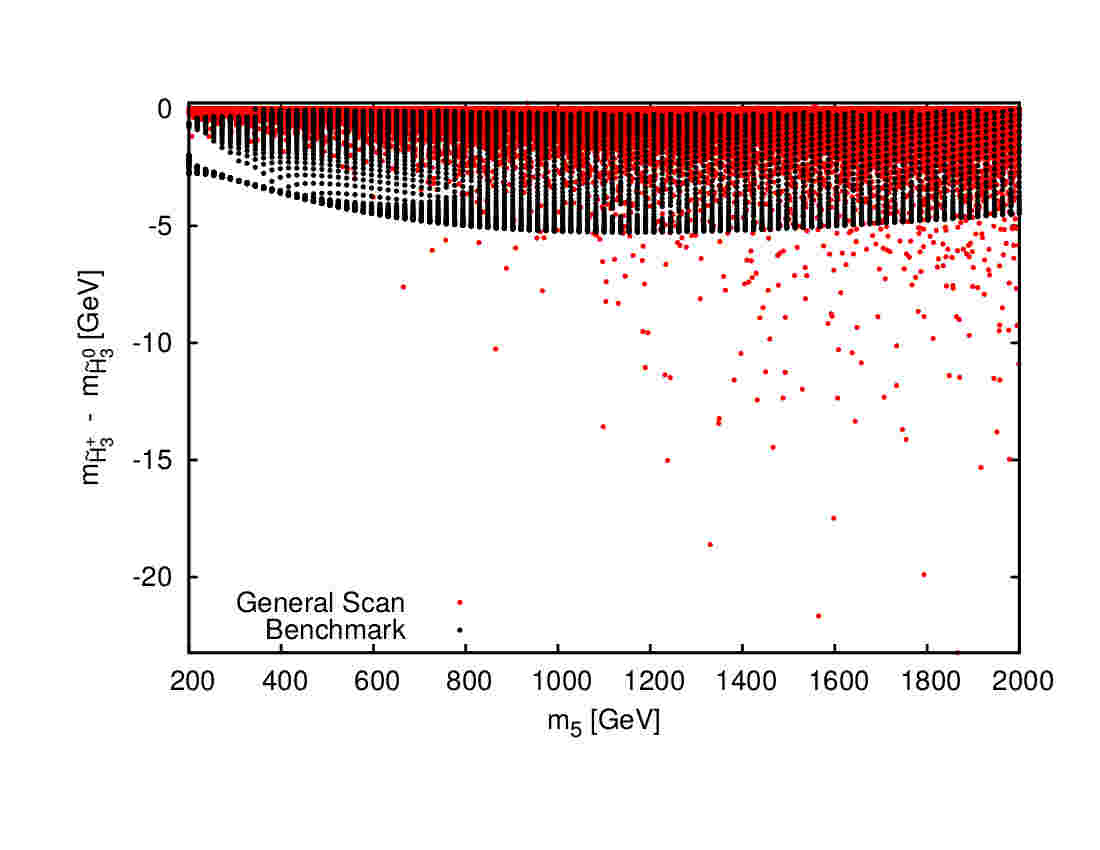}}%
	\resizebox{0.5\textwidth}{!}{\includegraphics{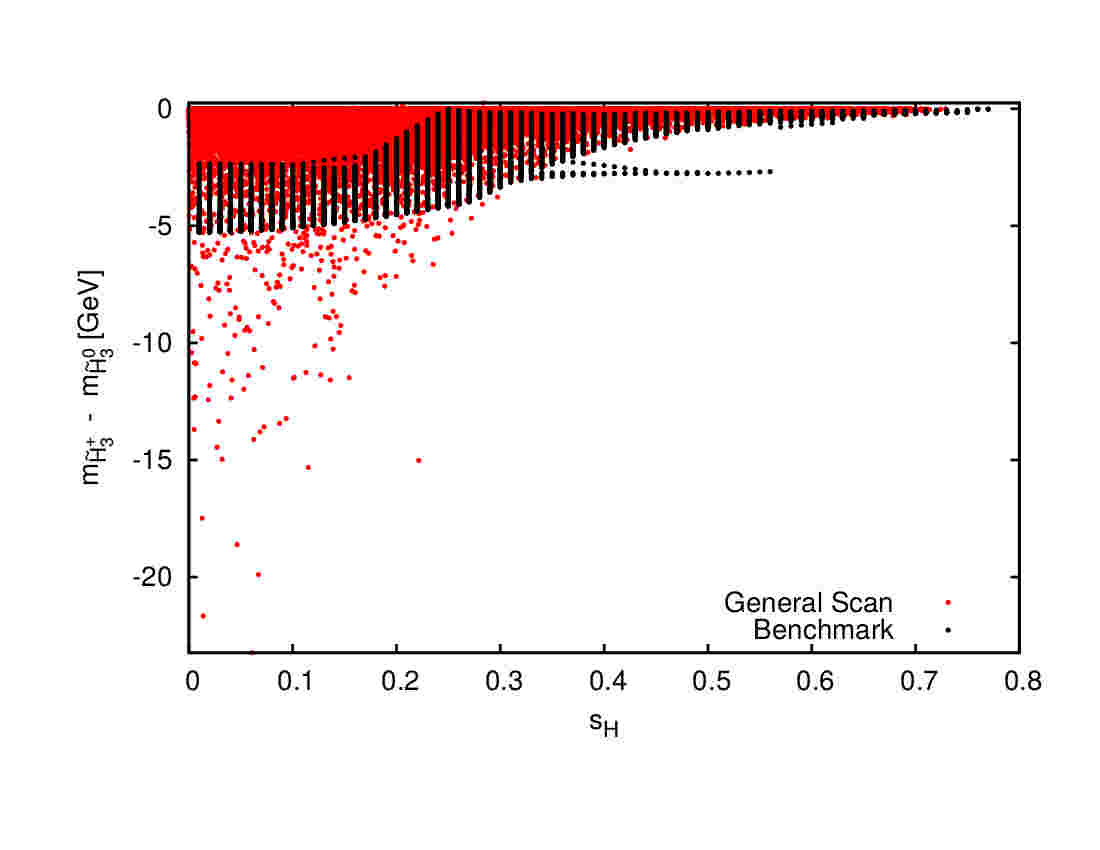}}
	\caption{Mass splitting $m_{\tilde H_3^+} - m_{\tilde H_3^0}$ evaluated with the maximum allowed cutoff scale in a general parameter scan (red) and in the H5plane benchmark (black), as a function of $m_5$ (left) and $s_H$ (right).  This quantity is negative because $\tilde H_3^+$ is lighter than $\tilde H_3^0$.  The mass splitting in the general scan ranges between $+0.25$~GeV and $-23$~GeV.}
	\label{fig:GSH3MassSplittings}
\end{figure}

In Fig.~\ref{fig:GSH3MassSplittings} we compare the mass splittings among the custodial triplet states in the H5plane benchmark (black points) to the results of a general scan (red points). In the left panel we plot $m_{\tilde H_3^+} - m_{\tilde H_3^0}$ versus $m_5$, and in the right panel we plot $m_{\tilde H_3^+} - m_{\tilde H_3^0}$ versus $s_H$.  The range of mass splittings obtained in the H5plane benchmark is generally typical of the results of the general scan, except that the general scan generates a small number of points with mass splittings up to four times as large as in the benchmark.  There are also a very small number of points in the general scan with the opposite mass hierarchy, for which $\tilde H_3^+$ becomes heavier than $\tilde H_3^0$ by up to 0.25~GeV.

\begin{figure}
\resizebox{0.5\textwidth}{!}{\includegraphics{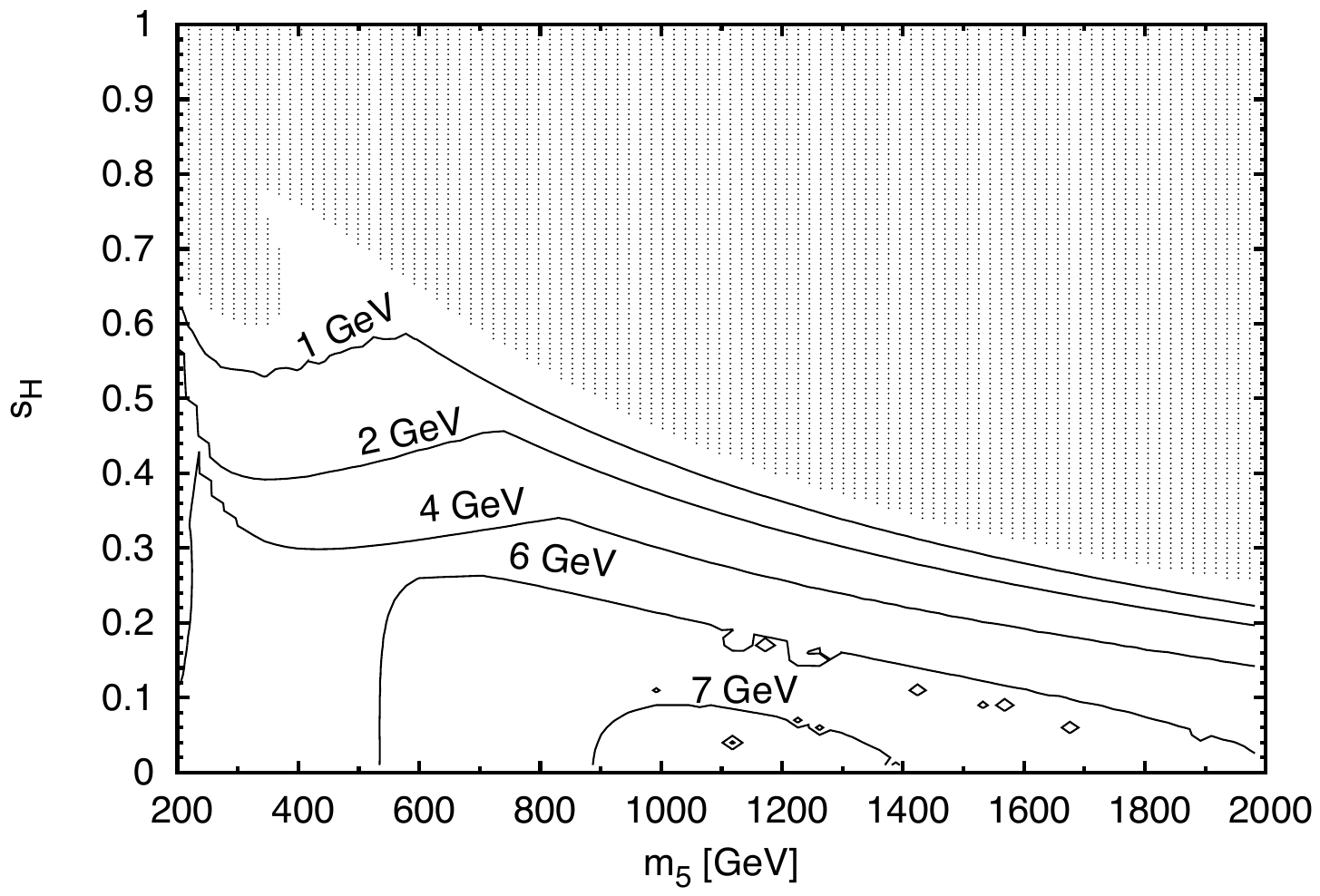}}%
\resizebox{0.5\textwidth}{!}{\includegraphics{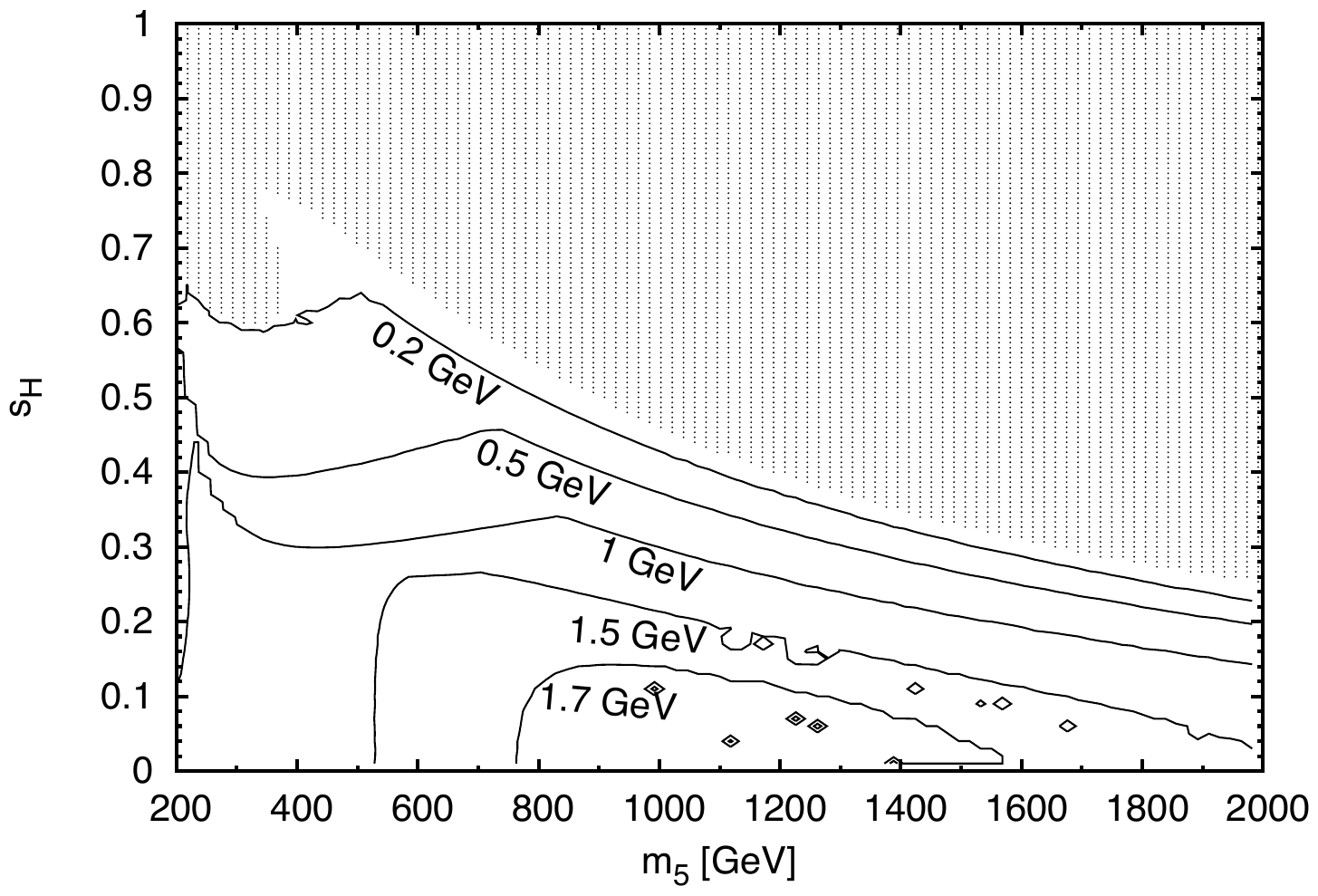}}
\resizebox{0.5\textwidth}{!}{\includegraphics{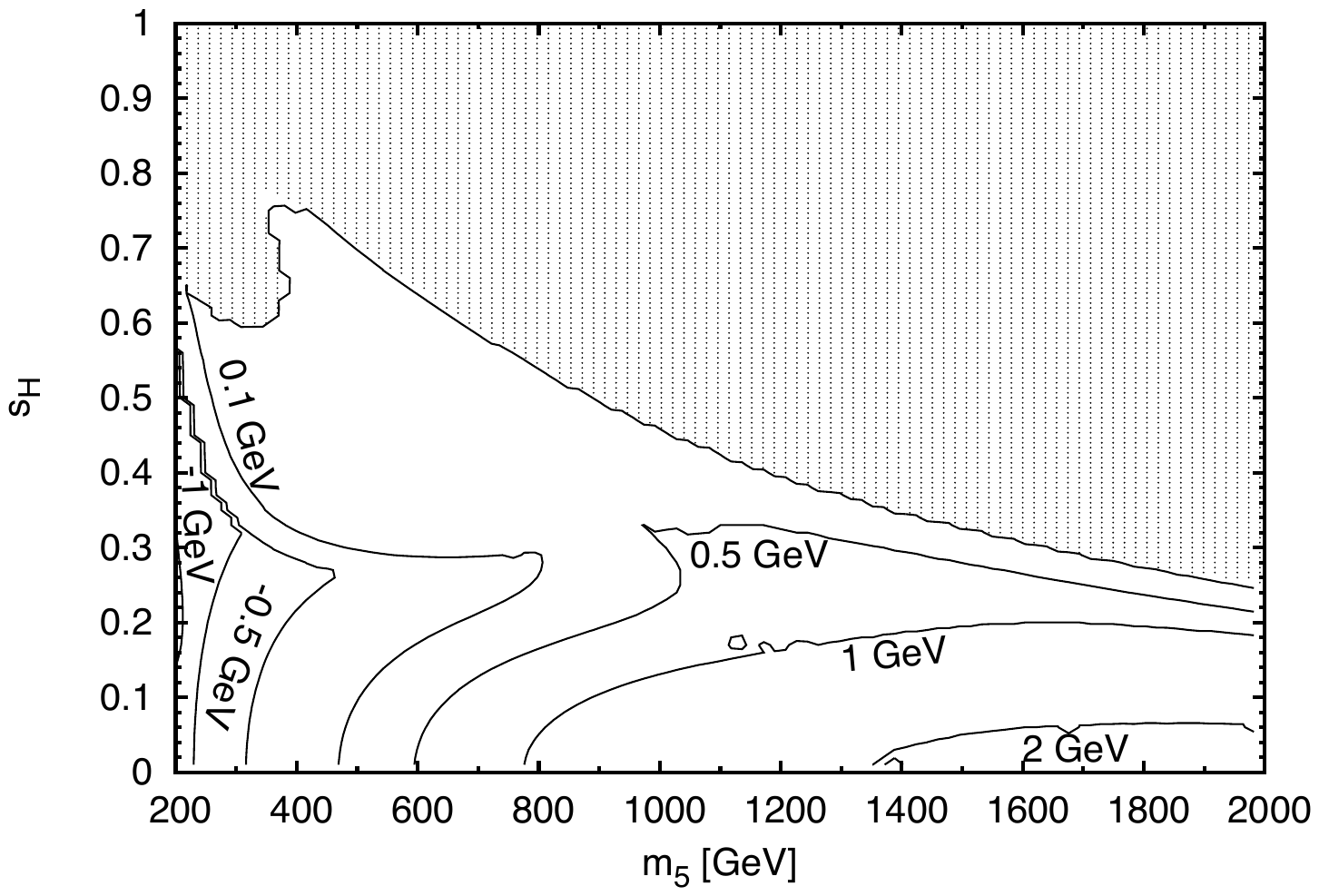}}%
\resizebox{0.5\textwidth}{!}{\includegraphics{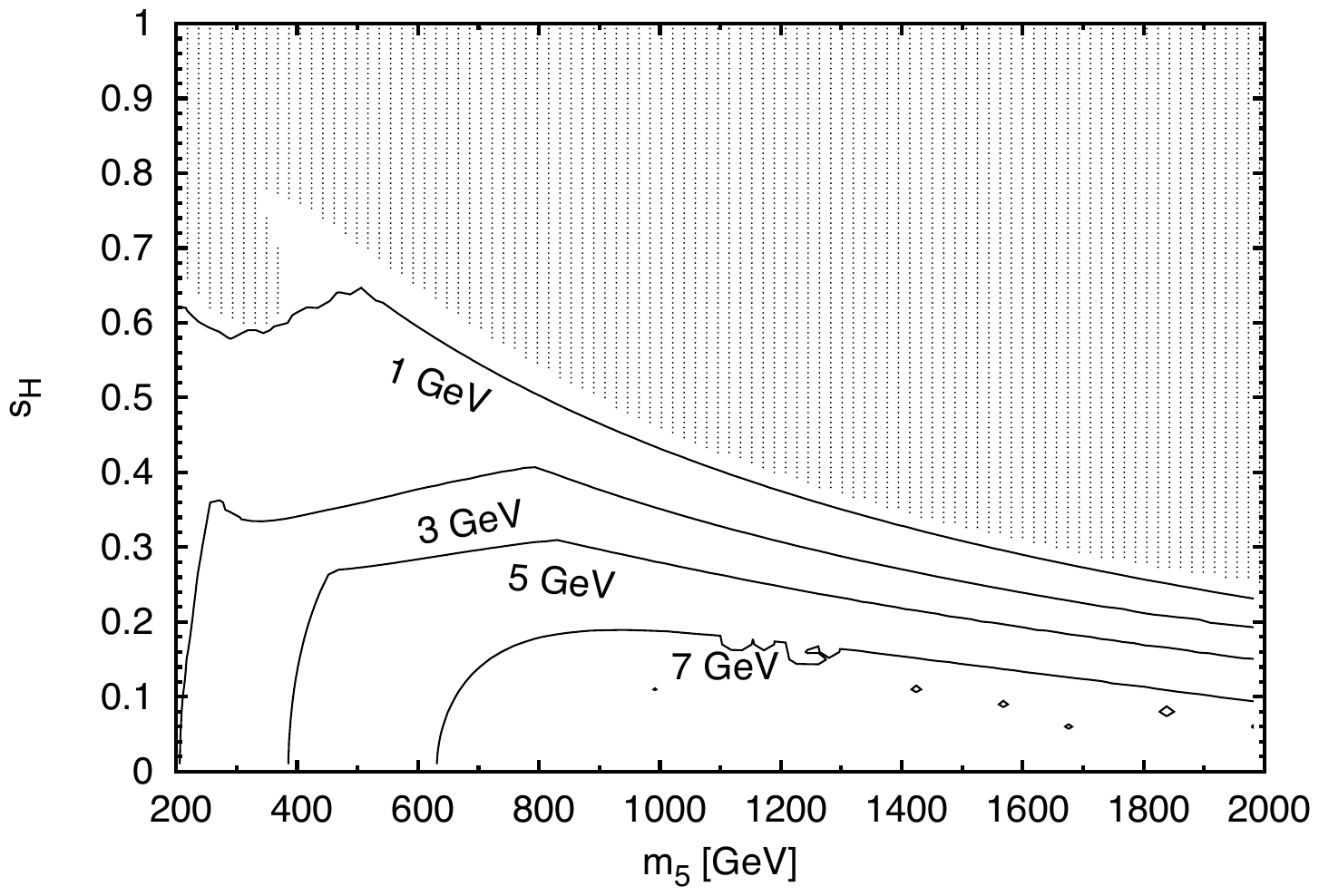}}
	\caption{The mass splittings within the custodial fiveplet in the H5plane benchmark, taking the scale of the custodial-symmetric theory to be as large as possible subject to perturbative unitarity and the $\rho$ parameter constraint. Top left: $m_{\tilde H_5^{++}} - m_{\tilde H_5^0}$.  This mass splitting ranges between $4.0$~MeV and $7.2$~GeV. Top right: $m_{\tilde H_5^{+}} - m_{\tilde H_5^0}$. This mass splitting ranges between $6.0$~MeV and $1.8$~GeV. Bottom left: $m_{\tilde H_5^0} - m_5$, where $m_5$ is the weak-scale custodial-symmetric input value of the custodial fiveplet mass. This mass difference ranges between $-1.5$~GeV and $2.3$~GeV.  Bottom right: $m_{\tilde H_5^{++}} - m_5$.  $m_{\tilde H_5^{++}}$ is always larger than $m_5$, with the difference ranging between $7.0$~MeV and $9.0$~GeV.}
	\label{fig:H5MassSplitting}
\end{figure}

Among the custodial-fiveplet mass eigenstates, $\tilde H_5^{++}$ is almost always the heaviest, followed by $\tilde H_5^+$ and then $\tilde H_5^0$.  Again the mass splittings are small, as shown in Fig.~\ref{fig:H5MassSplitting} for the H5plane benchmark.  The top left panel of Fig.~\ref{fig:H5MassSplitting} shows the mass difference between $\tilde H_5^{++}$ and $\tilde H_5^0$, which is at most 7.2~GeV.  The mass of $\tilde H_5^+$ falls between these two, but closer to the lighter $\tilde H_5^0$ state: the mass difference between $\tilde H_5^+$ and $\tilde H_5^0$ reaches at most 1.8~GeV, as shown in the top right panel of Fig.~\ref{fig:H5MassSplitting}.
The mass of $\tilde H_5^0$ remains within 2.3~GeV of the weak-scale custodial-symmetric input value of $m_5$, but can be heavier or lighter: this is plotted in the bottom left panel of Fig.~\ref{fig:H5MassSplitting}.  The mass of $\tilde H_5^{++}$ is always larger than $m_5$, with the difference reaching a maximum of 9.0~GeV, as shown in the bottom right panel of Fig.~\ref{fig:H5MassSplitting}.  The smallness of these shifts of the physical $\tilde H_5$ masses relative to the weak-scale custodial-symmetric input value of $m_5$ justifies our use of this input value on the $x$ axis of the plots.

\begin{figure}
	\resizebox{0.5\textwidth}{!}{\includegraphics{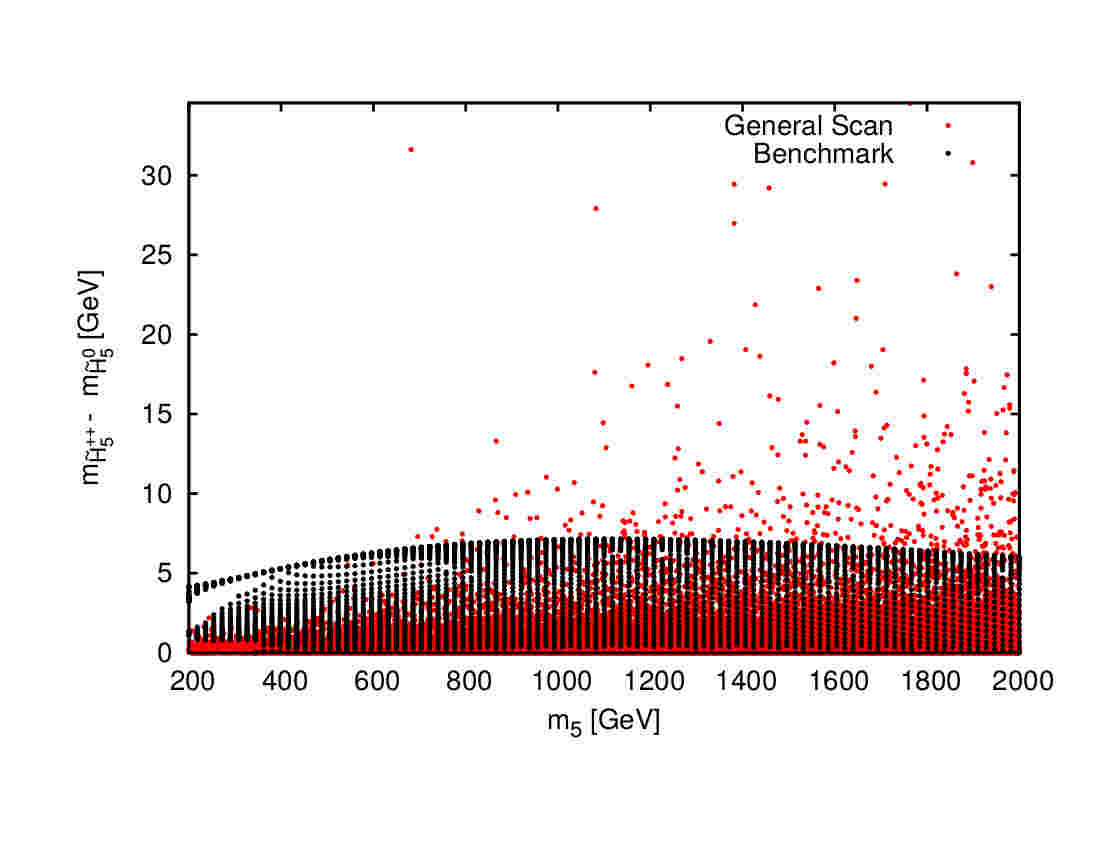}}%
	\resizebox{0.5\textwidth}{!}{\includegraphics{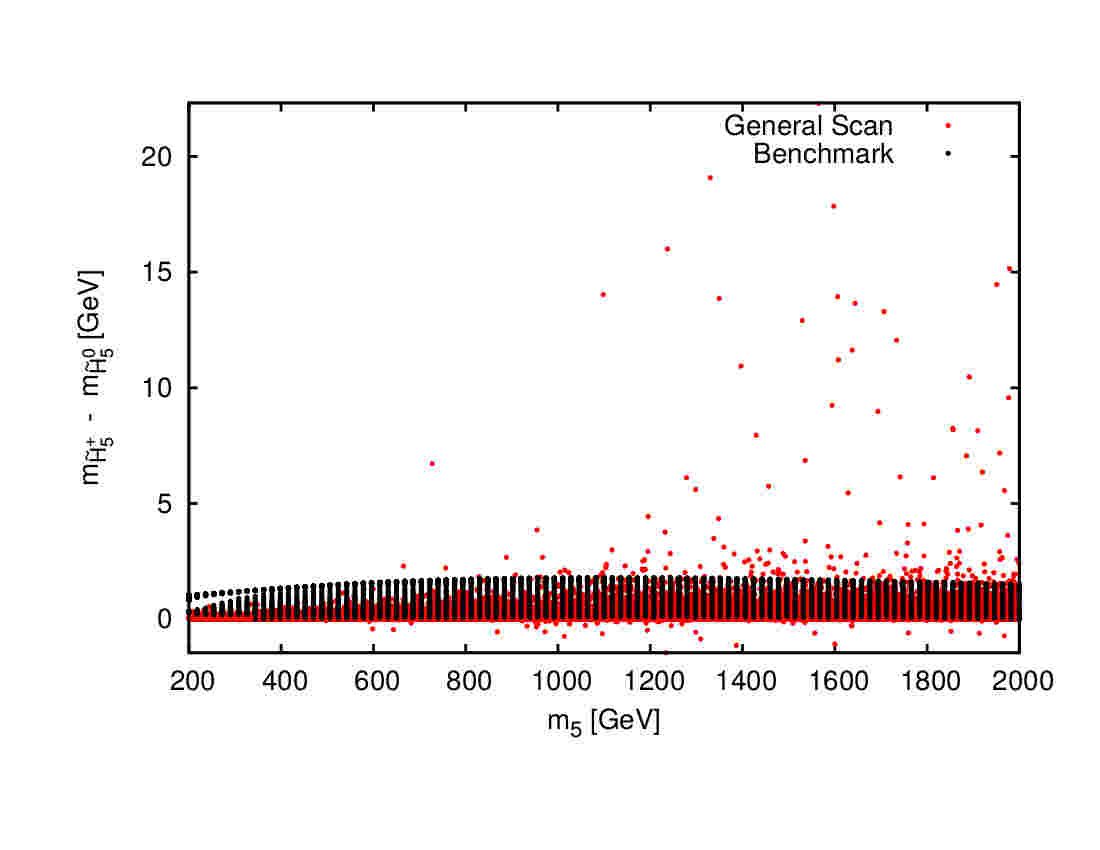}}
	\resizebox{0.5\textwidth}{!}{\includegraphics{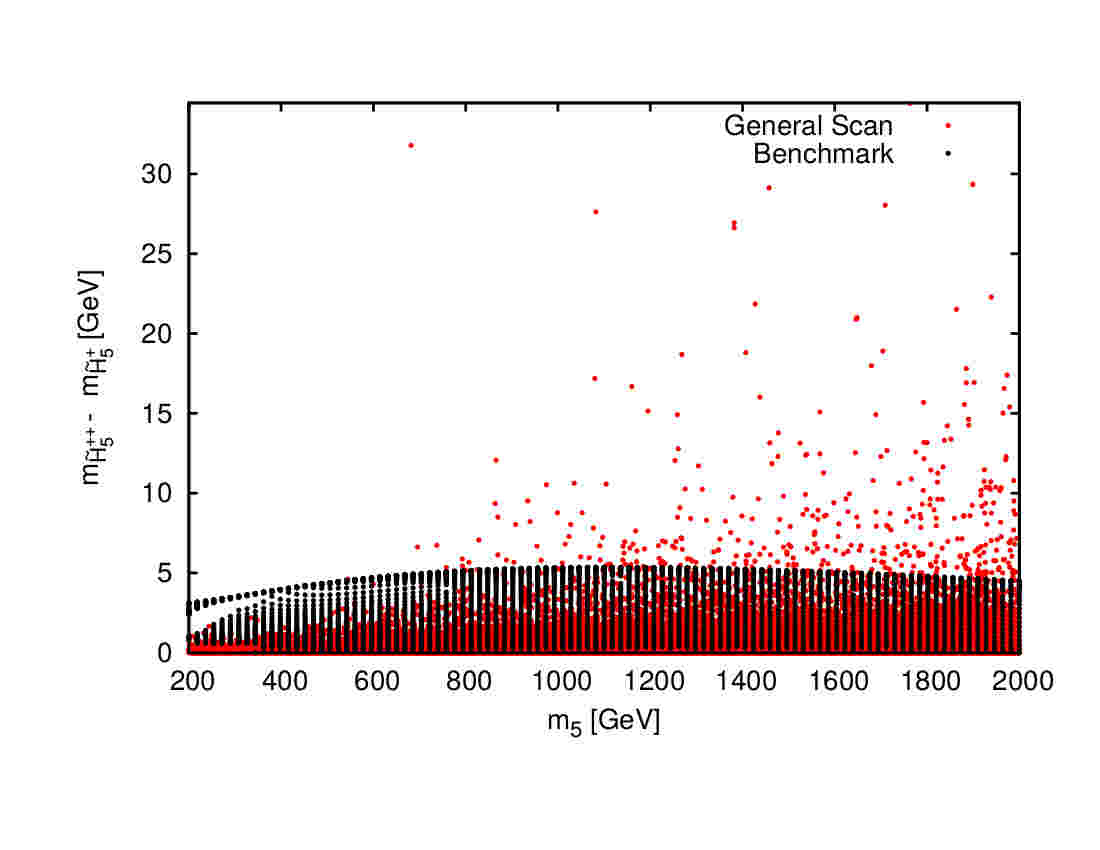}}%
		
	\caption{Mass splittings $m_{\tilde H_5^{++}} - m_{\tilde H_5^0}$ (top left), $m_{\tilde H_5^{+}} - m_{\tilde H_5^0}$ (top right), and $m_{\tilde H_5^{++}} - m_{\tilde H_5^{+}}$ (bottom) as a function of $m_5$, evaluated with the maximum allowed cutoff scale in a general parameter scan (red) and in the H5plane benchmark (black).  In the general scan $m_{\tilde H_5^{++}} - m_{\tilde H_5^0}$ ranges between zero and $34.5$~GeV, $m_{\tilde H_5^{+}} - m_{\tilde H_5^0}$ ranges between $-1.45$~GeV and $22.3$~GeV, and $m_{\tilde H_5^{++}} - m_{\tilde H_5^{+}}$ ranges between zero and $34.4$~GeV.}
	\label{fig:GSH5MassSplitting}
\end{figure}

In Fig.~\ref{fig:GSH5MassSplitting} we compare the mass splittings among the custodial fiveplet states in the H5plane benchmark (black points) to the results of a general scan (red points) as a function of $m_5$.  We show $m_{\tilde H_5^{++}} - m_{\tilde H_5^0}$ (top left), $m_{\tilde H_5^+} - m_{\tilde H_5^0}$ (top right), and $m_{\tilde H_5^{++}} - m_{\tilde H_5^+}$ (bottom).  Again the ranges of mass splittings obtained in the H5plane benchmark are generally typical of the results in the general scan, except that the general scan generates a small number of points with mass splittings up to six times as large as in the benchmark.  There are also a very small number of points in the general scan with the opposite mass hierarchy for which $\tilde H_5^{+}$ becomes lighter than $\tilde H_5^0$ by up to 1.5~GeV.  It is also possible in the general scan to have a large mass spitting between the $\tilde H_5^{++}$ and $\tilde H_5^0$ but a small mass splitting between $\tilde H_5^{+}$ and the $\tilde H_5^0$.

Examining the contours in Figs.~\ref{fig:H3MassSplittings} and~\ref{fig:H5MassSplitting}, it is apparent that within the H5plane benchmark the mass splittings within the fiveplet and within the triplet tend to follow a common pattern albeit with different scaling. In particular, the splitting $m_{\tilde H_5^{++}} - m_{\tilde H_5^0}$ is very close to four times that of $m_{\tilde H_5^{+}} - m_{\tilde H_5^0}$. To understand this behavior we expand the mass splittings to first order in the custodial violation, such that $\tilde x = x + \delta_x$. The mass splittings become: 
\begin{eqnarray}
m_{\tilde H_5^{++}} - m_{\tilde H_5^0} &=& \frac{1}{2 m_5} \left[ \left( v_\phi^2 d_2 + v_\phi^2 \frac{1}{6 v_\chi} d_3 + v_\phi^2 \frac{M_1}{6 v_\chi^2} d_1 - v_\chi \frac{32}{3} \lambda_3 d_1 + 16 M_2 d_1 \right) + \frac{2 v_\phi^2}{3 v_\chi} \lambda_5 d_1 + v_\chi^2 d_4 \right], \label{eq:splitexpansion1} \\
m_{\tilde H_5^{+}} - m_{\tilde H_5^0} &=& \frac{1}{2 m_5} \left[ \frac{1}{4} \left( v_\phi^2 d_2 + v_\phi^2 \frac{1}{6 v_\chi} d_3 + v_\phi^2 \frac{M_1}{6 v_\chi^2} d_1 - v_\chi \frac{32}{3} \lambda_3 d_1 + 16 M_2 d_1 \right) - \frac{v_\phi^2}{6 v_\chi} \lambda_5 d_1 + v_\chi^2 d_4^{\prime} \right], \label{eq:splitexpansion2} \\
m_{\tilde H_3^{+}} - m_{\tilde H_3^0} &=& \frac{1}{2 m_3} \left[ \frac{v^2}{4} d_2 - \frac{v^2}{8} d_3 + d_1 \left( -\frac{v^2}{v_\chi}\left(\frac{\lambda_5}{2} + \frac{M_1}{8 v_\chi}\right) +2 M_1 + \frac{\lambda_5}{4 v_\chi} ( 16 v_\chi^2 -v^2) \right) \right],
\end{eqnarray}
where $d_1$ through $d_4$ and $d_4^{\prime}$ are zero in the limit of exact custodial symmetry and are given by
\begin{eqnarray}
d_1 &=& \tilde v_\xi - \tilde v_\chi = \frac{v^2}{4 v_\chi} \Delta \rho , \nonumber \\
d_2 &=& \frac{\tilde \lambda_4}{\sqrt{2}} - \frac{\tilde \lambda_3}{2}, \nonumber \\ 
d_3 &=& \tilde M_1^\prime - \tilde M_1, \nonumber \\
d_4 &=& 4 \tilde \lambda_2 - \frac{4}{3} \tilde \lambda_7 -\frac{16}{3} \tilde \lambda_8 + \frac{8}{3} \tilde \lambda_{10}, \nonumber \\
d_4^\prime &=& 2 \tilde \lambda_9 - \frac{4}{3} \tilde \lambda_7 -\frac{16}{3} \tilde \lambda_8 + \frac{8}{3} \tilde \lambda_{10}.
\end{eqnarray}
The approximate relation $m_{\tilde H_5^{++}} - m_{\tilde H_5^0} \simeq 4 (m_{\tilde H_5^{+}} - m_{\tilde H_5^0})$ is therefore to be expected because only the last two terms in Eqs.~(\ref{eq:splitexpansion1}) and (\ref{eq:splitexpansion2}) break this proportionality. Simply from the generic size of the dimensionful parameters in the terms that break this relation, we naively expect that they will be sub-dominant contributors to the mass splittings. In the general scan shown in Fig.~\ref{fig:GSH5MassSplitting} this relation tends to hold to a good approximation throughout the parameter space but can be badly broken by the enhanced mixing caused by approximately degenerate charged eigenstates. 

The similarities in the patterns of the fiveplet and triplet mass splittings can also be explained by comparing their approximate forms, as they both depend on the same terms as the sources of custodial violation.  Although these terms come in with different coefficients, when a single term dominates one expression, it will generally dominate all of them. In the case of the custodial triplet, the splitting is always negative because the dominant terms (mainly the term proportional to $d_3$, but the term proportional to $d_1$ is usually significant and sometimes dominant) tend not to change sign throughout the whole parameter space. This remains true for the general parameter scan shown in Fig.~\ref{fig:GSH3MassSplittings} where the triplet mass splittings are overwhelmingly negative even when $\rho$ is less than 1.

\begin{figure}
\resizebox{0.5\textwidth}{!}{\includegraphics{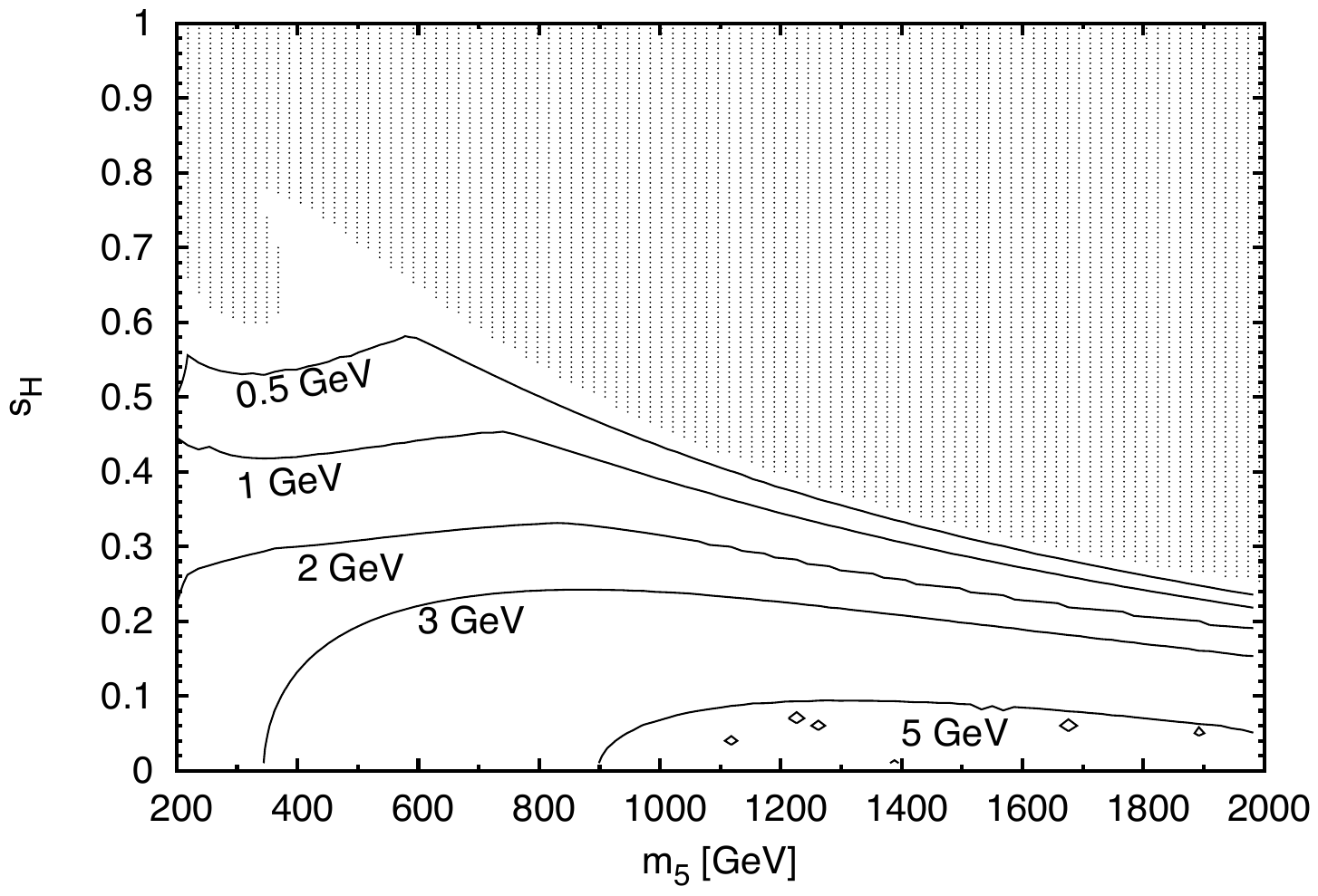}}
	\caption{Deviation $m_{\tilde H} - m_H$ of the physical $\tilde H$ mass from the mass of the heavier custodial singlet $H$ in the weak-scale custodial-symmetric theory, computed in the H5plane benchmark taking the scale of the custodial-symmetric theory to be as large as possible subject to perturbative unitarity and the $\rho$ parameter constraint.  This mass difference ranges between $-0.059$~GeV and $+5.6$~GeV. }
	\label{fig:HMassSplitting}
\end{figure}

In Fig.~\ref{fig:HMassSplitting} we plot the shift of the mass of the physical mass eigenstate $\tilde H$ relative to the weak-scale custodial-symmetric input value of $m_H$.  The $\tilde H$ mass is shifted upwards over almost all of the H5plane benchmark, and the shift is by at most 5.6~GeV.  We conclude that, within the H5plane benchmark and even allowing for custodial symmetry violation, the custodial-symmetric predictions for the masses of the scalars in the model are reliable to within better than 10~GeV.

Experimentally checking the mass degeneracy of the scalars within the custodial triplet and the custodial fiveplet has been proposed as a way to test the custodial symmetry in the GM model~\cite{Chiang:2012cn,Chiang:2015rva}.  At the LHC, mass reconstruction of the $H_3$ states relies on their decays to dijets, $H_3^+ \to c \bar s$, $H_3^0 \to b \bar b$~\cite{Chiang:2012cn}.  Considering that the dijet invariant mass resolution at the LHC is not sufficient to kinematically separate the hadronic decays of the $W$ and the $Z$ with their $11$~GeV mass difference, it will not be possible to resolve a custodial-symmetry-violation-induced mass splitting between $\tilde H_3^+$ and $\tilde H_3^0$ of at most 5.3~GeV within the H5plane benchmark.  Mass reconstruction of the $H_5$ states at the LHC relies on their decays to vector boson pairs $VV$.  Reference~\cite{Chiang:2012cn} studied the fully-leptonic final states, in which the masses of $H_5^{++}$, $H_5^+$, and $H_5^0$ could be determined from the endpoint of the transverse mass distribution of the $VV$ final state.  The resolution is worse than for a dijet resonance.  The ATLAS experiment has performed a search for $H_5^+ \to W^+Z \to jj\ell^+\ell^-$~\cite{Aad:2015nfa}, in which reconstruction of a mass peak for $\tilde H_5^+$ becomes possible; however, the mass resolution is still limited by the dijet invariant mass resolution of the LHC, which is too poor to resolve the custodial-symmetry-violation-induced mass splitting among the $\tilde H_5$ states spanning at most 7.2~GeV in the H5plane benchmark.  Larger mass splittings are possible for a small number of points in the general scan, but these tend to appear at relatively large $m_5$ values so that the splittings remain below a few percent of the overall scalar masses, smaller than the single jet energy resolution of the LHC experiments at these energies (see, e.g., Ref.~\cite{Jaeger:2661735}).

Prospects are somewhat better at the ILC, as studied in Ref.~\cite{Chiang:2015rva}.  $\tilde H_5^0$ and $\tilde H_5^{\pm}$ can be singly produced in $e^+e^-$ collisions via vector boson fusion, or in association with a $Z$ or $W^{\mp}$ boson, respectively.  In the clean lepton collider environment, the $H_5$ decays to dibosons can be reconstructed using the fully hadronic final states.  With the ILC target dijet energy resolution of $\sigma_E = 0.3 \times \sqrt{E_{jj}}$~GeV~\cite{Behnke:2013lya}, the dijet resolution will be $\sigma_E \simeq 3$~GeV for $E_{jj} \simeq 100$~GeV, famously allowing for $W$ and $Z$ bosons to be distinguished in the all-hadronic channel.  Unfortunately, even this excellent mass resolution is too poor to resolve the custodial-symmetry-violation-induced mass splitting between $\tilde H_5^+$ and $\tilde H_5^0$, which reaches at most 1.8~GeV in the H5plane benchmark.  One could hope to do better by using the leptonic decays of $H_5^0 \to ZZ \to 4 \ell$ and $H_5^{\pm} \to W^{\pm} Z \to \ell^{\pm} E_T^{miss} \ell^+ \ell^-$; these suffer from smaller branching fractions, but may offer good enough mass resolution to detect the mass splitting effect of the custodial symmetry violation.

\subsection{Direct search constraints}

\begin{figure}
	\resizebox{0.5\textwidth}{!}{\includegraphics{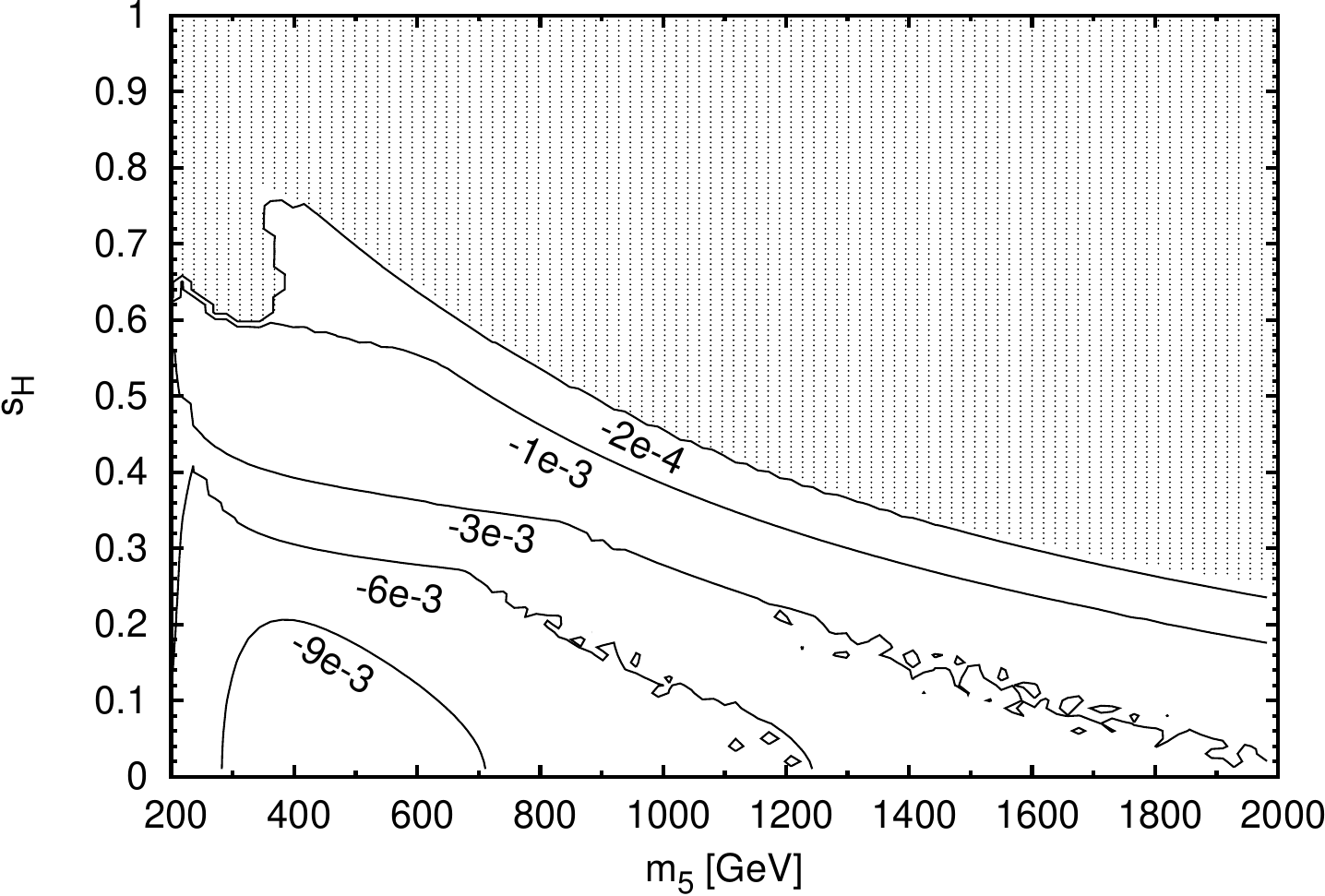}}
	\caption{The fractional change in $\tilde v_\chi$ relative to the weak-scale custodial-symmetric input $v_{\chi}$, defined as $\frac{\tilde v_\chi}{v_\chi} -1$, in the H5plane benchmark taking the scale of the custodial-symmetric theory to be as large as possible subject to perturbative unitarity and the $\rho$ parameter constraint.  The fractional change is always negative and its absolute value reaches a maximum of 1.0\%.}
	\label{fig:vChiChange}
\end{figure}

\begin{figure}
	\resizebox{0.5\textwidth}{!}{\includegraphics{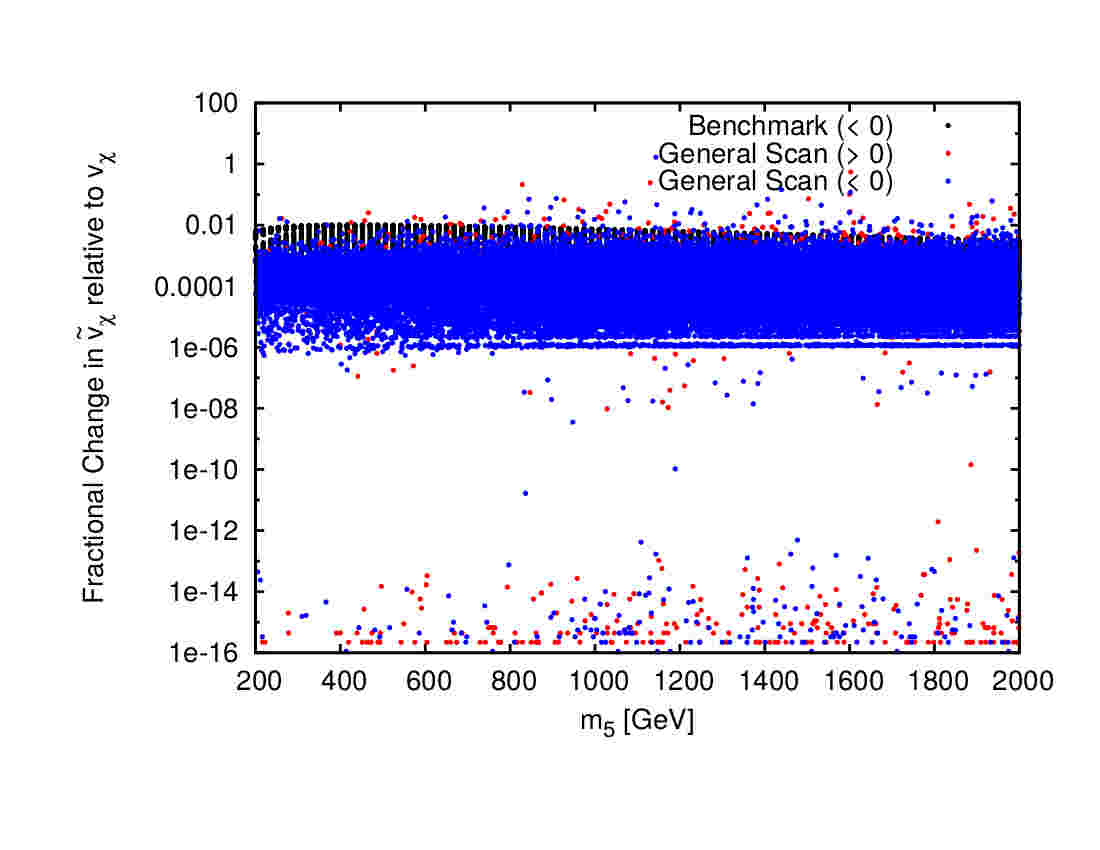}}%
	\resizebox{0.5\textwidth}{!}{\includegraphics{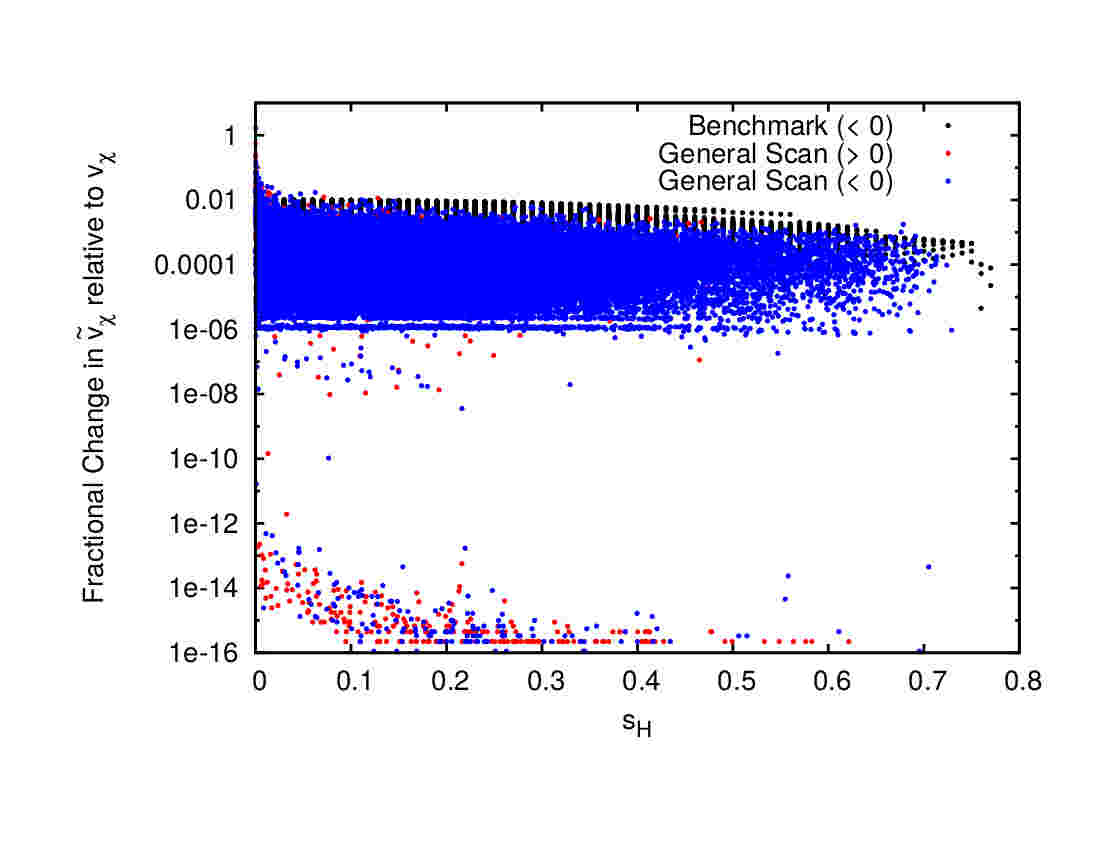}}
	\caption{The fractional change in $\tilde v_\chi$ relative to the weak-scale custodial-symmetric input $v_{\chi}$, defined as $\frac{\tilde v_\chi}{v_\chi} -1$, as a function of $m_5$ (left) and $s_H$ (right), evaluated with the maximum allowed cutoff scale in a general parameter scan (blue and red points) and in the H5plane benchmark (black points).  The fractional change can be positive (blue points) or negative (red points) in the general scan, but is always negative in the H5plane benchmark (black points).  The fractional change reaches maxima and minima of 0.55 and $-1.66$, respectively, at very low $s_H$, but for $s_H > 0.1$ its absolute value reaches at most 0.0114.}
	\label{fig:GSvChiChange}
\end{figure}

The most stringent direct search constraint on the custodial-symmetric H5plane benchmark comes from a CMS search for $H_5^{\pm\pm}$ produced in vector boson fusion and decaying to $W^{\pm}W^{\pm} \to \ell^{\pm} \ell^{\pm} E_T^{miss}$~\cite{Sirunyan:2017ret}.  This constraint excludes $s_H$ above 0.2 for $m_5 = 200$~GeV, rising to $s_H = 0.45$ at $m_5 = 1000$~GeV.  We can apply this straightforwardly to the model with custodial symmetry violation by noting the following.  First, as shown in the bottom right panel of Fig.~\ref{fig:H5MassSplitting}, the physical mass of $\tilde H_5^{++}$ is at most 5~GeV higher than $m_5$ in the region of interest in the H5plane benchmark (and not much different in the general scan).  Second, we show in Figs.~\ref{fig:vChiChange} and \ref{fig:GSvChiChange} the shift in $\tilde v_{\chi}$, which controls the $\tilde H_5^{\pm\pm}W^{\mp}W^{\mp}$ coupling and hence the vector boson fusion production cross section, relative to the value of $v_{\chi}$ in the weak-scale custodial-symmetric theory.  In the H5plane benchmark this shift is negative and amounts to less than a percent, so that the cross section is suppressed by no more than 2\% due to the custodial symmetry violation.  In the general scan this conclusion holds for $s_H$ values above 0.1 of interest to us here.  Finally, the custodial-symmetry-violation-induced mass splitting between $\tilde H_5^{++}$ and $\tilde H_5^+$ is less than 5~GeV in the region of interest, too small for the cascade decay $\tilde H_5^{\pm\pm} \to W^{\pm} \tilde H_5^{\pm}$ to compete significantly with the dominant $\tilde H_5^{\pm\pm} \to W^{\pm} W^{\pm}$ signal channel.  Thus we conclude that this direct search constraint on the custodial symmetry violating parameter space studied in this paper will be almost identical to that in the custodial-symmetric H5plane benchmark.\footnote{Very recent LHC searches for $H_3^0 \to Z h$ and $H \to hh$ may further constrain the custodial-symmetric H5plane benchmark~\cite{IsmailLogan}, and are worth examining more closely in future work.}

\subsection{Low-$m_5$ region}
\label{sec:lowm5}

Finally in this subsection we present the results of a dedicated general scan of the low-$m_5$ region, focusing on $m_5 < 200$~GeV.  As usual, we take the cutoff as large as allowed by perturbative unitarity and the $\rho$ parameter constraint to maximize the amount of custodial symmetry violation.

\begin{figure}
	\resizebox{0.5\textwidth}{!}{\includegraphics{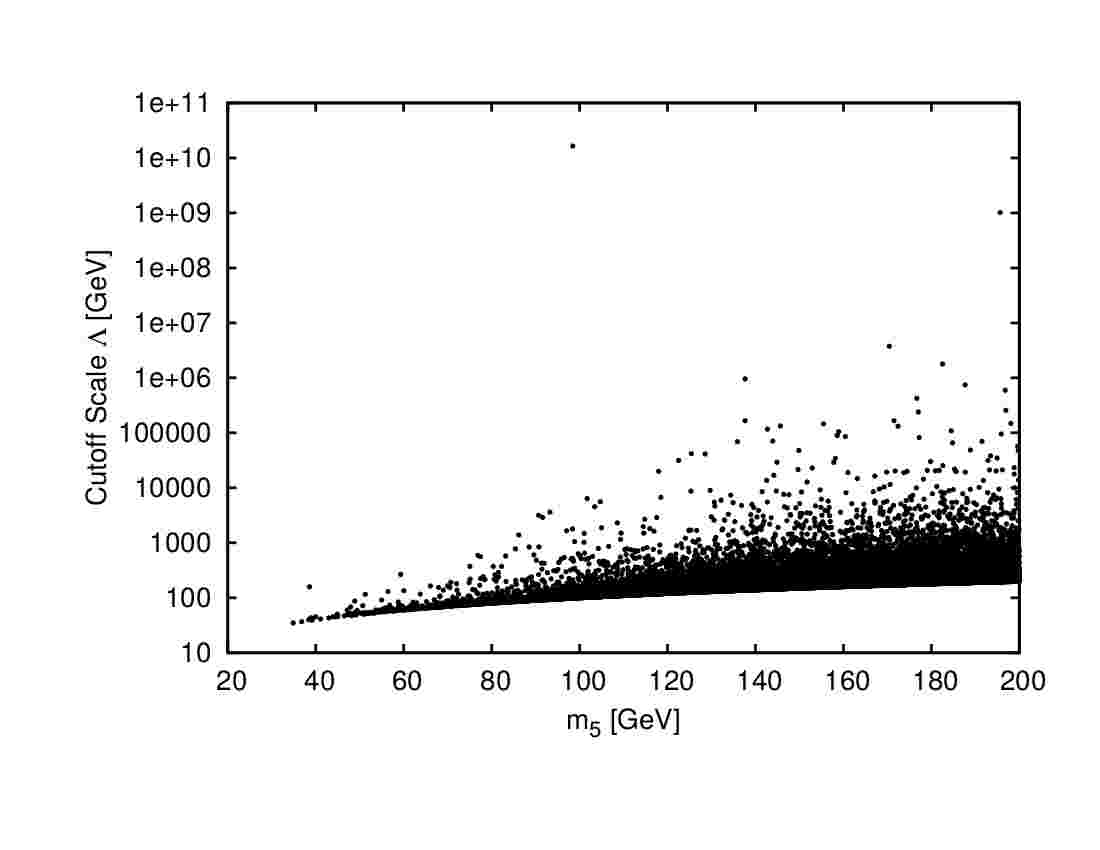}}%
	\resizebox{0.5\textwidth}{!}{\includegraphics{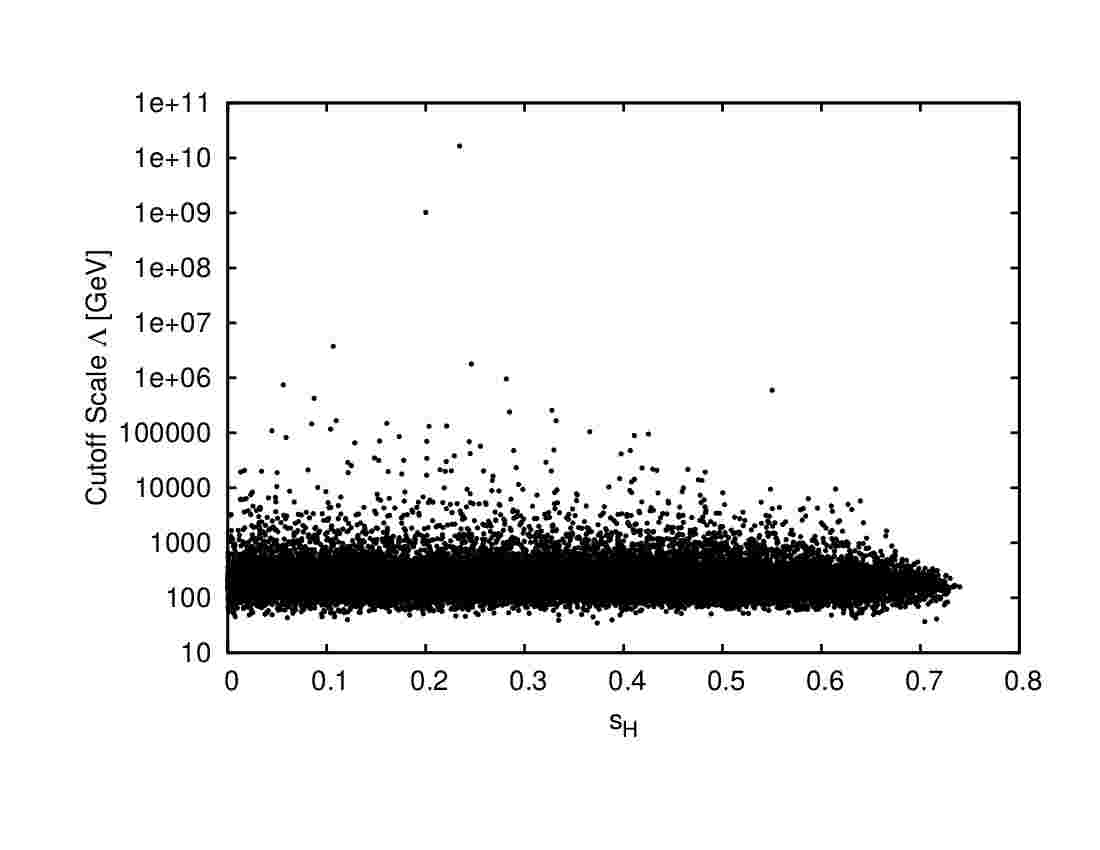}}
	\caption{Maximum value of the custodial-symmetric cutoff scale subject to perturbative unitarity and the experimental constraint on the $\rho$ parameter in a general scan of the low-$m_5$ region, as a function of $m_5$ (left) and $s_H$ (right).  The maximum cutoff scale ranges from 35~GeV (for very low $m_5$) to $1.6 \times 10^{10}$~GeV.}
	\label{fig:LMGScutoff}
\end{figure}

In Fig.~\ref{fig:LMGScutoff} we show the maximum allowed cutoff scale subject to perturbative unitarity of the quartic couplings in the custodial-symmetric theory and the $\rho$ parameter constraint, as a function of $m_5$ (left) and $s_H$ (right).  Compared to the general scan for higher $m_5$, the maximum allowed cutoff tends to be lower, but large cutoff scales on the order of 100~TeV are still somewhat common and the maximum cutoff scale found in our scan is of order $10^{10}$~GeV. 

\begin{figure}
	\resizebox{0.5\textwidth}{!}{\includegraphics{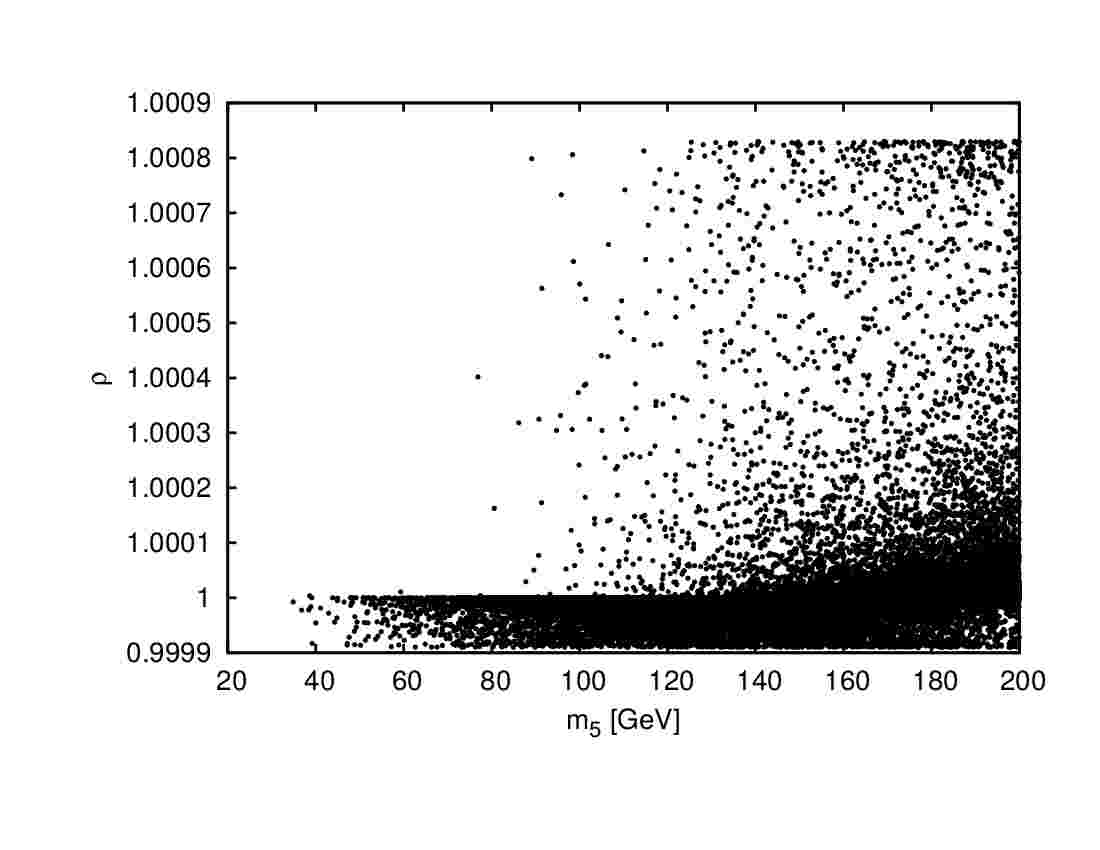}}%
	\resizebox{0.5\textwidth}{!}{\includegraphics{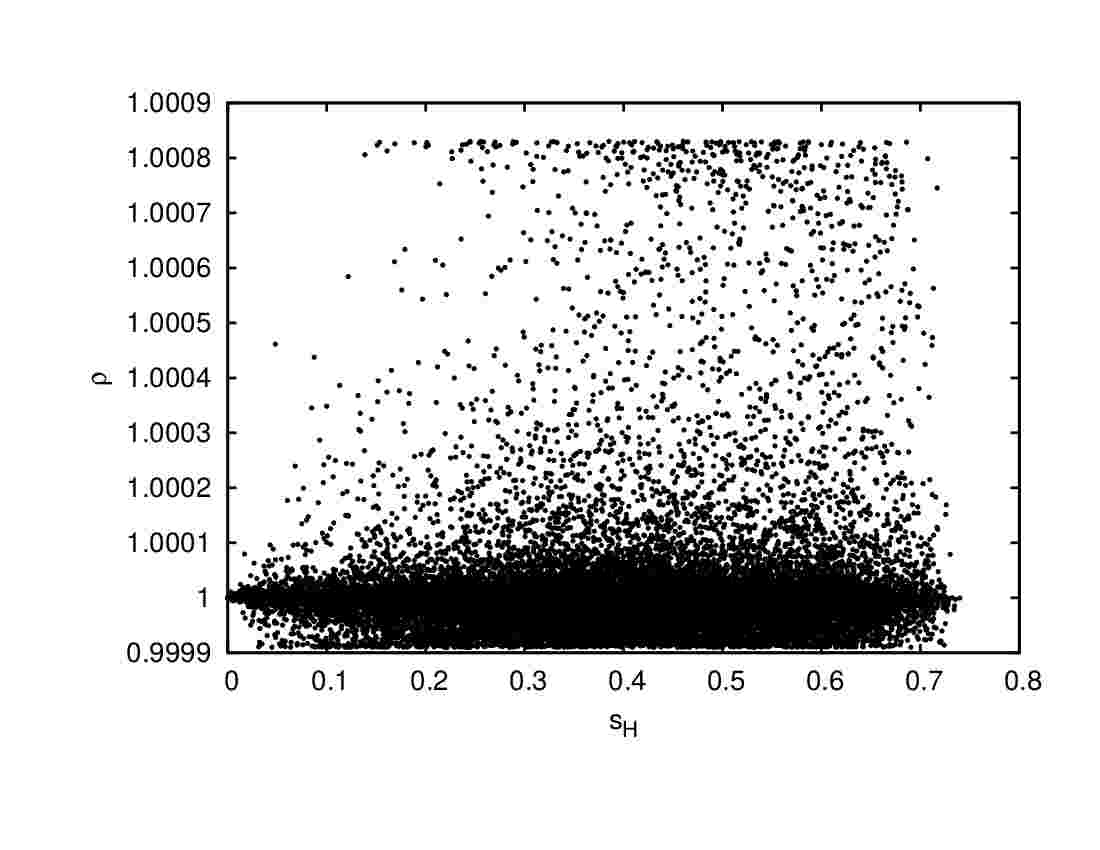}}
	\caption{Value of the $\rho$ parameter in a general scan of the low-$m_5$ region as a function of $m_5$ (left) and $s_H$ (right), taking the scale of the custodial-symmetric theory to be as large as possible subject to perturbative unitarity and the $\rho$ parameter constraint.}
	\label{fig:LMGSrho}
\end{figure}

Subject to these constraints, in Fig.~\ref{fig:LMGSrho} we show the value of the $\rho$ parameter in the weak-scale theory, again as a function of $m_5$ (left) and $s_H$ (right).  The scan populates the entirety of the allowed region; in particular, the small allowed region with $\rho < 1$ is heavily populated.  This is in contrast to the general scan at larger $m_5$, which strongly favours $\rho > 1$.

\begin{figure}
	\resizebox{0.5\textwidth}{!}{\includegraphics{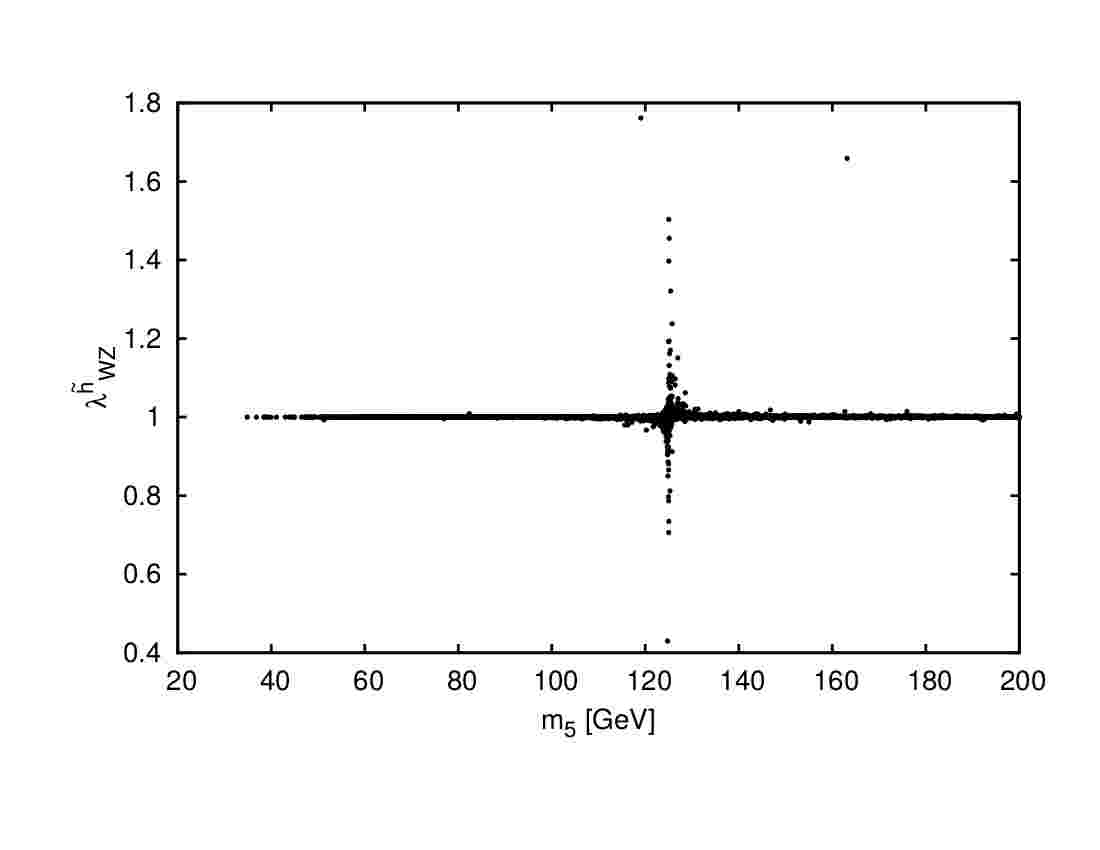}}%
	\resizebox{0.5\textwidth}{!}{\includegraphics{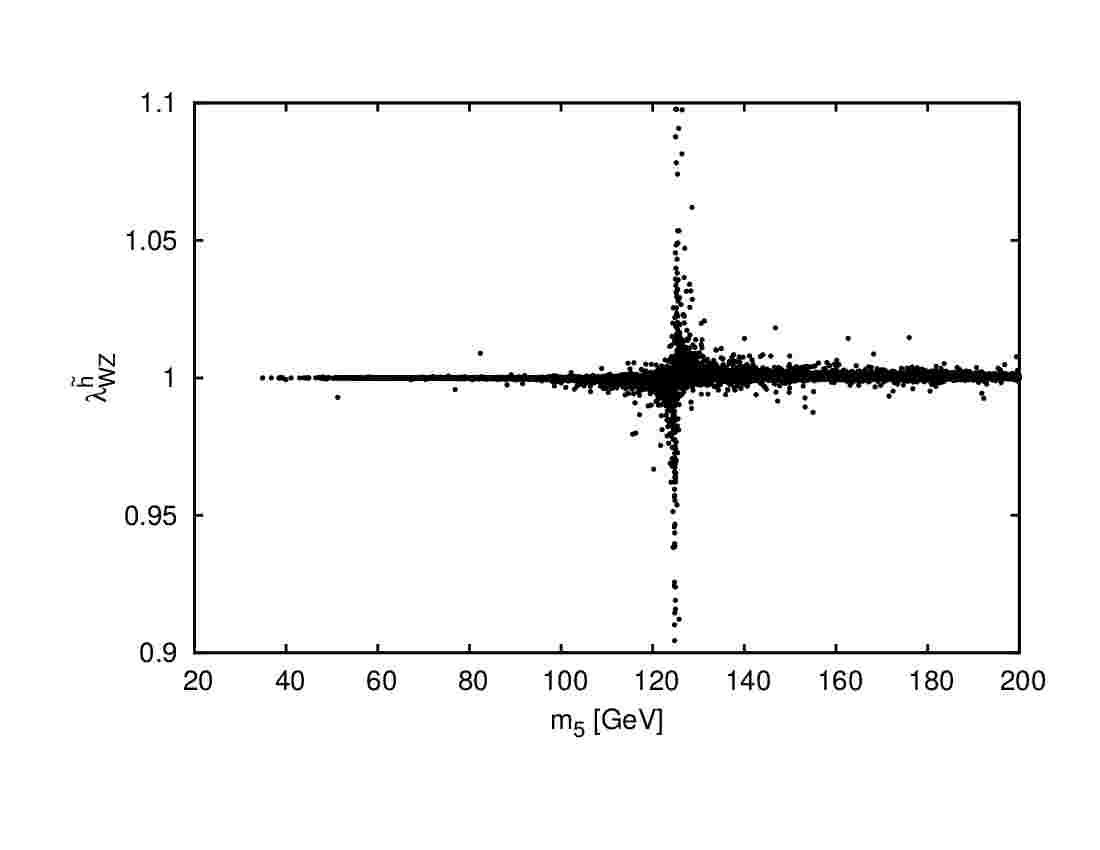}}
	\caption{$\lambda^{\tilde h}_{WZ}$ in a general scan of the low-$m_5$ region as a function of $m_5$, taking the scale of the custodial-symmetric theory to be as large as possible subject to perturbative unitarity and the $\rho$ parameter constraint.  The right panel is a zoom of the $y$-axis.  In the resonant mixing region $m_5 \simeq 125$~GeV we find values between 0.43 and 1.76, while away from this region $\lambda^{\tilde h}_{WZ}$ can deviate from one by as much as 1--2\%.}
	\label{fig:LMGSlambdahwz}
\end{figure}

We next consider custodial symmetry violation effects in couplings.
In Fig.~\ref{fig:LMGSlambdahwz} we show the ratio $\lambda^{\tilde h}_{WZ}$ of the 125~GeV Higgs boson's couplings to $WW$ and to $ZZ$.  In the left panel we plot versus $m_5$ while the right panel zooms in to $\lambda^{\tilde h}_{WZ}$ between 0.9 and 1.1.  The most dramatic feature is the resonant mixing when $\tilde H_5^0$ and $h$ become degenerate, for which deviations in $\lambda^{\tilde h}_{WZ}$ of tens of percent in either direction are possible.  Such large mixing also substantially modifies the other couplings of $h$.  Away from the resonant region, $\lambda^{\tilde h}_{WZ}$ can deviate from one by as much as 1--2\%, which is large enough to be probed at future $e+e-$ colliders.

\begin{figure}
	\resizebox{0.5\textwidth}{!}{\includegraphics{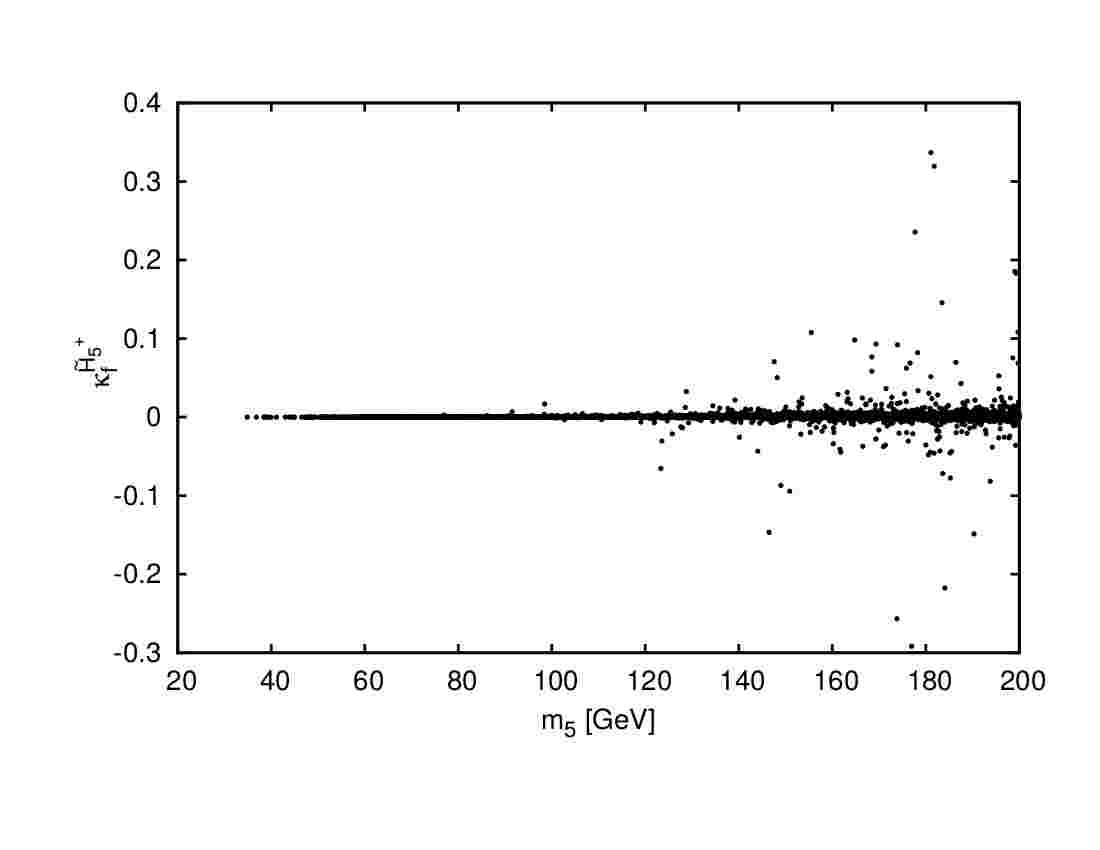}}%
	\resizebox{0.5\textwidth}{!}{\includegraphics{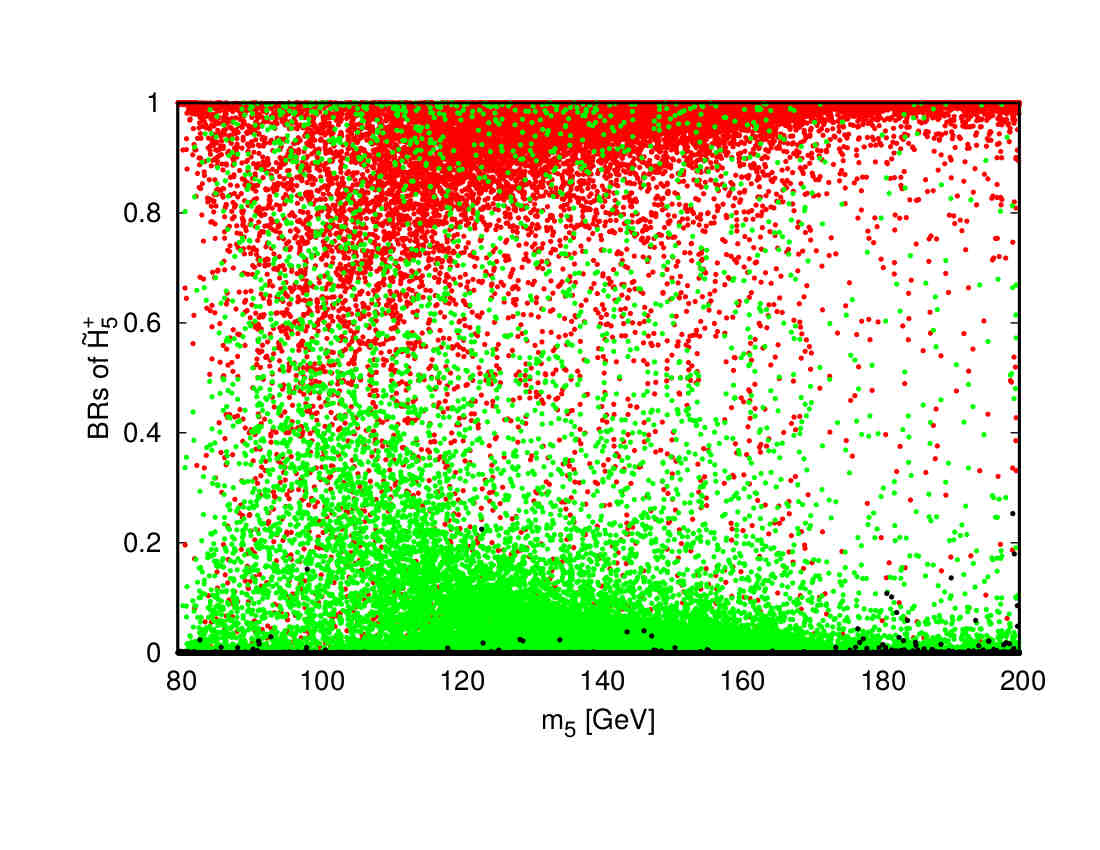}}
	\caption{Custodial-symmetry-violation-induced couplings of $\tilde H_5^+$ to fermions in a general scan of the low-$m_5$ region, taking the scale of the custodial-symmetric theory to be as large as possible subject to perturbative unitarity and the $\rho$ parameter constraint.  Left: $\kappa_f^{\tilde H_5^+}$ as a function of $m_5$.  The minimum and maximum values are $-0.29$ and $0.34$ respectively.  Right: decay branching ratios of $\tilde H_5^+$ to $WW$ (red), $W\gamma$ (green), and $f \bar f$ (black) as a function of $m_5$.  Decays to fermions are computed including only the dominant modes: $tb$ above the $tb$ threshold and $cs$ and $cd$ below.  The calculation of $H_5^+ \to W \gamma$ assumes an on-shell final-state $W$, so we plot the branching ratios only between 80 and 200~GeV.}
	\label{fig:LMGSkappaH5Pf}
\end{figure}

Of particular interest are decays of the would-be fermiophobic $H_5$ states to fermion pairs induced by custodial symmetry violation.  We study this for $\tilde H_5^+$ in Fig.~\ref{fig:LMGSkappaH5Pf}.  In the left panel we plot $\kappa_f^{\tilde H_5^+}$ as a function of $m_5$.  While the values are tiny for most scan points, they can reach values as large as about $\pm 0.3$ for $m_5$ close to 200~GeV.  However, this does not change the overall pattern of $H_5^+$ decays; as shown in the right panel of Fig.~\ref{fig:LMGSkappaH5Pf}, the branching ratio into $WZ$ (red) continues to dominate, with the loop-induced decay into $W\gamma$ (green) becoming important for $m_5$ below the $WZ$ kinematic threshold.  The branching ratio into fermions (black) remains small.  

\begin{figure}
	\resizebox{0.5\textwidth}{!}{\includegraphics{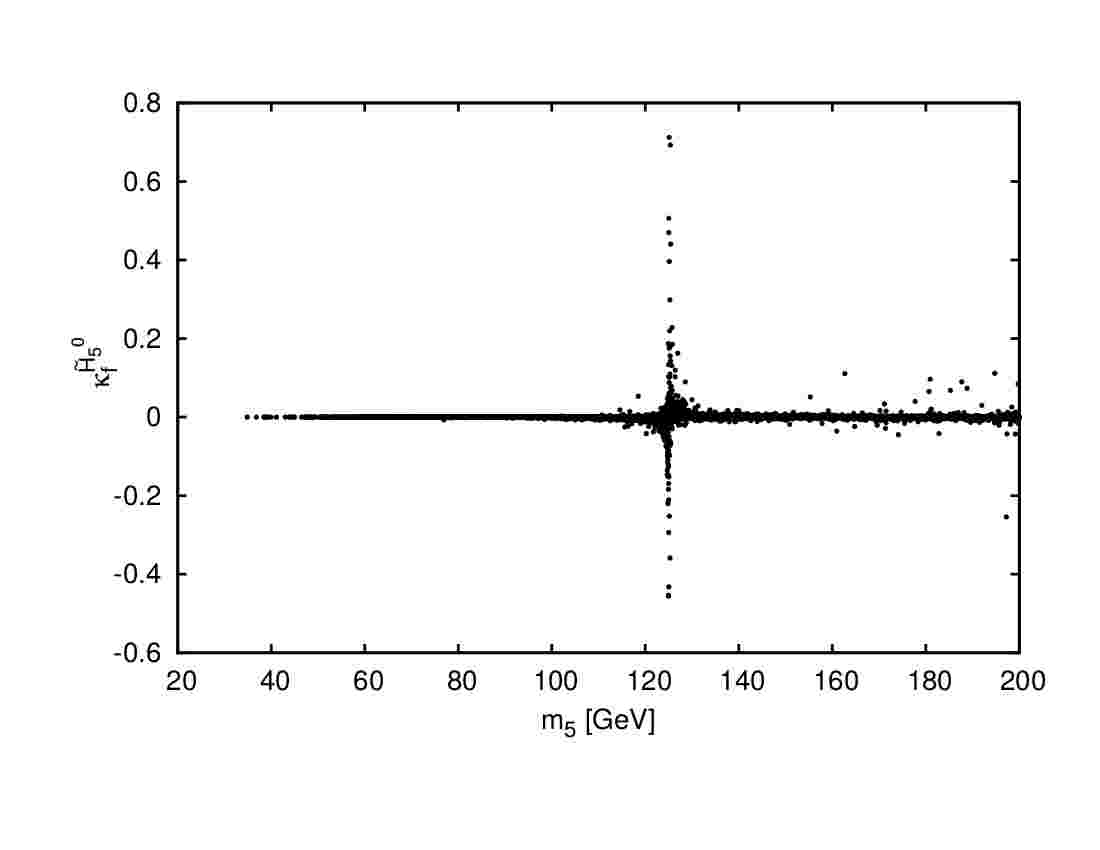}}%
	\resizebox{0.5\textwidth}{!}{\includegraphics{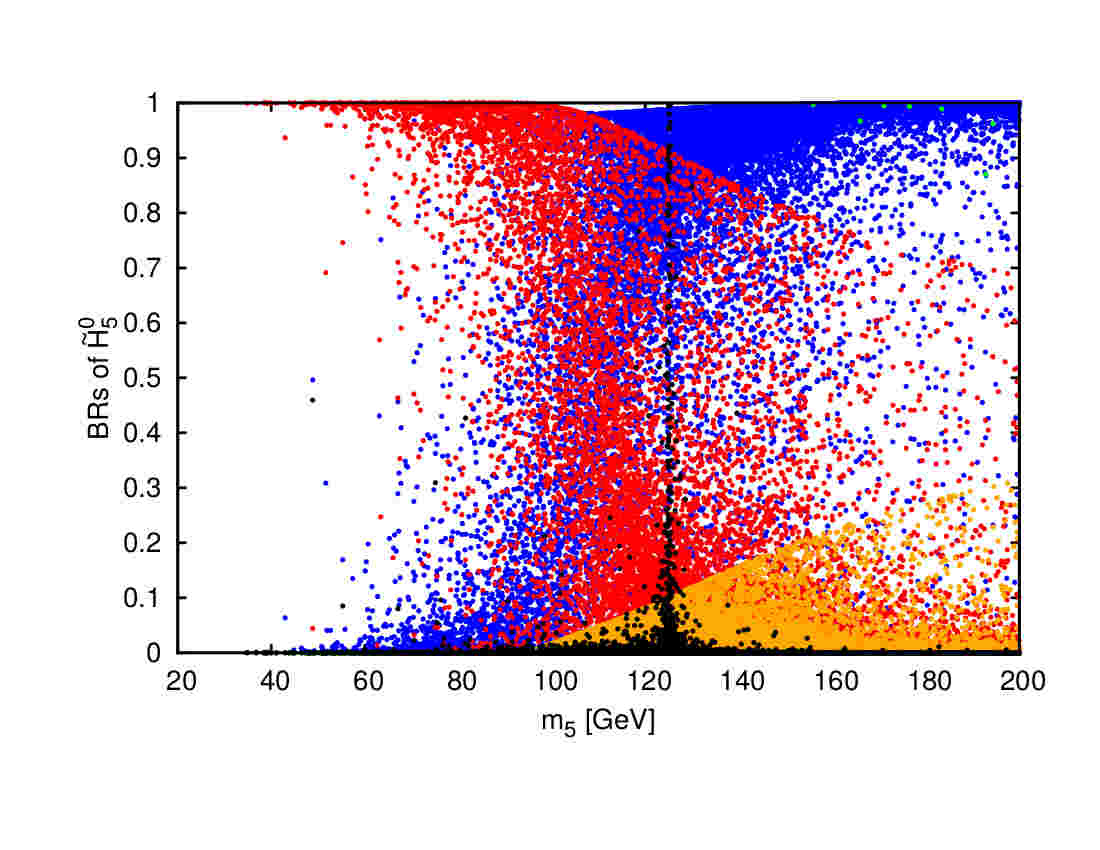}}
	\caption{Custodial-symmetry-violation-induced couplings of $\tilde H_5^0$ to fermions in a general scan of the low-$m_5$ region, taking the scale of the custodial-symmetric theory to be as large as possible subject to perturbative unitarity and the $\rho$ parameter constraint.  Left: $\kappa_f^{\tilde H_5^0}$ as a function of $m_5$.  The minimum and maximum values are $-0.46$ and $0.71$ respectively. Right: decay branching ratios of $\tilde H_5^0$ to $WW/ZZ$ (blue), $\gamma\gamma$ (red), $f \bar f$ (black), $Z\gamma$ (orange), and custodial-violating decays to pairs of other scalars (green---a few points in the upper right of the plot).}
	\label{fig:LMGSkappaH50f}
\end{figure}

Decays of $H_5^0$ to fermion pairs are studied in Fig.~\ref{fig:LMGSkappaH50f}.  In the left panel we plot $\kappa_f^{\tilde H_5^0}$ as a function of $m_5$.  The values are reasonably small except for $m_5$ around 125~GeV, where mixing between $H_5^0$ and $h$ becomes resonant.  In the right panel we show the branching ratios of $\tilde H_5^0$.  Decays to fermion pairs (black points) can become dominant only in the resonant-mixing region; away from $m_5 \simeq 125$~GeV the branching ratio to fermion pairs generally remain below 10\%, including at very low $m_5$ values where the branching ratio into $\gamma \gamma$ (red) remains dominant.  This is good news for the continued viability of the diphoton resonance search to constrain $\tilde H_5^0$ at low masses as proposed in Ref.~\cite{Delgado:2016arn}.  

\begin{figure}
	\resizebox{0.5\textwidth}{!}{\includegraphics{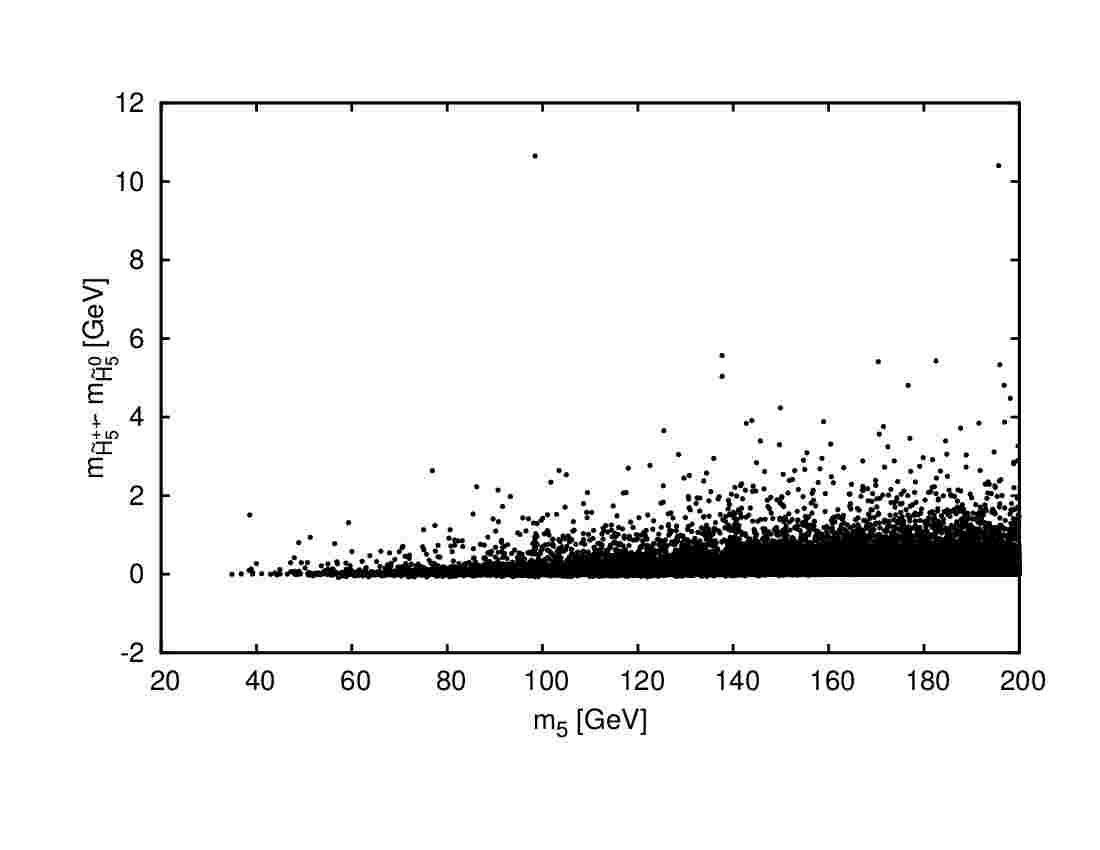}}%
	\resizebox{0.5\textwidth}{!}{\includegraphics{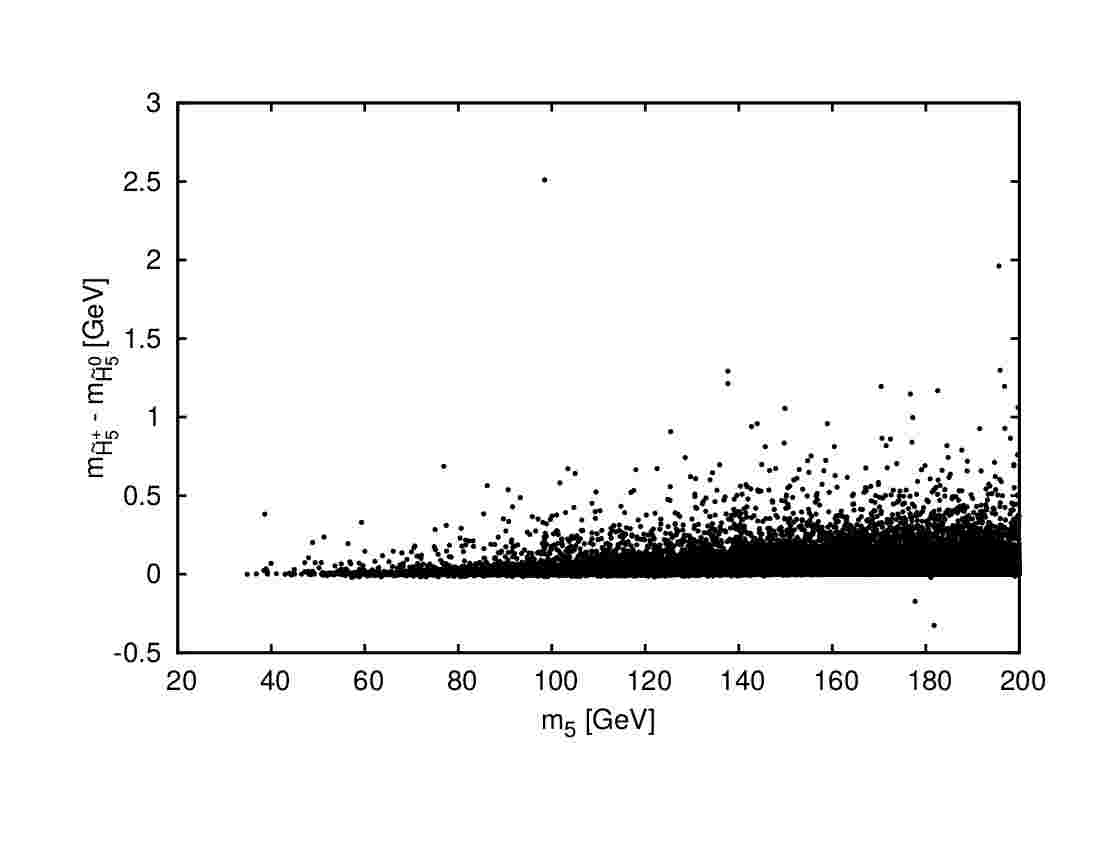}}
	\resizebox{0.5\textwidth}{!}{\includegraphics{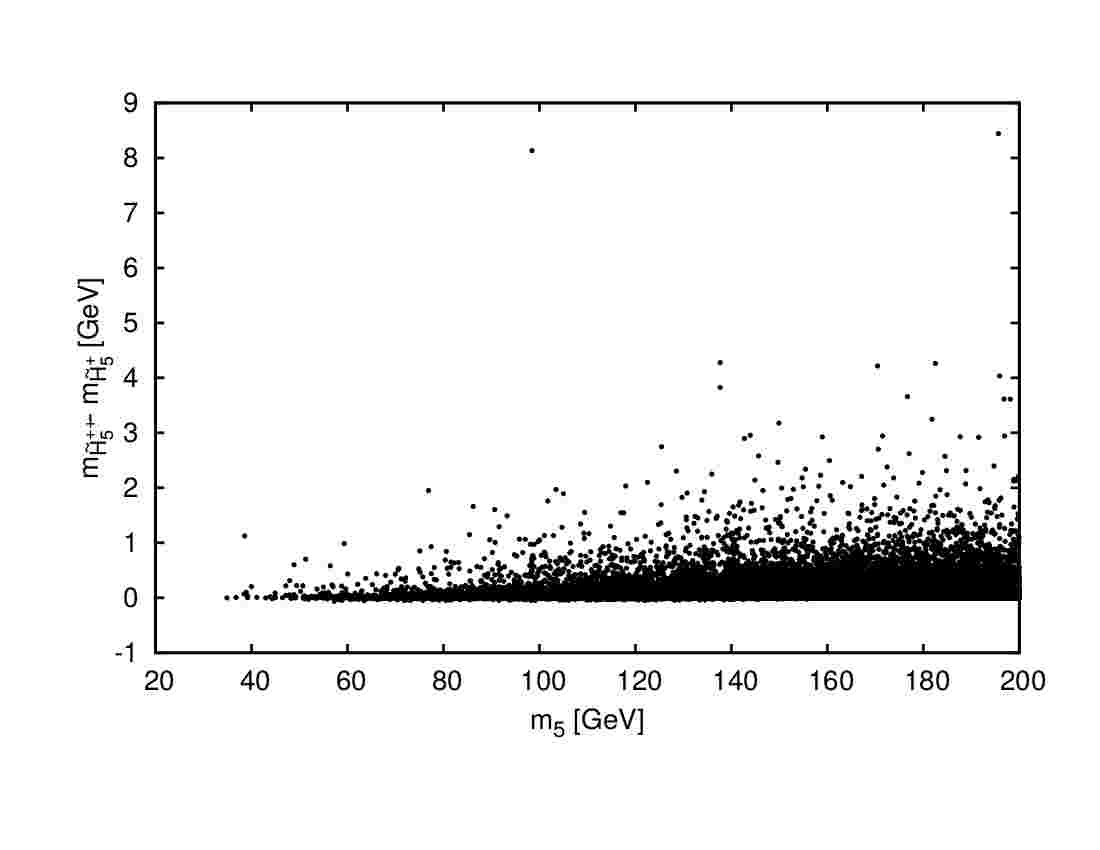}}%
	\resizebox{0.5\textwidth}{!}{\includegraphics{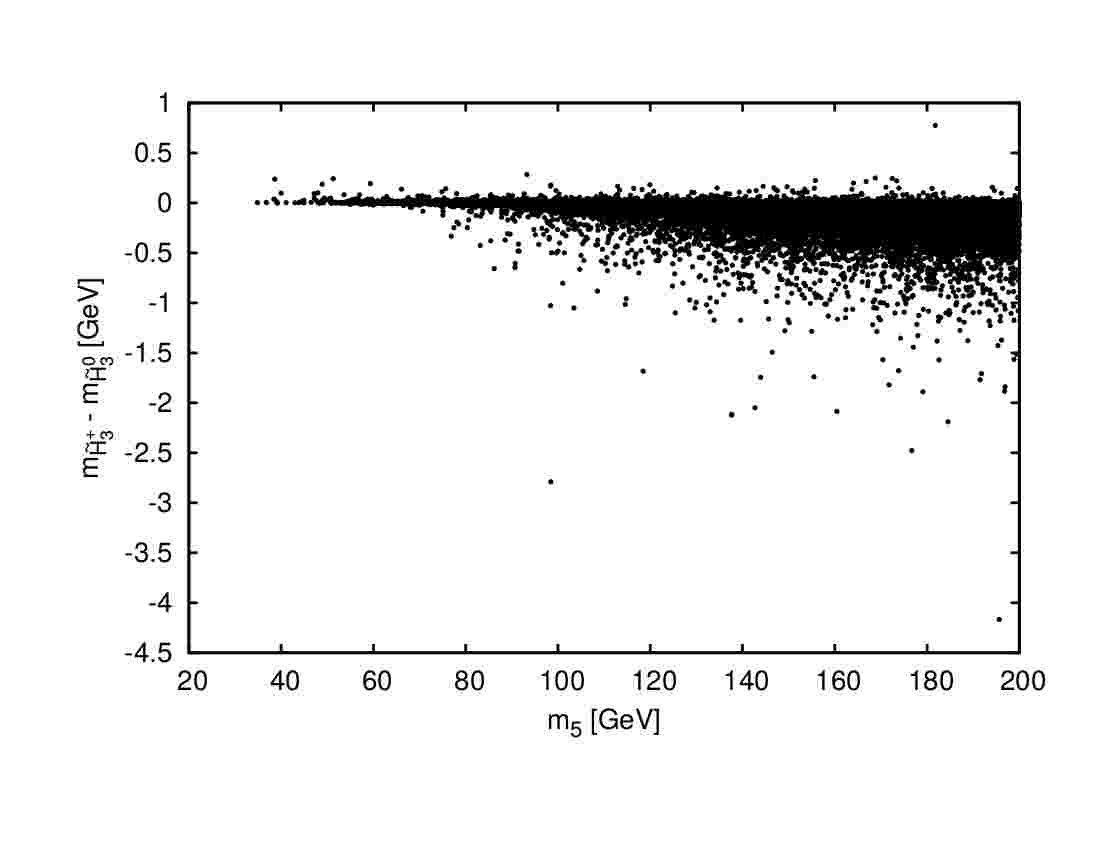}}
	\caption{Mass splittings among the members of the custodial fiveplet and triplet in a general scan of the low-$m_5$ region, taking the scale of the custodial-symmetric theory to be as large as possible subject to perturbative unitarity and the $\rho$ parameter constraint.  For the fiveplet we show  $m_{\tilde H_5^{++}} - m_{\tilde H_5^0}$ (top left, ranging between $-0.08$~GeV and $10.6$~GeV), $m_{\tilde H_5^+} - m_{\tilde H_5^0}$ (top right, ranging between $-0.33$~GeV and $2.51$~GeV), and $m_{\tilde H_5^{++}} - m_{\tilde H_5^+}$ (bottom left, ranging between $-0.06$~GeV and $8.44$~GeV), and for the triplet we show $m_{\tilde H_3^+} - m_{\tilde H_3^0}$ (bottom right, ranging between $-4.17$~GeV and $0.78$~GeV).}
	\label{fig:LSGSH5MassSplitting}
\end{figure}

Mass splittings among the members of the custodial fiveplet and triplet are shown in Fig.~\ref{fig:LSGSH5MassSplitting}.  As in the general scans for larger $m_5$, $\tilde H_5^{++}$ tends to be the heaviest of the fiveplet states, followed by $\tilde H_5^+$, with $\tilde H_5^0$ the lightest.  Likewise $\tilde H_3^0$ tends to be heavier than $\tilde H_3^+$, though this ordering can be reversed for a minority of the scan points.  The approximate relation $m_{\tilde H_5^{++}} - m_{\tilde H_5^0} \simeq 4 (m_{\tilde H_5^{+}} - m_{\tilde H_5^0})$ holds true in the low-$m_5$ scan as well.  The custodial-violating mass splittings are below 2~GeV in most of the parameter space, and less than about 10~GeV over the whole scan.

\begin{figure}
	\resizebox{0.5\textwidth}{!}{\includegraphics{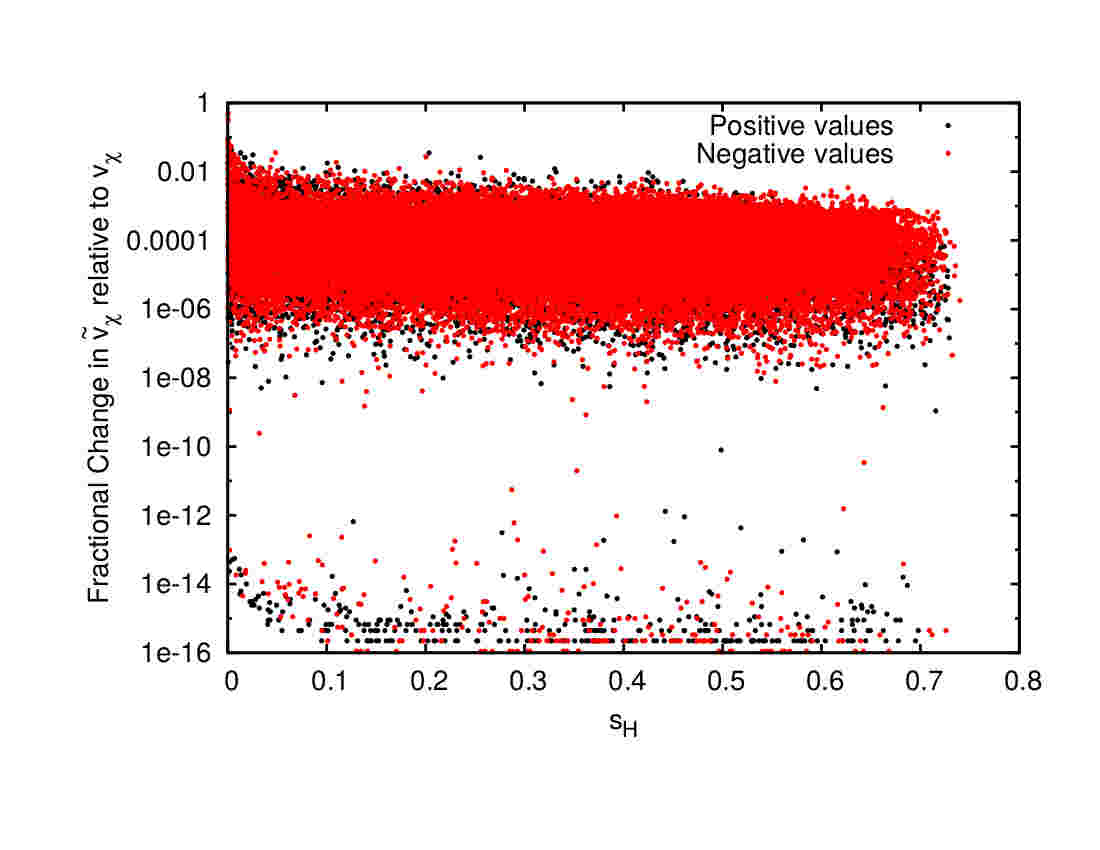}}%
	\resizebox{0.5\textwidth}{!}{\includegraphics{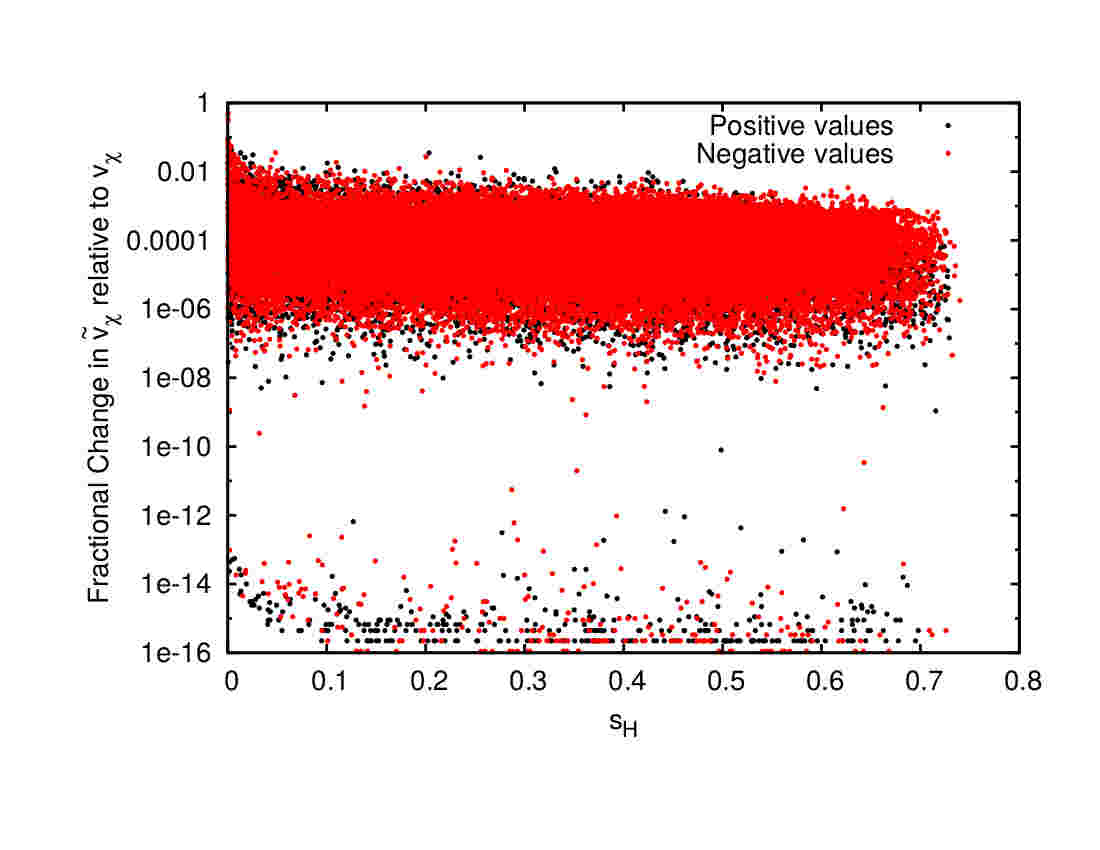}}
	\caption{Fractional deviation of $\tilde v_\chi$ relative to the weak-scale custodial-symmetric input $v_{\chi}$, defined as $\frac{\tilde v_\chi}{v_\chi} -1$, as a function of $m_5$ (left) and $s_H$ (right) in a general scan of the low-$m_5$ region, taking the scale of the custodial-symmetric theory to be as large as possible subject to perturbative unitarity and the $\rho$ parameter constraint.  Positive deviations are shown in black and negative in red so that both can be plotted on a log scale.  The fractional deviation ranges between $-1.66$ and $0.55$, with these large deviations appearing mainly at very small $s_H$.}
	\label{fig:LSGSvChiChange}
\end{figure}

Finally in Fig.~\ref{fig:LSGSvChiChange} we plot the fractional change in $\tilde v_\chi$ relative to the weak-scale custodial-symmetric input $v_{\chi}$, defined as $\frac{\tilde v_{\chi}}{v_{\chi}} - 1$.  This deviation can be positive (black points) or negative (red points), though negative deviations tend to occur only for $m_5$ above 100~GeV.  The absolute value of the deviation is again small, below the percent level unless $s_H$ is very small.

\section{Conclusions}
\label{sec:conclusions}

In this paper we studied the effects of custodial symmetry violation in the Georgi-Machacek model.  We considered the scenario in which the exactly custodial-symmetric GM model emerges at some high scale $\Lambda$ as an effective low energy theory of an unspecified ultraviolet completion, and then ran the model down to the weak scale, through which running the hypercharge interactions give rise to custodial symmetry violation at one loop.  The amount and pattern of custodial symmetry violation at the weak scale, as manifested through the couplings and masses of the physical scalars, is uniquely determined by the parameters of the high-scale custodial-symmetric theory and the value of the scale $\Lambda$ and hence can be meaningfully constrained by the measured value of the electroweak $\rho$ parameter.

To implement this program we used the most general gauge invariant scalar potential for the theory, from which we computed the minimization conditions for the vevs, and the expressions for the physical scalar mass eigenstates.  These allowed us to calculate the custodial symmetry violating couplings of the physical $\tilde H_5^0$ and $\tilde H_5^+$ states to fermions, as well as the parameter $\lambda^{\tilde h}_{WZ} \equiv \kappa_W^{\tilde h}/\kappa_Z^{\tilde h}$ for the 125~GeV Higgs boson.  We rederived the renormalization group equations for the parameters of the most general scalar potential including CP violation, and confirm the results of Ref.~\cite{Blasi:2017xmc} in the CP-conserving limit.  In our numerical implementation of the RGE running we self-consistently adjusted the custodial-symmetric inputs to obtain the correct values of the physical 125~GeV Higgs boson mass, top quark mass, and Fermi constant $G_F$ in the weak-scale custodial-violating theory.  

We presented numerical results in the H5plane benchmark (which helped make evident patterns like the relationship between the $\tilde H_5^{++}$--$\tilde H_5^+$ and the $\tilde H_5^+$--$\tilde H_5^0$ mass splittings) as well as a general scan over the full parameter space.  We showed that the results in the H5plane benchmark are broadly typical of the full parameter scan, though more extreme values can be obtained in small regions of parameter space in the general scan, particularly when the custodial-symmetric mass spectrum is such that the mixing among scalars in different custodial representations becomes resonant.  We also performed a dedicated general scan for low $m_5 < 200$~GeV which is not captured in the H5plane benchmark.

In each case, we determined the maximum allowed scale of the custodial-symmetric theory imposing perturbative unitarity of two-to-two scalar scattering amplitudes and the experimental constraint on the $\rho$ parameter.  This allowed us to quantify the maximum possible deviation of $\lambda^{\tilde h}_{WZ}$ from its SM value, as well as the branching ratios of the otherwise-fermiophobic $\tilde H_5^0$ and $\tilde H_5^{\pm}$ scalars into fermions and the mass splittings within the custodial triplet and fiveplet. We found that the scale of the custodial-symmetric theory could be as high as tens to hundreds of TeV, with an upper bound of 290~TeV in the H5plane benchmark.  We showed that $\lambda^{\tilde h}_{WZ}$ can deviate from its SM value by at most two per mille when $m_5 > 200$~GeV, though larger deviations at the percent level are possible in the low-$m_5$ region even away from the resonant mixing region $m_5 \simeq m_h$.  We also showed that the mass splittings within the custodial triplet and the custodial fiveplet are below 10~GeV over almost the entire parameter space, reaching larger values only for large scalar masses.  Both of these custodial-violating effects are too small to be probed at the LHC, but may be detectable at a future $e^+e^-$ collider.  Finally we showed that the fermionic branching ratios of $\tilde H_5^0$ and $\tilde H_5^+$ remain below the 10\% level, even for $\tilde H_5$ masses below the $WW$ and $WZ$ thresholds where they can compete directly with the loop-induced $\gamma\gamma$ and $W\gamma$ decay modes (with the exception of a narrow region of resonant mixing between $H_5^0$ and $h$ at 125~GeV).  This preserves the usefulness of the $\gamma\gamma$ decay mode to put strong constraints on $\tilde H_5^0$ at low masses. 

From these results, we can draw two important conclusions about the GM model. The generically small custodial-violating effects allows us to conclude that the use of the custodial-symmetric GM model as a benchmark model for LHC searches is justified. Furthermore, the large upper bound on the scale of the UV completion suggests that virtual effects from particles at the UV completion scale will be highly suppressed and their contribution to effective operators measured at the LHC will be too small to detect. This means that not only is the GM model a useful benchmark at the LHC but it is also a valid effective theory at the weak scale.

\begin{acknowledgments}
We thank Kei Yagyu for helpful discussions about the results of Ref.~\cite{Blasi:2017xmc}.
This work was supported by the Natural Sciences and Engineering Research Council of Canada.  H.E.L.\ was also supported by the grant H2020-MSCA-RISE-2014 no.~645722 (NonMinimalHiggs). B.K.\ was also supported by a Mitacs Globalink--Japan Society for the Promotion of Science Internship.  T.P.\ has been partially supported by National Science Centre, Poland,
under research grant no.\ 2017/26/D/ST2/00225.  B.K.\ thanks the Particle Physics Theory Group at Osaka University for hospitality while part of this work was performed.
\end{acknowledgments}

\appendix

\section{Renormalization group equations for Lagrangian parameters}
\label{sec:rge}

In order to run the parameters down from a custodial-symmetric high scale to the weak scale, we need the RGEs for the parameters of the most general gauge-invariant potential as given in Eq.~(\ref{eq:potential3}). RGEs can be calculated with the public codes PyR@TE~\cite{Lyonnet:2013dna}, a Python code that generates two-loop RGEs for non-supersymmetric models, and SARAH~\cite{Staub:2011dp}, a Mathematica package which can generate two-loop RGEs for supersymmetric and non-supersymmetric models. PyR@TE requires the user to supply their own GM model card while SARAH provides a GM model card in an alternate parameterization of the scalar potential. We determine the RGES using the formalism presented in Ref.~\cite{Cheng:1973nv}, some details of which are given in Appendix~\ref{app:CEL}. The resulting equations are then (with $t \equiv \log \mu$, where $\mu$ is the energy scale), 
\begin{equation}
	16\pi^2 \frac{d\left(\tilde{\mu}_2^2\right)}{dt} =
	\frac{3}{2}\tilde{M}_1^2 + 3 |\tilde{M}^{\prime}_1|^2
	+ \tilde{\mu}_2^2 \left( 6 y_b^2 + 6 y_t^2 + 2  y_{\tau}^2
	- \frac{9}{10} g_1^2 - \frac{9}{2} g_2^2
	+ 12 \tilde{\lambda}_1 \right) + 6 \tilde{\mu}_3^2 \tilde{\lambda}_6 
	+ 6 \tilde{\mu}^{\prime2}_3 \tilde{\lambda}_5,
	\label{eq:RGEtmu2sq}
\end{equation}
\begin{equation}
	16\pi^2 \frac{d\left(\tilde{\mu}^{\prime2}_3\right)}{dt} =
	|\tilde{M}^{\prime}_1|^2 + 144 \tilde{M}_2^2 + \tilde{\mu}^{\prime2}_3 \left( 8 \tilde{\lambda}_2 + 	16 \tilde{\lambda}_7 - \frac{18}{5} g_1^2 - 12  g_2^2
	\right) + 4 \tilde{\mu}_2^2 \tilde{\lambda}_5 + 2 \tilde{\mu}_3^2 \left( \tilde{\lambda}_9 
	+ 3 \tilde{\lambda}_{10} \right),
\end{equation}
\begin{equation}
	16\pi^2 \frac{d\left(\tilde{\mu}_3^2\right)}{dt} =
	\tilde{M}_1^2 + 144 \tilde{M}_2^2
	+ 4 \tilde{\mu}_3^2 \left( 10 \tilde{\lambda}_8 - 3 g_2^2 \right)
	+ 8 \tilde{\mu}_2^2 \tilde{\lambda}_6 + 4 \tilde{\mu}^{\prime2}_3 \left( \tilde{\lambda}_9 
	+ 3 \tilde{\lambda}_{10} \right),
\end{equation}
\begin{equation}\begin{aligned}
	16\pi^2 \frac{d\tilde{\lambda}_1}{dt} =
	- 6 y_b^4 - 6 y_t^4 - 2 y_{\tau}^4 + \tilde{\lambda}_1 \left( 12 y_b^2 + 12 y_t^2 + 4 y_{\tau}^2 
	- \frac{9}{5} g_1^2 - 9 g_2^2
	+ 24 \tilde{\lambda}_1 \right) &\\
	+ \frac{27}{200} g_1^4 + \frac{9}{8} g_2^4 + \frac{9}{20} g_1^2 g_2^2 
	+ \frac{1}{2} \tilde{\lambda}_3^2 + 2 |\tilde{\lambda}_4|^2 + 3 \tilde{\lambda}_5^2 
	+ 6 \tilde{\lambda}_6^2&,
\end{aligned}\end{equation}
\begin{equation}
	16\pi^2 \frac{d\tilde{\lambda}_2}{dt} =
	3 g_2^4 - \frac{36}{5} g_1^2 g_2^2 + 12 \tilde{\lambda}_2 \left( \tilde{\lambda}_2 
	+ 2 \tilde{\lambda}_7 - \frac{3}{5} g_1^2 - 2 g_2^2 \right) - \frac{1}{2}\tilde{\lambda}_3^2 
	+ \tilde{\lambda}_9^2,
\end{equation}
\begin{equation}
	16\pi^2 \frac{d\tilde{\lambda}_3}{dt} =
	\tilde{\lambda}_3 \left( 6 y_b^2 + 6 y_t^2 + 2 y_{\tau}^2 + 4 \tilde{\lambda}_1 - 8 \tilde{\lambda}_2 
	+ 8 \tilde{\lambda}_5 + 4 \tilde{\lambda}_7 - \frac{9}{2} g_1^2 - \frac{33}{2} g_2^2 \right) 
	+ \frac{36}{5} g_2^2 g_1^2 + 4 |\tilde{\lambda}_4|^2,
\end{equation}
\begin{equation}
	16\pi^2 \frac{d\tilde{\lambda}_4}{dt} =
	\tilde{\lambda}_4 \left( 6 y_b^2 + 6 y_t^2 + 2 y_{\tau}^2 - \frac{27}{10} g_1^2 - \frac{33}{2} g_2^2 
	+ 4 \tilde{\lambda}_1 + 2 \tilde{\lambda}_3 + 4 \tilde{\lambda}_5 + 8 \tilde{\lambda}_6 
	- 2 \tilde{\lambda}_9 + 4 \tilde{\lambda}_{10} \right),
\end{equation}
\begin{equation}\begin{aligned}
	16\pi^2 \frac{d\tilde{\lambda}_5}{dt} =
	\tilde{\lambda}_5 \left( 6 y_b^2 + 6 y_t^2 + 2 y_{\tau}^2 + 4 \tilde{\lambda}_5 
	+ 12 \tilde{\lambda}_1 + 8 \tilde{\lambda}_2 + 16 \tilde{\lambda}_7 - \frac{9}{2} g_1^2 
	- \frac{33}{2} g_2^2 \right)&\\  + \frac{27}{25} g_1^4 + 6 g_2^4 + 2 \tilde{\lambda}_3^2 
	+ 4 |\tilde{\lambda}_4|^2 + 4 \tilde{\lambda}_6 \tilde{\lambda}_9 
	+ 12 \tilde{\lambda}_6 \tilde{\lambda}_{10}&,
\end{aligned}\end{equation}
\begin{equation}
	16\pi^2 \frac{d\tilde{\lambda}_6}{dt} =
	\tilde{\lambda}_6 \left( 6 y_b^2 + 6 y_t^2 + 2 y_{\tau}^2 + 8 \tilde{\lambda}_6 
	+ 12 \tilde{\lambda}_1 + 40 \tilde{\lambda}_8 - \frac{9}{10} g_1^2 - \frac{33}{2} g_2^2 \right) 
	+ 3 g_2^4 + 4 |\tilde{\lambda}_4|^2 + 2 \tilde{\lambda}_5 \tilde{\lambda}_9 
	+ 6 \tilde{\lambda}_5 \tilde{\lambda}_{10},
\end{equation}
\begin{equation}
	16\pi^2 \frac{d\tilde{\lambda}_7}{dt} =
	\frac{54}{25} g_1^4 + 9 g_2^4 + \frac{36}{5} g_2^2 g_1^2 +\left( - \frac{36}{5} g_1^2 - 24 g_2^2 
	+ 16 \tilde{\lambda}_2 + 28 \tilde{\lambda}_7 \right) + 16 \tilde{\lambda}_2^2 
	+ \frac{1}{2}\tilde{\lambda}_3^2 + 2 \tilde{\lambda}_5^2 + \tilde{\lambda}_9^2 
	+ 2 \tilde{\lambda}_{10} \left( 3 \tilde{\lambda}_{10} + 2 \tilde{\lambda}_9 \right),
\end{equation}
\begin{equation}
	16\pi^2 \frac{d\tilde{\lambda}_8}{dt} =
	3 g_2^4 + 8 \tilde{\lambda}_8 \left(- 3 g_2^2 + 11 \tilde{\lambda}_8 \right) + 2 \tilde{\lambda}_6^2 
	+ \tilde{\lambda}_9 \left( \tilde{\lambda}_9 + 2 \tilde{\lambda}_{10} \right) 
	+ 3 \tilde{\lambda}_{10}^2,
\end{equation}
\begin{equation}
	16\pi^2 \frac{d\tilde{\lambda}_9}{dt} =
	6 g_2^4 + 2 \tilde{\lambda}_9 \left( - 12 g_2^2 - \frac{9}{5} g_1^2 + 5 \tilde{\lambda}_9 
	+ 4 \tilde{\lambda}_2 + 2 \tilde{\lambda}_7 + 8 \tilde{\lambda}_8 + 8 \tilde{\lambda}_{10} \right) 
	- 2 |\tilde{\lambda}_4|^2,
\end{equation}
\begin{equation}
	16\pi^2 \frac{d\tilde{\lambda}_{10}}{dt} =
	6 g_2^4 + 2 \tilde{\lambda}_{10} \left( - \frac{9}{5} g_1^2 - 12 g_2^2 + 4 \tilde{\lambda}_2 
	+ 8 \tilde{\lambda}_7 + 20 \tilde{\lambda}_8 + 4 \tilde{\lambda}_{10} \right) 
	+ 2 |\tilde{\lambda}_4|^2 + 2 \tilde{\lambda}_9^2 + 4 \tilde{\lambda}_5 \tilde{\lambda}_6 
	+ 4 \tilde{\lambda}_9 \left( \tilde{\lambda}_7 + 2 \tilde{\lambda}_8 \right),
\end{equation}
\begin{equation}
	16\pi^2 \frac{d\tilde{M}^{\prime}_1}{dt} =
	\tilde{M}^{\prime}_1 \left( 6 y_b^2 + 6 y_t^2 + 2  y_{\tau}^2 - \frac{27}{10} g_1^2 
	- \frac{21}{2}  g_2^2 + 4 \tilde{\lambda}_1 + 4 \tilde{\lambda}_3 + 4 \tilde{\lambda}_5 \right) 
	+ 4 \sqrt{2} \tilde\lambda_4^* \left( \tilde M_1 + 6 \tilde M_2 \right),
\end{equation}
\begin{equation}
	16\pi^2 \frac{d\tilde{M}_1}{dt} = 
	\tilde{M}_1 \left( 6 y_b^2 + 6 y_t^2 + 2 y_{\tau}^2
	- \frac{9}{10} g_1^2 - \frac{21}{2} g_2^2
	+ 4 \tilde{\lambda}_1 + 8 \tilde{\lambda}_6 \right) + 24 \tilde{M}_2 \tilde{\lambda}_3 
	+ 8 \sqrt{2} {\rm Re} \left[ \tilde{M}^{\prime}_1 \tilde{\lambda}_4 \right],
\end{equation}
\begin{equation}
	16\pi^2 \frac{d\tilde{M}_2}{dt} =
	\tilde{M}_2 \left( - \frac{18}{5} g_1^2 - 18 g_2^2 - 8 \tilde{\lambda}_2 + 4 \tilde{\lambda}_7 
	- 4  \tilde{\lambda}_9 + 8 \tilde{\lambda}_{10} \right) + \frac{1}{6} \tilde{M}_1 \tilde{\lambda}_3 
	+ \frac{1}{3} \sqrt{2} {\rm Re} \left[ \tilde{M}^{\prime}_1 \tilde{\lambda}_4 \right],
\label{eq:RGEtM2}
\end{equation}
where $g_1$ and $g_2$ are gauge couplings (see below) and $y_b$, $y_t$, and $y_{\tau}$ are Yukawa couplings, normalized according to $y_f = \sqrt{2} m_f/\tilde v_{\phi}$.
These RGEs agree with those of Ref.~\cite{Blasi:2017xmc} (for real $\tilde \lambda_4$ and $\tilde M_1^{\prime}$) after translating the notation for the Lagrangian parameters as follows:  

\begin{eqnarray}
\sigma_1 &=&  - \frac{\tilde \lambda_3}{2} + \tilde \lambda_5, \nonumber \\
\sigma_2 &=&\tilde \lambda_3, \nonumber \\
\sigma_3 &=&\tilde \lambda_6, \nonumber \\
\sigma_4 &=&\tilde \lambda_4, \nonumber \\
\lambda &=&\tilde \lambda_1, \nonumber \\
\rho_1 &=& 2 \tilde \lambda_2 + \tilde \lambda_7, \nonumber \\
\rho_2 &=& -2 \tilde \lambda_2, \nonumber \\
\rho_3 &=& 2 \tilde \lambda_8, \nonumber \\
\rho_4 &=& \tilde \lambda_{10}, \nonumber \\ 
\rho_5 &=& \tilde \lambda_9, \nonumber \\
\mu_1 &=& \frac{ \tilde M_1}{\sqrt{2}}, \nonumber \\
\mu_2 &=& \frac{\tilde M_1^\prime}{2}, \nonumber \\ 
\mu_3 &=& -6 \sqrt{2} \tilde M_2, \nonumber \\
m_\phi^2 &=& \tilde \mu_2^2, \nonumber \\
m_\chi^2 &=& \tilde \mu_3^{\prime 2}, \nonumber \\
m_\xi^2 &=& \frac{\tilde \mu_3^2}{2}. \nonumber \\
\end{eqnarray} 

A few possible symmetries are apparent in these RGEs.  Setting $\tilde M_1^{\prime} = \tilde M_1 = \tilde M_2 = 0$, the potential becomes invariant under $(\chi, \xi) \to (-\chi, -\xi)$ and therefore these three parameters are not regenerated by the running.  Setting instead $\tilde \lambda_4 = \tilde M_1^{\prime} = 0$, the potential becomes invariant under $\chi \to -\chi$ and therefore these two parameters are not regenerated by the running.  Setting $\tilde \lambda_4 = \tilde M_1 = \tilde M_2 = 0$, the potential becomes invariant under $\xi \to -\xi$ and therefore these three parameters are not regenerated by the running.  Finally, if all the Lagrangian parameters are taken to be real at some scale, as will be the case when the most general potential is matched onto the intrinsically CP-conserving custodial-symmetric Georgi-Machacek model, they remain real at all scales.

Throughout we use the GUT normalization $g^{\prime} = \sqrt{\frac{3}{5}}\,g_1$, $g = g_2$, and $g_s = g_3$. The renormalization group equations for the electroweak gauge couplings, including all the particle content of the GM model in the spectrum, are~\cite{Hamada:2015bra},
\begin{eqnarray}
16\pi^2 \frac{d g_1}{dt} &=& \frac{47}{10}g_1^3 \qquad {\rm or \ equivalently} \qquad
16 \pi^2 \frac{d g^{\prime}}{dt} = \frac{47}{6} g^{\prime 3}, 
	\label{eq:RGEg1} \\
16\pi^2 \frac{d g_2}{dt} &=& -\frac{13}{6}g_2^3,
\end{eqnarray}
and that for the strong gauge coupling is the same as in the SM (including the top quark contribution),
\begin{equation}
16\pi^2 \frac{d g_3}{dt} = -7 g_3^3.
\end{equation}
The RGEs for the Yukawa couplings are identical to those of the SM~\cite{Goudelis:2013uca},
\begin{equation}
16\pi^2 \frac{d y_t}{dt} = \left(-\frac{17}{20}g_1^2 - \frac{9}{4}g_2^2 - 8g_3^2 + \frac{3}{2}y_b^2 + \frac{9}{2}y_t^2 + y_{\tau}^2\right)y_t,
	\label{eq:RGEyt}
\end{equation}
\begin{equation}
16\pi^2 \frac{d y_b}{dt} = \left(-\frac{1}{4}g_1^2 - \frac{9}{4}g_2^2 - 8g_3^2 + \frac{9}{2}y_b^2 + \frac{3}{2}y_t^2 + y_{\tau}^2\right)y_b,
\end{equation}
\begin{equation}
16\pi^2 \frac{d y_{\tau}}{dt} = \left(-\frac{9}{4}g_1^2 - \frac{9}{4}g_2^2 + 3y_b^2 + 3y_t^2 + \frac{5}{2}y_{\tau}^2\right)y_{\tau}.
\end{equation}
In our numerical work we will ignore $y_b$ and $y_{\tau}$.

As a consistency check, we can turn off the custodial-violating parts of the RGEs by setting $g_1 = 0$ and substituting the relations given in Eq.~(\ref{eq:nogprime}). We then find a self-consistent set of RGEs for the custodial-preserving Lagrangian parameters:
\begin{equation}
	16\pi^2 \frac{d\left(\mu_2^2\right)}{dt} = 
	 \frac{9}{2} M_1^2
 	+ \mu_2^2 \left( 6 y_b^2 + 6 y_t^2 + 2 y_{\tau}^2 - \frac{9}{2} g_2^2 + 48 \lambda_1 \right) 
	+ 36 \mu_3^2 \lambda_2,
\end{equation}
\begin{equation}
	16\pi^2 \frac{d\left(\mu_3^2\right)}{dt} = 
 	M_1^2 + 144 M_2^2
	+ 16 \mu_2^2 \lambda_2 
	+ \mu_3^2 \left( - 12 g_2^2 + 56 \lambda_3 + 88 \lambda_4 \right),
\end{equation}
\begin{equation}
	16\pi^2 \frac{d\lambda_1}{dt} = 
	 -\frac{3}{2} y_b^4 - \frac{3}{2} y_t^4 - \frac{1}{2} y_{\tau}^4 
	 + \lambda_1 \left( 12 y_b^2 + 12 y_t^2 + 4 y_{\tau}^2 - 9 g_2^2 + 96 \lambda_1 \right)
	  + \frac{9}{32} g_2^4 + 18 \lambda_2^2 + \frac{3}{2} \lambda_5^2,
\end{equation}
\begin{equation}
	16\pi^2 \frac{d\lambda_2}{dt} = 
 	\lambda_2 \left( 6 y_b^2 + 6 y_t^2 + 2 y_{\tau}^2 - \frac{33}{2} g_2^2 
	+ 48 \lambda_1 + 16 \lambda_2 + 56 \lambda_3 + 88 \lambda_4 \right)
 	+ \frac{3}{2} g_2^4 + 4 \lambda_5^2,
\end{equation}
\begin{equation}
	16\pi^2 \frac{d\lambda_3}{dt} = 
 	\frac{3}{2} g_2^4 
	+ \lambda_3 \left( - 24 g_2^2 + 80 \lambda_3 + 96 \lambda_4 \right) - \lambda_5^2,
\end{equation}
\begin{equation}
	16\pi^2 \frac{d\lambda_4}{dt} = 
	 \frac{3}{2} g_2^4 
	+ \lambda_4 \left( - 24 g_2^2 + 136 \lambda_4 + 112 \lambda_3 \right)
 	+ 8 \lambda_2^2 + 24 \lambda_3^2  + \lambda_5^2,
\end{equation}
\begin{equation}
	16\pi^2 \frac{d\lambda_5}{dt} = 
	\lambda_5 \left( 6 y_b^2 + 6 y_t^2 + 2 y_{\tau}^2
	 - \frac{33}{2} g_2^2
	 + 16 \lambda_1 + 32 \lambda_2 - 8 \lambda_3 + 16 \lambda_4 - 4 \lambda_5 \right),
\end{equation}
\begin{equation}
	16\pi^2 \frac{d M_1}{dt} = 
	M_1 \left( 6 y_b^2 + 6 y_t^2 + 2 y_{\tau}^2
 	- \frac{21}{2} g_2^2
 	+ 16 \lambda_1 + 16 \lambda_2 - 16 \lambda_5 \right)
	 - 48 M_2 \lambda_5,
\end{equation}
\begin{equation}
	16\pi^2 \frac{d M_2}{dt} = 
	- M_1 \lambda_5 
	+ M_2 \left( - 18 g_2^2 - 24 \lambda_3 + 48 \lambda_4 \right).
\end{equation}

\section{Scalar couplings of the custodial violating states}
\label{ap:SCOUP}

The custodial violating couplings of the custodial symmetric eigenstates are included below:

\subsection{Couplings of the $H_5^0$}

The modified couplings for decays to scalars allowed by custodial symmetry:
\begin{equation}
	\begin{split}
		g_{H_5^0 H_3^+ H_3^-} & = \tilde \lambda_3 \left( \frac{s_H c_H \tilde v_\phi}{2 \sqrt{3}} + \frac{s_H^2 \tilde v_\chi}{\sqrt{6}} \right) + \tilde \lambda_4 \frac{1}{\sqrt{6}} s_H c_H \tilde v_\phi - \tilde \lambda_5 \frac{2 \sqrt{2}}{\sqrt{3}} s_H^2 \tilde v_\chi + \tilde \lambda_6 \frac{4 \sqrt{2}}{\sqrt{3}} s_H^2 \tilde v_\xi - \tilde \lambda_7 \frac{\sqrt{2}}{\sqrt{3}} c_H^2 \tilde v_\chi + \frac{4 \sqrt{2}}{\sqrt{3}} \tilde \lambda_8 c_H^2 \tilde v_\xi \nonumber \\ &  + \tilde \lambda_9 \frac{\sqrt{2}}{\sqrt{3}} \left( c_H^2 \tilde v_\xi - \frac{c_H^2 \tilde v_\chi}{2} +\frac{c_H^2 \tilde v_\xi}{2} - c_H^2 \tilde v_\chi \right) + \tilde \lambda_{10} \frac{\sqrt{2}}{\sqrt{3}} c_H^2 \left( \tilde v_\xi - \tilde v_\chi \right) +\tilde M_1 \frac{\sqrt{2}}{\sqrt{3}} s_H^2 -\sqrt{6} c_H^2 \tilde M_2		
	\end{split},
\end{equation}
\begin{equation}
	\begin{split}
		g_{H_5^+ \tilde H_3^0 \tilde H_3^0} & = - \tilde \lambda_3 \frac{ 8 \tilde v_\chi^3}{\sqrt{6} A^2} + \frac{\tilde \lambda_4}{\sqrt{3} A^2} \left( 8 \tilde v_\xi \tilde v_\chi^2 - 16 \tilde v_\chi^3- 8 \tilde v_\chi \tilde v_\phi^2 \right) - \tilde \lambda_5 \frac{16 \tilde v_\chi^3}{ \sqrt{6} A^2} + \tilde \lambda_6 \frac{16 \sqrt{2} \tilde v_\chi^2 \tilde v_\xi}{\sqrt{3} A^2} - \tilde \lambda_7 \frac{4 \tilde v_\chi \tilde v_\phi^2}{\sqrt{6} A^2} + \tilde \lambda_{10} \frac{2 \sqrt{2} \tilde v_\phi^2 \tilde v_\xi}{\sqrt{3} A^2} \nonumber \\ & - \tilde M_1^\prime \frac{8 \tilde v_\chi^2}{\sqrt{6} A^2} - \tilde M_1 \frac{8 \tilde v_\chi^2}{\sqrt{6} A^2} - \tilde M_2 \frac{6 \sqrt{2} \tilde v_\phi^2}{\sqrt{3} A^2}
		\end{split},	
\end{equation}
 
\begin{equation}
g_{H_5^+ H_3^+ \tilde H_3^0} = - \frac{i}{A} \left( \tilde \lambda_3 \left( s_H \tilde v_\chi^2 +\frac{s_H \tilde v_\phi^2}{4} \right) + \tilde \lambda_4 \left( \sqrt{2} s_H \tilde v_\chi^2 + 2 c_H \tilde v_\phi \tilde v_\chi + \frac{s_H \tilde v_\phi^2}{2 \sqrt{2}} \right) - \tilde \lambda_9 \frac{c_H \tilde v_\phi \tilde v_\xi}{\sqrt{2}} + \tilde M_1^\prime  \tilde v_\chi s_H - \tilde M_1 s_H \tilde v_\chi - \tilde M_2 \frac{6 c_H \tilde v_\phi}{\sqrt{2}} \right)
\end{equation}

\begin{equation}
g_{H_5^{++} H_3^-  H_3^-} =  -2 \left( - \tilde \lambda_2 c_H^2 \tilde v_\chi - \tilde \lambda_3 \frac{s_H c_H \tilde v_\phi}{2 \sqrt{2}} + \tilde \lambda_4 \left( \frac{s_H^2 \tilde v_\xi}{\sqrt{2}} - \frac{s_H c_H \tilde v_\phi}{2} \right) - \tilde M_1^\prime  \frac{s_H^2}{2} - 3 \tilde M_2 c_H^2 \right),
\end{equation}

\begin{equation}
g_{H_5^{++} H_5^-  H_5^-} =  -2 \left( - \tilde \lambda_2 \tilde v_\chi + 3 \tilde M_2 \right),
\end{equation}

where $A^2 = \tilde v_\phi^2 + 8 \tilde v_\chi^2$. 

The modified couplings for loop decays mediated by a $H_5$ loop:

\begin{equation}
	g_{H_5^0 H_5^+ H_5^-} = - \tilde \lambda_7 \frac{\tilde v_\chi}{\sqrt{6}} + \tilde \lambda_8 \frac{4 \sqrt{2} \tilde v_\xi}{\sqrt{3}} + \tilde \lambda_9 \left( \frac{\sqrt{2} \tilde v_\xi}{\sqrt{3}} - \frac{\tilde v_\chi}{6} - \frac{\tilde v_\xi}{6} + \frac{\sqrt{2} \tilde v_\chi}{\sqrt{3}} \right) + \tilde \lambda_{10} \left( \frac{\sqrt{2} \tilde v_\xi}{\sqrt{3}} - \frac{2 \tilde v_\chi}{6} \right) - \sqrt{6} \tilde M_2
\end{equation}

\begin{equation}
g_{H_5^0 H_5^{++} H_5^{--}} = - \tilde \lambda_2 \frac{8 \tilde v_\chi}{\sqrt{6}} - \tilde \lambda_7 \frac{4 \tilde v_\chi}{\sqrt{6}} + \tilde \lambda_{10} \frac{2 \sqrt{2} \tilde v_\xi}{\sqrt{3}} - \frac{6 \sqrt{2}}{\sqrt{3}} \tilde M_2.
\end{equation}

The couplings for decays to scalars that violate custodial symmetry:
\begin{equation}
	\begin{split}
		g_{H_5^0 h_{\tilde \alpha} h_{\tilde \alpha}} & = \frac{\tilde \lambda_3}{\sqrt{2}} \left( -\frac{\tilde v_\chi c_{\tilde \alpha}^2 }{2 \sqrt{3}} + \frac{\tilde v_\phi c_{\tilde \alpha} s_{\tilde \alpha}}{3} \right) + \tilde \lambda_4 \left( -\frac{\tilde v_\phi c_{\tilde \alpha} s_{\tilde \alpha}}{3} + c_{\tilde \alpha}^2 \left( \frac{\tilde v_\chi}{\sqrt{3}} - \frac{\tilde v_\xi}{2 \sqrt{3}} \right) \right) + \tilde \lambda_5 \left( - \frac{\tilde v_\chi c_{\tilde \alpha}^2}{\sqrt{6}} + \frac{\sqrt{2} c_{\tilde \alpha} s_{\tilde \alpha} \tilde v_\phi}{3} \right) \nonumber \\ & + \tilde \lambda_6 \left( \frac{\sqrt{2} \tilde v_\xi c_{\tilde \alpha}^2}{\sqrt{3}} -\frac{2 \sqrt{2} \tilde v_\phi c_{\tilde \alpha} s_{\tilde \alpha}}{3} \right) - \tilde \lambda_7 \frac{2 \sqrt{2} \tilde v_\chi s_{\tilde \alpha}^2}{\sqrt{3}} + \tilde \lambda_8 \frac{4 \sqrt{2} \tilde v_\xi s_{\tilde \alpha}^2}{\sqrt{3}} + \tilde \lambda_{10} \frac{\sqrt{2} \tilde v_\chi s_{\tilde \alpha}^2}{\sqrt{3}} +\tilde M_1^\prime \frac{c_{\tilde \alpha}^2}{2 \sqrt{6}} -\tilde M_1 \frac{c_{\tilde \alpha}^2}{2 \sqrt{6}}
	\end{split}
\end{equation}

\begin{equation}
\begin{split}
g_{H_5^0 H_{\tilde \alpha} H_{\tilde \alpha}} & = -\frac{\tilde \lambda_3}{\sqrt{2}} \left( \frac{\tilde v_\chi s_{\tilde \alpha}^2 }{2 \sqrt{3}} + \frac{\tilde v_\phi c_{\tilde \alpha} s_{\tilde \alpha}}{3} \right) + \tilde \lambda_4 \left( \frac{\tilde v_\phi c_{\tilde \alpha} s_{\tilde \alpha}}{3} + s_{\tilde \alpha}^2 \left( \frac{\tilde v_\chi}{\sqrt{3}} - \frac{\tilde v_\xi}{2 \sqrt{3}} \right) \right) + \tilde \lambda_5 \left( - \frac{\tilde v_\chi s_{\tilde \alpha}^2}{\sqrt{6}} - \frac{\sqrt{2} c_{\tilde \alpha} s_{\tilde \alpha} \tilde v_\phi}{3} \right) \nonumber \\ & + \tilde \lambda_6 \left( \frac{\sqrt{2} \tilde v_\xi s_{\tilde \alpha}^2}{\sqrt{3}} +\frac{2 \sqrt{2} \tilde v_\phi c_{\tilde \alpha} s_{\tilde \alpha}}{3} \right) - \tilde \lambda_7 \frac{2 \sqrt{2} \tilde v_\chi c_{\tilde \alpha}^2}{\sqrt{3}} + \tilde \lambda_8 \frac{4 \sqrt{2} \tilde v_\xi c_{\tilde \alpha}^2}{\sqrt{3}} + \tilde \lambda_{10} \frac{\sqrt{2} \tilde v_\chi c_{\tilde \alpha}^2}{\sqrt{3}} + \tilde M_1^\prime \frac{s_{\tilde \alpha}^2}{2 \sqrt{6}} -\tilde M_1 \frac{s_{\tilde \alpha}^2}{2 \sqrt{6}}
\end{split}
\end{equation}

\begin{equation}
\begin{split}
g_{H_5^0 h_{\tilde \alpha} H_{\tilde \alpha}} & = -\frac{\tilde \lambda_3}{\sqrt{2}} \left( \frac{\tilde v_\chi s_{\tilde \alpha} c_{\tilde \alpha}}{\sqrt{3}} + \frac{\tilde v_\phi c_{ 2 \tilde \alpha} }{3} \right) + \tilde \lambda_4 \left( \frac{\tilde v_\phi c_{2 \tilde \alpha} }{3} + 2 s_{\tilde \alpha} c_{\tilde \alpha} \left( \frac{\tilde v_\chi}{\sqrt{3}} - \frac{\tilde v_\xi}{2 \sqrt{3}} \right) \right) + \tilde \lambda_5 \left( - 2 \frac{\tilde v_\chi s_{\tilde \alpha} c_{\tilde \alpha}}{\sqrt{6}} - \frac{\sqrt{2} c_{2 \tilde \alpha} \tilde v_\phi}{3} \right) \nonumber \\ & + \tilde \lambda_6 \left( \frac{2 \sqrt{2} \tilde v_\xi s_{\tilde \alpha} c_{\tilde \alpha}}{\sqrt{3}} +\frac{2 \sqrt{2} \tilde v_\phi c_{2 \tilde \alpha}}{3} \right) - \tilde \lambda_7 \frac{4 \sqrt{2} \tilde v_\chi c_{\tilde \alpha} s_{\tilde \alpha}}{\sqrt{3}} - \tilde \lambda_8 \frac{8 \sqrt{2} \tilde v_\xi c_{\tilde \alpha} s_{\tilde \alpha}}{\sqrt{3}} - \tilde \lambda_{10} \frac{2 \sqrt{2} \tilde v_\chi c_{\tilde \alpha} s_{\tilde \alpha}}{\sqrt{3}} + \tilde M_1^\prime \frac{s_{\tilde \alpha} c_{\tilde \alpha}}{\sqrt{6}} -\tilde M_1 \frac{s_{\tilde \alpha} c_{\tilde \alpha}}{\sqrt{6}}
\end{split}
\end{equation}

\section{Calculating the renormalization group equations}
\label{app:CEL}

We calculate the one-loop renormalization group equations (RGEs) in this paper using the formalism of Cheng, Eichten, and Lee~\cite{Cheng:1973nv}.  They considered a Lagrangian for nonabelian gauge fields $A_{\mu}^a$, real scalar fields $\phi_i$, and fermionic fields $\psi_{\alpha}$ of the form
\begin{equation}
	\mathcal{L} = -\frac{1}{4}F^a_{\mu\nu}F^{a\mu\nu} 
	+ \frac{1}{2}\left(\mathcal{D}_\mu \phi\right)_i\left(\mathcal{D}^\mu \phi\right)_i 
	+ i\,\overline{\psi}\,\gamma^\mu\mathcal{D}_\mu \psi - \overline{\psi}\,m_0\,\psi - \overline{\psi}\,h_i\,\psi\,\phi_i - V(\phi)\;,
	\label{eq:CELlagrangian}
\end{equation}
where the gauge field strength tensor and covariant derivatives are
\begin{align}
	\label{eq:gaugederivative}F^a_{\mu\nu} &= \partial_\mu A^a_\nu - \partial_\nu A^a_\mu - g\,C^{abc}\,A^b_\mu\,A^c_\nu,\\
	\label{eq:scalarderivative}\left(\mathcal{D}_\mu \phi\right)_i &= \partial_\mu \phi_i + i\,g\,\theta^a_{ij}\,\phi_j\,A^a_{\mu},\\
	\label{eq:fermionderivative}\left(\mathcal{D}_\mu \psi\right)_\alpha &= \partial_\mu \psi_\alpha + i\,g\,t^a_{\alpha\beta}\,\psi_\beta\,A^a_\mu.
\end{align}
Here $g$ is the gauge coupling, $\theta^a_{ij}$ and $t^a_{\alpha\beta}$ are the generators of the gauge group acting on the scalar and fermion representations, respectively, and $C^{abc}$ are the structure constants of the gauge group.
The fermion masses $m_0$ and Yukawa couplings $h_i$ are matrices in the space of fermions.  
The (quartic) scalar potential is given by
\begin{equation}
	V(\phi) = \sum_{ijkl} \frac{1}{4!}f_{ijkl}\,\phi_i\,\phi_j\,\phi_k\,\phi_l\;.
	\label{eq:CELpotential}
\end{equation}
The quartic scalar couplings $f_{ijkl}$ are defined to be symmetric under interchange of any pair of indices; after collecting terms in the scalar potential, they can be extracted using
\begin{equation}
	f_{ijkl} = 4! \times \frac{\mbox{coefficient of $\phi_i\,\phi_j\,\phi_k\,\phi_l$ in $V$}}{\mbox{number of permutations of ($ijkl$)}}\;.
\end{equation}

The trilinear couplings and quadratic mass-squared coefficients in Eq.~(\ref{eq:potential3}) can be integrated into this formalism by inserting one or two factors of a nondynamical scalar field $\phi_0$ that has no gauge or fermion couplings, e.g., $\mu^2 \phi_i \phi_i \to \mu^2 \phi_0 \phi_0 \phi_i \phi_i$.  The trilinear and quadratic coefficients can then be treated in the same way as the quartic coupling coefficients $f_{ijkl}$, setting one or two of $ijkl$ equal to $0$.

The RGEs for the quartic scalar couplings are given by Eq.~(2.8) of Ref.~\cite{Cheng:1973nv},
\begin{equation}\label{eq:CELrge}
	16\pi^2\,\frac{d f_{ijkl}}{dt} = \beta_{ijkl},
\end{equation}
with $t = \log \mu$ where $\mu$ is the energy scale and
\begin{equation}\label{eq:CELbeta}
	\beta_{ijkl} \equiv f_{ijmn}f_{mnkl} + f_{ikmn}f_{mnjl} + f_{il mn}f_{mnjk} - 12 g^2 S_2(S) f_{ijkl} + 3 g^4 A_{ijkl} + 8\Tr{h_i\,h_m}f_{mjkl} - 12H_{ijkl}\;.
\end{equation}
Repeated indices are to be summed over.  In this expression the first three terms come from one-loop diagrams with two quartic scalar vertices, the fourth term comes from diagrams in which an external leg is decorated with a gauge boson loop, the fifth term is a four-scalar coupling induced by a closed loop of gauge bosons, the sixth term comes from diagrams in which an external leg is decorated with a fermion loop, and the last term is a four-scalar coupling induced by a closed box of fermions (see Fig.~3 in Ref.~\cite{Cheng:1973nv}).
The new symbols in Eq.~(\ref{eq:CELbeta}) are defined as~\cite{Cheng:1973nv}:
\begin{equation}
	S_2(S)\delta_{ij} \equiv \left[\theta^a\theta^a\right]_{ij}\;,
	\label{eq:S2}
\end{equation}
\begin{equation}
	A_{ijkl} \equiv \{\theta^a,\,\theta^b\}_{ij}\{\theta^a,\,\theta^b\}_{kl} + \{\theta^a,\,\theta^b\}_{ik}\{\theta^a,\,\theta^b\}_{jl} + \{\theta^a,\,\theta^b\}_{il}\{\theta^a,\,\theta^b\}_{jk}\;,
	\label{eq:Aijkl}
\end{equation}
with repeated gauge indices summed over, and
\begin{equation}
	H_{ijkl} \equiv \frac{1}{3!}\Tr{h_ih_j\{h_k,\,h_l\} + h_ih_k\{h_j,\,h_l\} + h_ih_l\{h_j,\,h_k\}}.
\end{equation}

The formalism in Ref.~\cite{Cheng:1973nv} assumes a single gauge group and a single representation containing all the scalars.  This can be straightforwardly generalized to our theory in which the scalars transform under SU(2)$_L \times$U(1)$_Y$ as a doublet and two triplets as follows.  We first write out all the scalar fields in terms of their real components, using $\varphi_1 = (\phi_1 + i \phi_2)/\sqrt{2}$ for the complex scalars.  The covariant derivative for the scalars can then be written as
\begin{equation}
	\left(\mathcal{D}_\mu \phi\right)_i = \partial_\mu \phi_i + i g \, \theta^a_{ij} \phi_j W^a_{\mu} \
		+ i g^{\prime} \frac{Y_{ii}}{2} \phi_i B_{\mu},
\end{equation}
where $g$ and $g^{\prime}$ are now the SU(2)$_L$ and U(1)$_Y$ gauge couplings and $\theta^a_{ij}$ and $Y_{ii}/2$ are the SU(2)$_L$ and U(1)$_Y$ generators written as big matrices in the space of the 13 real scalars $\phi_i$ in our model (plus one nondynamical scalar field $\phi_0$).

Equation~(\ref{eq:CELbeta}) must then be modified slightly to take into account the two gauge groups:
\begin{eqnarray}
	\beta_{ijkl} &=& f_{ijmn}f_{mnkl} + f_{ikmn}f_{mnjl} + f_{il mn}f_{mnjk}
	- 12 g^{\prime 2} S^{\prime}_2(S) f_{ijkl} - 12 g^2 S_2(S) f_{ijkl} \nonumber \\
	&& + 3 \bar A_{ijkl} + 8\Tr{h_i\,h_m}f_{mjkl} - 12H_{ijkl}.
\label{eq:newbeta}
\end{eqnarray}
The new gauge terms are given as follows.  The $S_2^{\prime}(S)$ term comes from diagrams in which a U(1)$_Y$ gauge boson loop decorates one of the external scalar legs.  Using Eq.~(\ref{eq:S2}) with $\theta^a_{ij} = (Y_i/2) \delta_{ij}$, this term is given for each $ijkl$ by
\begin{eqnarray}
	-12 g^{\prime 2} S^{\prime}_2(S)
	&=& -3 g^{\prime 2} \sum_{\rm legs} \left[\frac{Y}{2}\frac{Y}{2}\right]_{\rm leg} \nonumber\\
	&=& -3 g^{\prime 2} \left[\left(\frac{Y_i}{2}\right)^2 + \left(\frac{Y_j}{2}\right)^2 + \left(\frac{Y_k}{2}\right)^2 + \left(\frac{Y_l}{2}\right)^2\right].
\end{eqnarray}
The $S_2(S)$ term comes from diagrams in which an SU(2)$_L$ gauge boson loop decorates one of the external scalar legs.  It will have different values depending on the SU(2)$_L$ representation of the scalar on each leg.  Using the SU(2)$_L$ generators for doublets and triplets, we obtain from Eq.~(\ref{eq:S2}) for each leg
\begin{equation}
	S_2(S)_{\rm leg} \delta_{ij} = \left[\theta^a\theta^a\right]_{ij} = \left\{ 
	\begin{array}{ll} 
	(3/4) \delta_{ij} & {\rm doublet} \\
	2 \delta_{ij} & {\rm triplet} \\
	{}[(n^2-1)/4] \delta_{ij} & n{\rm plet}
	\end{array} \right.
\end{equation}
Summing over the four legs then gives, for each $ijkl$,
\begin{equation}
	-12 g^2 S_2(S) = -3 g^2 \left[S_2(S)_i + S_2(S)_j + S_2(S)_k + S_2(S)_l\right] 
	= -\frac{3}{4} g^2 \left(n_i^2 + n_j^2 + n_k^2 + n_l^2 - 4\right),
\end{equation}
where $n_i = 2T_i + 1$ is the dimensionality of the SU(2)$_L$ representation of the $i$th leg.

The $3 \bar A_{ijkl}$ term in Eq.~(\ref{eq:newbeta}) yields terms in the RGEs of order $g^4$, $g^{\prime 4}$, and $g^2 g^{\prime 2}$.  The couplings that give rise to these terms are the quartic scalar-scalar-vector-vector vertices, which can be found by examining the anti-commutation relations among the generators of the relevant gauge groups. We derive the form of $\bar A_{ijkl}$ as follows. First, starting from Eq.~(\ref{eq:Aijkl}) we absorb the gauge coupling into the generators and define
\begin{equation}
	\bar \theta^1 = g t^1, \qquad
	\bar \theta^2 = g t^2, \qquad
	\bar \theta^3 = g t^3, \qquad
	\bar \theta^4 = g^{\prime} \frac{Y}{2} I_{n\times n},
\end{equation}
where $t^a$ are the appropriate SU(2)$_L$ generators acting on the relevant subspaces of the scalars and $I_{n \times n}$ is the unit matrix on the subspace of scalars with a common hypercharge.  
Then,
\begin{equation}
	\bar A_{ijkl} \equiv \{\bar \theta^a,\,\bar \theta^b\}_{ij}\{\bar \theta^a,\,\bar \theta^b\}_{kl} 
	+ \{\bar \theta^a,\,\bar \theta^b\}_{ik}\{\bar \theta^a,\,\bar \theta^b\}_{jl} 
	+ \{\bar \theta^a,\,\bar \theta^b\}_{il}\{\bar \theta^a,\,\bar \theta^b\}_{jk}\;,
	\label{eq:barAijkl}
\end{equation}
To actually calculate this, we write 
\begin{equation}
	\bar A_{ijkl} = \sum_{a,b = 1}^4 \alpha^{ab}_{ij} \alpha^{ab}_{kl} 
		+ \alpha^{ab}_{ik} \alpha^{ab}_{jl} + \alpha^{ab}_{il} \alpha^{ab}_{kj}\;,
\end{equation}
where for a real scalar multiplet the gauge-covariant terms yield,
\begin{equation}
	\Phi^T \left(\bar \theta^a \bar \theta^b + \bar \theta^b \bar \theta^a\right)\Phi = \sum_{i,j} \phi_i \phi_j \alpha^{ab}_{ij}\;,
\end{equation}
and for a complex scalar multiplet they give,
\begin{equation}
	2\,\Phi\ct\left(\bar \theta^a \bar \theta^b + \bar \theta^b \bar \theta^a\right)\Phi = \sum_{i,j} \phi_i \phi_j \alpha^{ab}_{ij}\;.
\end{equation}
Note that $\alpha^{ab}_{ij}$ is symmetric under interchange of $i$ and $j$; care must be taken with factors of two in extracting the $\alpha^{ab}_{ij}$ from terms involving two different real scalar fields.

Finally for the fermion contributions, it is most straightforward to separate the contributions into a sum of terms each involving only leptons, down-type quarks, or up-type quarks.  In our model only the SU(2)$_L$ doublet couples to fermions, as in the SM, and we can write the Yukawa matrices in the fermion mass basis as 
\begin{equation}
\begin{aligned}
	Y_i^u &= \left(\begin{matrix}y_u & 0 & 0\\0 & y_c & 0\\ 0 & 0 & y_t\end{matrix}\right),&
	Y_i^d &= \left(\begin{matrix}y_d & 0 & 0\\0 & y_s & 0\\ 0 & 0 & y_b\end{matrix}\right),&
	Y_i^e &= \left(\begin{matrix}y_e & 0 & 0\\0 & y_\mu & 0\\ 0 & 0 & y_\tau\end{matrix}\right)\;,
\end{aligned}
\end{equation}
for $i$ being one of the four real fields in the scalar doublet, and
\begin{equation}
	Y_i^f = \left(\begin{matrix}0 & 0 & 0\\0 & 0 & 0\\ 0 & 0 & 0\end{matrix}\right)
\end{equation}
for $f \in \{u, d, e\}$ and $i$ being any other scalar field. 

The contribution from diagrams in which an external leg is decorated with a fermion loop is then given for each $ijkl$ by
\begin{equation}
	8\Tr{h_i\,h_m} f_{mjkl} = \left(\Upsilon_i + \Upsilon_j + \Upsilon_k + \Upsilon_l\right) f_{ijkl}\;,
\end{equation}
where
\begin{equation}
	\Upsilon_m = \Tr{\sum_{f \in \{u, d, e\}} N_c^f Y^f_m Y^f_m}\;,
\end{equation}
with $N_c^f$ being the number of colors of fermion type $f$. 

The contribution from the fermion box diagram will be
\begin{equation}
	-12 H_{ijkl} = -4\left(\delta_{ij}\delta_{kl} + \delta_{ik}\delta_{jl} + \delta_{il}\delta_{jk}\right)\frac{1}{\mbox{\footnotesize\# of permutations of ($ijkl$)}}\Tr{\sum_{f \in \{u, d, e\}}\sum_{\substack{\mbox{\footnotesize permutations}\\\mbox{\footnotesize of}\\(ijkl)}} N_c^f Y_i^f Y_j^f Y_k^f Y_l^f}\;.
\end{equation}

This yields the RGEs for the coefficients $f_{ijkl}$ defined in Eq.~(\ref{eq:CELpotential}).  To obtain the RGEs for the individual quartic couplings $\tilde \lambda_i$ in Eq.~(\ref{eq:potential3}), one can write the $f_{ijkl}$ as linear combinations of the $\tilde \lambda_i$ and solve the set of linear equations.  The multiple redundant solutions for each $\tilde \lambda_i$ can be used as a check of the algebraic implementation.



\end{document}